\documentclass[aps,pra,onecolumn,showpacs,prabib]{revtex4}
\usepackage{graphicx}
\usepackage{amsmath}
\usepackage{amssymb,wasysym}
\usepackage{amsfonts}
\usepackage{bm,verbatim}

\newcommand{\bea}{\begin{eqnarray}}
\newcommand{\eea}{\end{eqnarray}}
\newcommand{\be}{\begin{equation}}
\newcommand{\ee}{\end{equation}}
\newcommand{\ds}{\displaystyle}
\newcommand{\rr}{\mathbf{r}}
\newcommand{\kk}{\mathbf{k}}
\newcommand{\KK}{\mathbf{K}}
\newcommand{\qq}{\mathbf{q}}
\newcommand{\uu}{\mathbf{u}}
\newcommand{\RR}{\mathbf{R}}
\newcommand{\CC}{\mathbf{C}}
\newcommand{\vn}{\mathbf{0}}
\newcommand{\Oom}{\mathbf{\Omega}}
\newcommand{\ra}{\rangle}
\newcommand{\la}{\langle}
\newcommand{\si}{\sigma}
\newcommand{\sip}{{\sigma'}}
\newcommand{\ktyp}{k_{\rm typ}}
\newcommand{\up}{\uparrow}
\newcommand{\down}{\downarrow}
\newcommand{\rhob}{\mbox{\boldmath$\rho$}}
\newcommand{\rhobs}{\mbox{\scriptsize\boldmath$\rho$}}
\newcommand{\epsk}{\epsilon_{\mathbf{k}}}
\newcommand{\epsq}{\epsilon_{\mathbf{q}}}
\newcommand{\cA}{\mathcal{A}}

\newcounter{fnnumberbis}
\newcounter{fnnumberter}
\newcounter{fnnumberthesechaptroissectrois}

\begin{document}

\title{Exact relations for quantum-mechanical few-body and many-body problems
\\ with short-range interactions in two and three dimensions}

\author{F\'elix Werner}
\affiliation{Department of Physics, University of Massachusetts,
Amherst, MA 01003, USA}

\author{Yvan Castin}
\affiliation{Laboratoire Kastler Brossel, \'Ecole Normale
Sup\'erieure, UPMC and CNRS, 24 rue Lhomond, 75231 Paris Cedex 05, France}

\begin{abstract}
We derive relations between various observables for $N$ particles 
with zero-range or short-range interactions, in continuous space or on a lattice, in two or three dimensions, in an arbitrary external potential.
Some of our results generalize known relations between large-momentum behavior of the momentum distribution,
short-distance behavior of the pair correlation function and of the one-body density matrix, derivative of the energy with respect to the scattering length or to 
time, and the norm of the regular part of the wavefunction; in the case of finite-range interactions, the interaction energy
is also related to $dE/da$.  The expression relating the energy to a functional of the momentum distribution
is also generalized, and is found to break down for Efimov states with zero-range interactions, due to 
a subleading oscillating tail in the momentum distribution.
We also obtain new expressions for the derivative of the energy of a universal state with respect to the effective range,
the derivative of the energy of an efimovian state with respect to the three-body parameter,
and the second order derivative of the energy with respect to the inverse (or the logarithm in the two-dimensional case) of the scattering length.
The latter is negative at fixed entropy.  We use exact relations to compute corrections to exactly solvable three-body problems
and find agreement with available numerics.  For the unitary gas, we compare exact relations to existing fixed-node Monte-Carlo data, 
and we test, with existing Quantum Monte Carlo results on different finite range models, our prediction that the leading deviation of the critical temperature 
from its zero range value is linear in the interaction effective range $r_e$ with a model independent numerical coefficient.
\end{abstract}

\pacs{}
\date{\today}

\maketitle

\section{Introduction}

The experimental breakthroughs of 1995 having led to the first realization
of a Bose-Einstein condensate in an atomic vapor
\cite{Cornell_bec,Ketterle_bec,Hulet_bec}
have opened the era of experimental studies of ultracold gases with 
non-negligible or even strong interactions, in dimension lower or equal to three~\cite{RevueBlochDalibard,RevueTrentoFermions, HouchesBEC,HouchesLowD,Varenna}.
In these systems, the thermal de Broglie wavelength and the mean distance
between atoms are much larger than the range of the interaction potential. 
This so-called {\sl zero-range} 
limit has interesting
universal properties: Several quantities such as the thermodynamic 
functions of the gas depend on the interaction potential only through
the scattering length $a$, a length characterizing the low-energy
scattering amplitude of two atoms.

This universality property holds for the weakly repulsive
Bose gas in three dimensions \cite{LHY} up to the order of expansion
in $(n a^3)^{1/2}$ corresponding to Bogoliubov theory 
\cite{Wu1959}, $n$ being the gas density. 
It is also true for the weakly repulsive 
Bose gas in two dimensions \cite{Schick,Popov,Lieb2D}, even at the next order
beyond Bogoliubov theory \cite{MoraCastin2D}. For $a$ much larger than the range of the interaction potential, the ground state of $N$ 
bosons in two dimensions is 
a universal $N$-body bound state~\cite{BruchTjon3bosons2D,Fedorov3bosons2D,Platter4bosons2D,HammerSon,Lee2D}.
In one dimension, the universality holds for 
any scattering length,
and 
the Bose gas with zero-range interaction
 is exactly solvable
by the Bethe ansatz both in the repulsive case \cite{LiebLiniger}
and in the attractive case \cite{Herzog,Caux}.

For spin 1/2 fermions, the universality properties are expected to be even stronger.
The weakly interacting regimes in 3D 
\cite{LeeYang,HuangYang,Abrikosov,Galitski,
Seiringer_fermions}
and in 2D \cite{Bloom} are universal, and the 1D case is also
solvable by Bethe ansatz for an arbitrary interaction strength
\cite{GaudinArticle,GaudinLivre}. 
Universality is expected to hold for an arbitrary scattering length 
even in $3D$ (see however \cite{Teta}),
as was recently tested by experimental studies on the BEC-BCS crossover
using a Feshbach resonance, see e.~g.~\cite{Varenna,HuletClosedChannel, HuletPolarized,ThomasVirielExp,thomas_entropie_PRL,thomas_entropie_JLTP,JinPhotoemission,KetterleGap,GrimmModesTfinie},
and in agreement with unbiased Quantum Monte Carlo calculations \cite{bulgacQMC,zhenyaPRL,Juillet,BulgacCrossover,zhenyas_crossover,ChangAFMC}. 
A similar universal crossover from BEC to BCS is expected in $2D$ when the parameter $\ln (k_F a)$ varies from $-\infty$ to $+\infty$~\cite{Petrov2D,Miyake2D,Randeria2D,Zwerger2D,Leyronas4corps}.
Universality is also expected for mixtures
in $2D$~\cite{PetrovCristal,Leyronas4corps,LudoPedri}, and in $3D$ for Fermi-Fermi mixtures below a critical mass ratio~\cite{Efimov73,PetrovCristal,BaranovLoboShlyapMassesDiff}.

In the zero-range regime, it is intuitive that the short-range or high-momenta properties of the gas are dominated by two-body physics. 
For example the pair distribution function $g^{(2)}(\mathbf{r}_{12})$ 
of particles at distances $r_{12}$
much smaller than the de Broglie wavelength is expected 
to be proportional to the modulus squared of the zero-energy two-body
scattering wavefunction $\phi(r_{12})$, with a proportionality factor
$\Lambda_g$ depending on the many-body state of the gas.
Similarly the large momentum tail of the momentum distribution
$n(\mathbf{k})$, at 
wavevectors much larger than the inverse de Broglie wavelength, 
is expected to be proportional to the modulus squared of
the Fourier component of the zero energy scattering state 
$\tilde{\phi}(\mathbf{k})$, 
with a proportionality factor $\Lambda_n$ depending on the many-body
state of the gas: 
Whereas two colliding atoms in the gas
have a center of mass wavevector of the order of the inverse de Broglie
wavelength, their relative wavevector can access much larger values,
up to the inverse of the interaction range,
simply because the interaction potential has a width in the space
of relative momenta of the order of the inverse of its range in real space.

For these intuitive reasons, and with the notable exception of one-dimensional
systems, one expects that the mean interaction energy $E_{\rm int}$
of the gas, being
sensitive to the shape of $g^{(2)}$ at distances of the order
of the interaction range, is not universal, but diverges
in the zero-range limit; one also expects
that, apart from the 1D case,
the mean kinetic energy, being dominated by the large-momentum tail
of the momentum distribution, is not universal and diverges
in the zero-range limit, a well known fact
in the context of Bogoliubov theory for Bose gases
and of BCS theory for Fermi gases.
Since the total energy of the gas is universal, and $E_{\rm int}$
is proportional to $\Lambda_g$ while $E_{\rm kin}$ is proportional
to $\Lambda_n$, one expects that there exists a simple relation
between $\Lambda_g$ and $\Lambda_n$.

The precise link between the pair distribution function, the tail
of the momentum distribution and the energy of the gas
was first established for one-dimensional systems.
In \cite{LiebLiniger} the value of the
pair distribution function for $r_{12}=0$
was expressed in terms of the derivative of the gas energy with respect
to the one-dimensional scattering length, thanks to the Hellmann-Feynman
theorem. In \cite{Olshanii_nk} the large momentum tail
of $n(k)$ was also related to this derivative of the energy,
by using a simple and general property of the Fourier transform of a function
having discontinuous derivatives in isolated points.

In three dimensions, results in these directions were first obtained
for weakly interacting gases. For the weakly interacting Bose gas,
Bogoliubov theory contains the expected properties, in particular
on the short distance behavior of the pair distribution function
\cite{Huang_article,Holzmann,Glauber} 
and the fact that the momentum distribution
has a slowly decreasing tail.
For the weakly interacting two-component Fermi gas, it was shown that 
the BCS anomalous average (or pairing field) 
$\langle \hat{\psi}_\uparrow(\mathbf{r}_1) 
\hat{\psi}_\downarrow(\mathbf{r}_2)\rangle$ behaves at short
distances as the zero-energy two-body scattering wavefunction $\phi(r_{12})$
\cite{Bruun},
 resulting in a $g^{(2)}$ function indeed proportional
to $|\phi(r_{12})|^2$ at short distances. It was however understood
later that the corresponding proportionality factor $\Lambda_g$
predicted by BCS theory is incorrect \cite{CarusottoCastin},
e.g.\ at zero temperature the BCS prediction drops exponentially with $1/a$
in the non-interacting limit $a\to 0^-$, whereas the correct result
drops as a power law in $a$.

More recently, in a series of two articles \cite{TanEnergetics,TanLargeMomentum},
explicit expressions for the proportionality factors
$\Lambda_g$ and $\Lambda_n$ were obtained in terms of the derivative
of the gas energy with respect to the inverse scattering length, for a two-component interacting Fermi
gas in three dimensions, for an arbitrary value of the scattering length,
that is, not restricting to the weakly interacting limit. 
Later on, these results were rederived in \cite{Braaten,BraatenLong,ZhangLeggettUniv},
and also in \cite{WernerTarruellCastin} with very elementary methods
building on the intuition that $g^{(2)}\propto |\phi(r_{12})|^2$
at short distances and $n(k)\propto |\tilde{\phi}(k)|^2$
at large momenta. These relations were recently tested by
numerical four-body calculations \cite{BlumeRelations}.
An explicit relation between $\Lambda_g$ and the interaction energy was derived in \cite{ZhangLeggettUniv}. Another fundamental relation discovered in \cite{TanEnergetics} and recently generalized in \cite{CombescotC} to bosons, to Fermi-Bose mixtures and to fermions in 2D,
expresses the total energy as a functional of the momentum distribution and the spatial density. 

In the present work we derive generalizations of the relations
of \cite{LiebLiniger,Olshanii_nk,TanEnergetics,TanLargeMomentum,ZhangLeggettUniv,CombescotC} to two dimensional gases, 
to the general
case of a mixture of an arbitrary number of atomic species and spin
component, and to the case of a small but non-zero interaction range (both on a lattice and in continuous space).
We also find entirely new results for the first order derivative of the energy with respect to the effective range and, in presence of the Efimov effect, with respect to the three-body parameter, as well as the second order derivative with respect to the scattering length.

The article is organized as follows.
In Section \ref{sec:fermions} we treat
in detail
the case of two-component Fermi gases.
Relations holding for any system eigenstate for zero-range interactions are derived in Section \ref{sec:ZR} and summarized in Table~\ref{tab:fermions}. We then consider lattice models (Tab.~\ref{tab:latt}, Sec.~\ref{sec:latt}) and finite-range models in continuous space (Tab.~\ref{tab:V(r)}, Sec.~\ref{sec:V(r)}). In Section~\ref{sec:re} we derive a model-independent expression for the correction to the energy due to a finite range or a finite effective range of the interaction. The generalization to 
thermodynamic equilibrium, where the system is in a statistical mixture of eigenstates, is discussed in Section \ref{subsec:finiteT}.
In Section \ref{sec:bosons} we turn to the case of spinless bosons.
We focus on the case of zero-range interactions where, in $3D$, the Efimov effect leads to modifications or even breakdown of some relations, and to the appearance of a new relation.
Then we show briefly in Section \ref{sec:melange}
how to treat the case of an arbitrary mixture and present results for zero-range interactions (Tab.~\ref{tab:melange}).
Finally we present applications of exact relations: For three particles we compute corrections to exactly solvable cases and compare them to numerics (Sec. \ref{sec:appl_3body}), and we check that exact relations are satisfied by existing fixed-node Monte-Carlo data for correlation functions of the unitary gas.
We expect from our expression for the leading finite-range correction to the energy that the leading finite-range correction to the critical temperature 
in the BEC-BCS crossover depends only on the effective range of the interaction, an expectation that we test
against the Quantum Monte Carlo calculations of \cite{zhenyaPRL,zhenyas_crossover}.
We conclude in Section \ref{sec:conclusion}.

\section{Two-component fermions} \label{sec:fermions}

In this Section we consider spin-$1/2$ fermions. For a fixed number $N_\sigma$ of particles in each spin state $\sigma=\uparrow,\downarrow$, one can consider that particles $1,\ldots,N_\uparrow$ have a spin~$\uparrow$ and particles $N_\uparrow+1,\ldots,N_\uparrow+N_\downarrow=N$ have a spin~$\downarrow$, i.e. the wavefunction $\psi(\rr_1,\ldots,\rr_N)$ changes sign when one exchanges the positions of two particles having the same spin~\footnote{ The corresponding state vector is $|\Psi\rangle=
[N!/(N_\uparrow!N_\downarrow!)]^{1/2} \hat{A} \left( |\uparrow,\ldots,\uparrow,\downarrow,\ldots,\downarrow\rangle \otimes |\psi\rangle \right)$ where there are $N_\uparrow$ spins $\uparrow$ and $N_\downarrow$ spins $\downarrow$,
and the operator $\hat{A}$ antisymmetrizes with respect to all particles. The wavefunction $\psi(\rr_1,\ldots,\rr_N)$ is then proportional to $\left( \langle \uparrow,\ldots,\uparrow,\downarrow,\ldots,\downarrow | \otimes \langle \rr_1,\ldots,\rr_N | \right) |\Psi\rangle$.}.

\begin{table}
\begin{tabular}{|c|c|}
\hline   
Three dimensions & Two dimensions \\
\hline &  \\
$\ds \psi(\rr_1,\ldots,\rr_N)\underset{r_{ij}\to0}{=}
\left( \frac{1}{r_{ij}}-\frac{1}{a} \right) \, A_{ij}\left(
\mathbf{R}_{ij}, (\mathbf{r}_k)_{k\neq i,j}
\right)
+O(r_{ij})$
&
$\ds \psi(\rr_1,\ldots,\rr_N)\underset{r_{ij}\to0}{=}
\ln( r_{ij}/a ) \, A_{ij}\left(
\mathbf{R}_{ij}, (\mathbf{r}_k)_{k\neq i,j}
\right)
+O(r_{ij})$
\\
& \\
\hline 
\multicolumn{2}{|c|}{} \\
\multicolumn{2}{|c|}{$\ds
( A^{(1)},A^{(2)} )\equiv \sum_{i<j} \int \Big(\prod_{k\neq i,j} d^d\! r_k \Big) d^d\! R_{ij}
A^{(1)}_{ij}(\mathbf{R}_{ij}, (\mathbf{r}_k)_{k\neq i,j})^*
A^{(2)}_{ij}(\mathbf{R}_{ij}, (\mathbf{r}_k)_{k\neq i,j})$ }
\\ \multicolumn{2}{|c|}{} \\
\hline
\end{tabular}
\caption{Notation for the regular part $A$ of the many-body wavefunction appearing in the contact conditions (first line) and for the scalar product between such regular parts (second line).
\label{tab:notations}}
\begin{tabular}{|c|c|c|}
\hline  
&  Three dimensions & Two dimensions  \\
 \hline & & \\
 1 &
$\displaystyle\frac{dE}{d(-1/a)} = \frac{4\pi\hbar^2}{m} (A,A)$ &
$\ds\frac{dE}{d(\ln a)} = \frac{2\pi\hbar^2}{m} (A,A)$ 
  \\
& & \\
\hline & & \\
2 &
$\ds C\equiv {\displaystyle \lim_{k\to +\infty}} 
k^4 n_\sigma(\kk) = \frac{4\pi m}{\hbar^2}
\frac{dE}{d(-1/a)} $&
$\ds C\equiv {\displaystyle\lim_{k\to +\infty}} k^4 n_\sigma(\kk) = \frac{2\pi m}{\hbar^2}
\frac{dE}{d(\ln a)}$ \\
& & \\
\hline & & 
\\
3 &
$\ds
\int d^3R \,
 g_{\uparrow \downarrow}^{(2)} \left(\mathbf{R}+\frac{\mathbf{r}}{2},
\mathbf{R}-\frac{\mathbf{r}}{2}\right) 
\underset{r\to0}{\sim}
\frac{C}{(4\pi)^2}
\frac{1}{r^2}
$
&
$\ds
\int d^2R \,
 g_{\uparrow \downarrow}^{(2)} \left(\mathbf{R}+\frac{\mathbf{r}}{2},
\mathbf{R}-\frac{\mathbf{r}}{2}\right)
\underset{r\to0}{\sim}
\frac{C}{(2\pi)^2}
\ln^2 r$
\\
& & \\
\hline & &  \\
4 &
$\ds E - E_{\rm trap}  = \frac{\hbar^2 C}{4\pi m a}  $ &
$\ds E - E_{\rm trap}  = \lim_{\Lambda\to\infty}\left[-\frac{\hbar^2 C}{2\pi m} \ln \left(\frac{a \Lambda e^\gamma}{2}\right) \right.
$
\\
& & \\ &
$\ds +\sum_{\sigma} \int \frac{d^3\!k}{(2\pi)^3}  \frac{\hbar^2 k^2}{2m} 
\left[n_\sigma(\kk) - \frac{C}{k^4}\right]$
&
$\ds  +\sum_{\sigma} \left. \int_{k<\Lambda} \frac{d^2\!k}{(2\pi)^2}  \frac{\hbar^2 k^2}{2m} 
n_\sigma(\kk) \right]$
\\
&&\\
\hline &&  \\
5 &
$\ds \int d^3R \,
 g_{\sigma \sigma}^{(1)} \left(\mathbf{R}+\frac{\mathbf{r}}{2},
\mathbf{R}-\frac{\mathbf{r}}{2}\right)
\underset{r\to0}{=}
N_\sigma -\frac{C}{8\pi}\, r + O(r^2)$
&
$\ds \int d^2R \,
 g_{\sigma \sigma}^{(1)} \left(\mathbf{R}+\frac{\mathbf{r}}{2},
\mathbf{R}-\frac{\mathbf{r}}{2}\right)
\underset{r\to0}{=}
N_\sigma +\frac{C}{4\pi}\, r^2\ln r + O(r^2)$
\\ && \\
\hline
&
&
\\ 6 &
$\ds \frac{1}{3} \sum_{i=1}^3  \sum_\si \int d^3R \,
 g_{\sigma \sigma}^{(1)} \left(\mathbf{R}+\frac{r {\bf u_i}}{2},
\mathbf{R}-\frac{r {\bf u_i}}{2}\right)
\underset{r\to 0}{=}
N$
&
$\ds \frac{1}{2} \sum_{i=1}^2 \sum_\si
 \int d^2R \,
 g_{\sigma \sigma}^{(1)} \left(\mathbf{R}+\frac{r {\bf u_i}}{2},
\mathbf{R}-\frac{r {\bf u_i}}{2}\right)
\underset{r\to 0}{=}
N
$
\\&
$\ds -\frac{C}{4\pi}r -\frac{m}{3\hbar^2}\left(
E-E_{\rm trap} - \frac{\hbar^2 C}{4\pi m a}
\right) r^2 + o(r^2)$
&
$\ds +\frac{C}{4\pi}r^2
\left[\ln\left(\frac{r}{a}\right)+\frac{\mathcal{F}}{32}\right]
-\frac{m}{2\hbar^2}\left(
E-E_{\rm trap}
\right) r^2 + o(r^2)$
\\& &
\\ \hline
& &
 \\7 &
$\ds\frac{1}{2} \frac{d^2E}{d(-1/a)^2}
= \left(\frac{4\pi\hbar^2}{m}\right)^2 \sum_{n,E_{n}\neq E} 
\frac{|(A^{(n)},A^{})|^2}{E-E_{n}}$
&
$\ds\frac{1}{2} \frac{d^2E}{d(\ln a)^2}
= \left(\frac{2\pi\hbar^2}{m}\right)^2 \sum_{n,E_n\neq E} 
\frac{|(A^{(n)},A^{})|^2}{E-E_{n}} $
\\&
& \\
\hline
 & 
& \\
8 &
$\ds\left(\frac{d^2F}{d(-1/a)^2}\right)_T < 0 $,\ \ \ 
$\ds\left(\frac{d^2E}{d(-1/a)^2}\right)_{\!S} < 0 $
&
$\ds\left(\frac{d^2F}{d(\ln a)^2}\right)_T < 0 $,\ \ \ 
$\ds\left(\frac{d^2E}{d(\ln a)^2}\right)_{\!S} < 0 $

\\
& &\\
\hline
 & & \\
9 &
$\ds\frac{dE}{dt} = \frac{\hbar^2 C}{4\pi m} \frac{d(-1/a)}{dt}+
\Big<\sum_{i=1}^N\partial_t U(\rr_i,t)\Big>$
&
$\ds\frac{dE}{dt} = \frac{\hbar^2 C}{2\pi m} \frac{d(\ln a)}{dt}+
\Big<\sum_{i=1}^N\partial_t U(\rr_i,t)\Big>$
\\
& & \\
\hline
\end{tabular}

\caption{Relations for two-component fermions with zero-range interactions. The regular part $A$ is defined in Table \ref{tab:notations}. Lines 1-7
hold for any eigenstate, and can be generalized to finite temperature by taking a thermal average in the canonical ensemble and by taking the derivatives of $E$ with respect to $a$ at constant entropy $S$.
Line 8 holds in the canonical ensemble.
Line 9 holds for any time-dependence of scattering length and trapping potential and any corresponding time-dependent statistical mixture .
\label{tab:fermions}}
\end{table}

\begin{table}
\begin{tabular}{|c|c|c|}
\hline
& & \\
1 & $\ds C\equiv\frac{4\pi m}{\hbar^2}\frac{dE}{d(-1/a)}$
& $\ds C\equiv\frac{2\pi m}{\hbar^2}\frac{dE}{d(\ln a)}$
\\
& & \\
\hline
& & \\
2
&$\ds\frac{dE}{d(-1/a)}=\frac{4\pi\hbar^2}{m}(A,A)$
&$\ds\frac{dE}{d(\ln a)}=\frac{2\pi\hbar^2}{m}(A,A)$
\\
& & \\
\hline
&  \multicolumn{2}{|c|}{} \\
3
& \multicolumn{2}{|c|}{$\ds E_{\rm int}=\left(\frac{\hbar^2}{m}\right)^2 \frac{C}{g_0}$}
\\
&  \multicolumn{2}{|c|}{} \\
\hline
& & \\
4
&$\ds E - E_{\rm trap}  = \frac{\hbar^2 C}{4\pi m a}$
&$\ds E - E_{\rm trap}  = \lim_{q\to0}\Bigg\{-\frac{\hbar^2 C}{2\pi m} \ln \left(\frac{a q e^\gamma}{2}\right)$
\\
&$\ds +\sum_{\sigma} \int_D \frac{d^3\!k}{(2\pi)^3}  \epsk
\left[n_\sigma(\kk) - C\left(\frac{\hbar^2}{2m\epsk}\right)^2\right]$ &
$\ds +\sum_{\sigma} \int_D \frac{d^2\!k}{(2\pi)^2}  \epsk
\left[n_\sigma(\kk) - C\frac{\hbar^2}{2m\epsk}\mathcal{P}\frac{\hbar^2}{2m(\epsk-\epsq)}\right]\Bigg\}$
 \\
& & \\
\hline
& \multicolumn{2}{|c|}{} \\
5
&
 \multicolumn{2}{|c|}{$\ds\frac{1}{2} \frac{d^2E}{dg_0^2}
= |\phi(\vn)|^4 \sum_{n,E_{n}\neq E} 
\frac{|(A^{(n)},A^{})|^2}{E-E_{n}}$}
\\
&  \multicolumn{2}{|c|}{} \\
\hline
& \multicolumn{2}{|c|}{} \\
6
&
 \multicolumn{2}{|c|}{$\ds \left(\frac{d^2F}{dg_0^2}\right)_T <0$,\ \ \ 
$\ds \left(\frac{d^2E}{dg_0^2}\right)_S <0$}
\\
&  \multicolumn{2}{|c|}{} \\
\hline
& & \\
7
&$\ds\sum_\RR b^3 g^{(2)}_{\up\down}(\RR,\RR)=\frac{C}{(4\pi)^2}|\phi(\vn)|^2$
& $\ds\sum_\RR b^2 g^{(2)}_{\up\down}(\RR,\RR)=\frac{C}{(2\pi)^2}|\phi(\vn)|^2$
\\
& & \\
\hline
  \multicolumn{3}{|c|}{In the zero-range regime $\ktyp b\ll1$}
\\ \hline
& & \\
 8
 & $\ds\sum_{\RR} b^3
 g_{\uparrow \downarrow}^{(2)} \left(\mathbf{R}+\frac{\mathbf{r}}{2},
\mathbf{R}-\frac{\mathbf{r}}{2}\right)
\simeq
\frac{C}{(4\pi)^2}|\phi(\rr)|^2$\ \ \ for $r\ll \ktyp^{-1}$
& $\ds\sum_{\RR} b^2
 g_{\uparrow \downarrow}^{(2)} \left(\mathbf{R}+\frac{\mathbf{r}}{2},
\mathbf{R}-\frac{\mathbf{r}}{2}\right)
\simeq
\frac{C}{(2\pi)^2}|\phi(\rr)|^2$\ \ \ for $r\ll \ktyp^{-1}$
\\
& &
\\ \hline
& \multicolumn{2}{|c|}{} \\
9 &
\multicolumn{2}{|c|}{$\ds n_\sigma(\kk)\simeq C \left(\frac{\hbar^2}{2m\epsk}\right)^2\ \ \ \ $ for $k\gg\ktyp$}
\\
&\multicolumn{2}{|c|}{}
\\ 
\hline
\end{tabular}
\caption{Relations for two-component fermions in a lattice model. $C$ is defined in line 1.\label{tab:latt}}
\end{table}
\begin{table}
\begin{tabular}{|c|c|c|}
\hline
& Three dimensions & Two dimensions \\
\hline
& & \\
1 & $\ds C\equiv\frac{4\pi m}{\hbar^2}\frac{dE}{d(-1/a)}$
& $\ds C\equiv\frac{2\pi m}{\hbar^2}\frac{dE}{d(\ln a)}$
\\
& & \\
\hline
& & \\
2 & $\ds E_{\rm int}=\frac{C}{(4\pi)^2}\int d^3r\,V(r) |\phi(r)|^2$
& $\ds E_{\rm int}=\frac{C}{(2\pi)^2}\int d^2r\,V(r) |\phi(r)|^2$
\\
& & \\
\hline
& & \\
3
& $\ds E-E_{\rm trap}=\frac{\hbar^2 C}{4\pi m a}$
& $\ds E-E_{\rm trap}=\lim_{R\to\infty}\Bigg\{\frac{\hbar^2 C}{2\pi m a}\ln\left(\frac{R}{a}\right)$
\\
& $\ds +\sum_{\si}\int \frac{d^3 k}{(2\pi)^3}\,\frac{\hbar^2 k^2}{2m}\left[n_\si(\kk)-\frac{C}{(4\pi)^2}|\tilde{\phi}'(k)|^2\right]$
& $\ds  +\sum_{\si}\int \frac{d^2 k}{(2\pi)^2}\,\frac{\hbar^2 k^2}{2m}\left[n_\si(\kk)-\frac{C}{(2\pi)^2}|\tilde{\phi}'_R(k)|^2\right]\Bigg\}$
\\
& & \\
\hline
 & \multicolumn{2}{|c|}{In the zero-range regime $\ktyp b\ll1$}
\\ \hline
& & \\
 5
 & $\ds\int d^3R\,
 g_{\uparrow \downarrow}^{(2)} \left(\mathbf{R}+\frac{\mathbf{r}}{2},
\mathbf{R}-\frac{\mathbf{r}}{2}\right)
\simeq
\frac{C}{(4\pi)^2}|\phi(\rr)|^2$\ \ \ for $r\ll \ktyp^{-1}$
& $\ds\int d^2R\,
 g_{\uparrow \downarrow}^{(2)} \left(\mathbf{R}+\frac{\mathbf{r}}{2},
\mathbf{R}-\frac{\mathbf{r}}{2}\right)
\simeq
\frac{C}{(2\pi)^2}|\phi(\rr)|^2$\ \ \ for $r\ll \ktyp^{-1}$
\\
& &
\\ \hline
& & \\
6 &
$\ds n_\si(\kk)\simeq\frac{C}{(4\pi)^2}|\tilde{\phi}(\kk)|^2$\ \ \ for $k\gg\ktyp$
&
$\ds n_\si(\kk)\simeq\frac{C}{(2\pi)^2}|\tilde{\phi}(\kk)|^2$\ \ \ for $k\gg\ktyp$
\\
& & 
\\ 
\hline
\end{tabular}
\caption{Relations for two-component fermions with a finite-range interaction potential $V(r)$ in continuous space. $C$ is defined in line 1.
\label{tab:V(r)}}
\end{table}

\begin{table}
\begin{tabular}{|c|c|}
\hline   & \\
Three dimensions & Two dimensions \\
& \\
\hline & \\
$\displaystyle \left(\frac{\partial E}{\partial(-1/a)}\right)_{\!\!R_t} = \frac{4\pi\hbar^2}{m} (A,A)$ &
$\ds\frac{dE}{d(\ln a)} = \frac{2\pi\hbar^2}{m} (A,A)$  \\
&\\
\hline & \\
$\ds C\equiv {\displaystyle \lim_{k\to +\infty}} 
k^4 n(k) = \frac{8\pi m}{\hbar^2}
\left(\frac{\partial E}{\partial(-1/a)}\right)_{\!\!R_t} $&
$\ds C\equiv {\displaystyle\lim_{k\to +\infty}} k^4 n(k) = \frac{4\pi m}{\hbar^2}
\frac{dE}{d(\ln a)}$ \\
&\\
\hline &  \\
$\ds
\int d^3\!R \,
 g^{(2)} \!\left(\mathbf{R}+\frac{\mathbf{r}}{2},
\mathbf{R}-\frac{\mathbf{r}}{2}\right) 
\underset{r\to0}{\sim}
\frac{C}{(4\pi)^2}
\frac{1}{r^2}$
&
$\ds
\int d^2\!R \,
 g^{(2)}\! \left(\mathbf{R}+\frac{\mathbf{r}}{2},
\mathbf{R}-\frac{\mathbf{r}}{2}\right)
\underset{r\to0}{\sim}
\frac{C}{(2\pi)^2}
\ln^2 r$
\\
&\\
\hline &  \\
$\ds E - E_{\rm trap} \stackrel{\mathrm{if}\, \exists\, \mathrm{lim}}{=}
\frac{\hbar^2 C}{8\pi m a}$
 &
$\ds E - E_{\rm trap}  = \lim_{\Lambda\to\infty}\left[-\frac{\hbar^2 C}{4\pi m} \ln \left(\frac{ a\Lambda e^\gamma}{2}\right) \right.
$
\\
$\ds +\lim_{\Lambda\to +\infty}  
\int_{k<\Lambda} \frac{d^3\!k}{(2\pi)^3}  \frac{\hbar^2 k^2}{2m}
\left[n(k)-\frac{C}{k^4}\right] $
&
$\ds \left. + \int_{k<\Lambda} \frac{d^2\!k}{(2\pi)^2}  \frac{\hbar^2 k^2}{2m} 
n(k)\right] $
\\
&\\
\hline &  \\
$\ds\frac{1}{2} \left(\frac{\partial^2E}{\partial(-1/a)^2}\right)_{\!\!R_t}
= \left(\frac{4\pi\hbar^2}{m}\right)^2 \sum_{n,E_n\neq E} 
\frac{|(A^{(n)},A)|^2}{E-E_n}$
&
$\ds\frac{1}{2} \frac{d^2E}{d(\ln a)^2}
= \left(\frac{2\pi\hbar^2}{m}\right)^2 \sum_{n,E_n\neq E} 
\frac{|(A^{(n)},A)|^2}{E-E_n} $
\\
& \\
\hline 
& \\
$\ds \left(\frac{\partial E}{\partial \ln R_t}\right)_a=\frac{\hbar^2}{m}\frac{\sqrt{3}}{32}|s_0|^2 N(N-1)(N-2)$
&
$\ds\left(\frac{d^2F}{d(\ln a)^2}\right)_T < 0 $
\\
& \\
& \\
$\ds \times\int d\CC\int d\rr_4\ldots d\rr_N\,|B(\CC,\rr_4,\ldots,\rr_N)|^2$
&
$\ds\left(\frac{d^2E}{d(\ln a)^2}\right)_S < 0 $
\\
& \\
\hline
\end{tabular}
\caption{Main results for spinless bosons in the limit of a zero range interaction. 
In three dimensions, the derivatives are taken for a fixed three-body parameter $R_t$.
As discussed in the text, in three dimensions, the relation between energy and momentum distribution is valid if the  large cut-off limit $\Lambda\to +\infty$ exists, which is not the case for Efimovian states
(i.e.\ eigenstates whose energy depends on $R_t$).
In the last relation in three-dimensions, $B$ is the three-body regular part defined in (\ref{eq:danilov}).
\label{tab:bosons}}
\end{table}

\begin{table}
\begin{tabular}{|c|c|}
\hline   & \\
Three dimensions & Two dimensions \\
& \\
\hline & \\
$\displaystyle\frac{\partial E}{\partial(-1/a_{\si\sip})} = \frac{2\pi\hbar^2}{\mu_{\si\sip}} (A,A)_{\si\sip}$ &
$\ds\frac{\partial E}{\partial(\ln a_{\si\sip})} = \frac{\pi\hbar^2}{\mu_{\si\sip}} (A,A)_{\si\sip}$  \\
&\\
\hline & \\
$\ds C_\si\equiv {\displaystyle \lim_{k\to +\infty}} 
k^4 n_\sigma(\kk) = 
\sum_{\sip}(1+\delta_{\si\sip})\frac{8\pi \mu_{\si\sip}}{\hbar^2}
\frac{\partial E}{\partial(-1/a_{\si \sip})} $&
$\ds C_\si\equiv {\displaystyle\lim_{k\to +\infty}} k^4 n_\sigma(\kk) = \sum_\sip(1+\delta_{\si\sip})\frac{4\pi \mu_{\si\sip}}{\hbar^2}
\frac{\partial E}{\partial(\ln a_{\si\sip})}$ \\
&\\
\hline &  \\
$\ds
\int d^3R \,
 g_{\si\sip}^{(2)} \left(\mathbf{R}+\frac{m_\sip}{m_\si+m_\sip}\mathbf{r},
\mathbf{R}-\frac{m_\si}{m_\si+m_\sip}\mathbf{r}\right) 
\underset{r\to0}{\sim}(1+\delta_{\si\sip})$

&
$\ds
\int d^2R \,
 g_{\si\sip}^{(2)} \left(\mathbf{R}+\frac{m_\sip}{m_\si+m_\sip}\mathbf{r},
\mathbf{R}-\frac{m_\si}{m_\si+m_\sip}\mathbf{r}\right)
\underset{r\to0}{\sim}(1+\delta_{\si\sip})
$
\\ & \\
$\ds \times \frac{\mu_{\si\sip}}{2\pi\hbar^2}\frac{\partial E}{\partial(-1/a_{\si\sip})}
\frac{1}{r^2}$ &
$\ds \times \frac{\mu_{\si\sip}}{\pi\hbar^2}\frac{\partial E}{\partial(\ln a_{\si\sip})}
\ln^2 r$
\\
&\\
\hline &  \\
$\ds E - E_{\rm trap}  = \sum_{\si\leq\sip} \frac{1}{a_{\si\sip}} \frac{\partial E}{\partial(-1/a_{\si\sip})}  $ &
$\ds E - E_{\rm trap}  = \lim_{\Lambda\to\infty}\left[-\sum_{\si\leq\sip}\frac{\partial E}{\partial(\ln a_{\si\sip})} \ln \left(\frac{a_{\si\sip} \Lambda e^\gamma}{2}\right) \right.
$
\\
& \\
$\ds +\sum_{\sigma} \int \frac{d^3\!k}{(2\pi)^3}  \frac{\hbar^2 k^2}{2m_\si} 
\left[n_\sigma(\kk) - \frac{C_\si}{k^4}\right]$
&
$\ds \left. +\sum_{\sigma} \int_{k<\Lambda} \frac{d^2\!k}{(2\pi)^2}  \frac{\hbar^2 k^2}{2m_\si} 
n_\sigma(\kk)\right]$
\\
&\\
\hline &  \\
$\ds\frac{1}{2} \frac{\partial^2E_n}{\partial(-1/a_{\si\sip})^2}
= \left(\frac{2\pi\hbar^2}{\mu_{\si\sip}}\right)^2 \sum_{n',E_{n'}\neq E_n} 
\frac{|(A^{(n')},A^{(n)})_{\si\sip}|^2}{E_n-E_{n'}}$
&
$\ds\frac{1}{2} \frac{\partial^2E_n}{\partial(\ln a_{\si\sip})^2}
= \left(\frac{\pi\hbar^2}{\mu_{\si\sip}}\right)^2 \sum_{n',E_{n'}\neq E_n} 
\frac{|(A^{(n')},A^{(n)})_{\si\sip}|^2}{E_n-E_{n'}} $
\\
& \\
\hline 
& \\
$\ds\left(\frac{\partial^2F}{\partial(-1/a_{\si\sip})^2}\right)_T < 0 $
&
$\ds\left(\frac{\partial^2F}{\partial(\ln a_{\si\sip})^2}\right)_T < 0 $
\\
& \\
\hline
& \\
$\ds\left(\frac{\partial^2E}{\partial(-1/a_{\si\sip})^2}\right)_S < 0 $
&
$\ds\left(\frac{\partial^2E}{\partial(\ln a_{\si\sip})^2}\right)_S < 0 $
\\
& \\
\hline
\end{tabular}
\caption{Main results for an arbitrary mixture with zero-range interactions.
In three dimensions, if the Efimov effect occurs, the derivatives must be taken for fixed three-body parameter(s) and the expression for $E$ in line~4 breaks down.
\label{tab:melange}}
\end{table}

\subsection{Models} \label{subsec:models}

Here we introduce the three models used in this work to model interparticle interactions.

\subsubsection{Zero-range model}

In this well-known model (see e.g. \cite{AlbeverioLivre,YvanHouchesBEC,PetrovJPhysB,Efimov, MaximLudo2D,RevueBraaten,YvanVarenna,WernerThese} and refs. therein)
the interaction potential is replaced by contact conditions on the many-body wavefunction: For any pair of particles $i\neq j$, there exists a function $A_{ij}$, hereafter called regular part of $\psi$, such that in $3D$
\be
 \psi(\rr_1,\ldots,\rr_N)\underset{r_{ij}\to0}{=}
\left( \frac{1}{r_{ij}}-\frac{1}{a} \right) \, A_{ij}\left(
\mathbf{R}_{ij}, (\mathbf{r}_k)_{k\neq i,j}
\right)
+O(r_{ij}),
\label{eq:CL_3D}
\ee
and in $2D$
\be
\ds \psi(\rr_1,\ldots,\rr_N)\underset{r_{ij}\to0}{=}
\ln( r_{ij}/a ) \, A_{ij}\left(
\mathbf{R}_{ij}, (\mathbf{r}_k)_{k\neq i,j}
\right)
+O(r_{ij}),
\label{eq:CL_2D}
\ee
where the limit of vanishing distance $r_{ij}$ between particles $i$ and $j$ is taken for a fixed position 
of their center of mass $\mathbf{R}_{ij}=(\rr_i+\rr_j)/2$
and fixed positions  of the remaining particles $(\rr_k)_{k\neq i,j}$.
Fermionic symmetry of course imposes $A_{ij}=0$ if particles $i$ and $j$ have the same spin.
When none of the $\rr_i$'s coincide, there is no interaction potential and Schr\"odinger's equation reads 
\be
H\,\psi(\rr_1,\ldots,\rr_N)=E\,\psi(\rr_1,\ldots,\rr_N)
\label{eq:schroZR}
\ee
with
\be
H=\sum_{i=1}^N \left[
-\frac{\hbar^2}{2 m}\Delta_{\rr_i} + U(\rr_i)
\right] \psi
\label{eq:HZR}
\ee
where $m$ is the atomic mass and $U$ is an external potential.
The crucial difference between the Hamiltonian $H$ and the non-interacting Hamiltonian is the boundary condition (\ref{eq:CL_3D},\ref{eq:CL_2D}).

\subsubsection{Lattice models} \label{sec:models:lattice}

These models were used for quantum Monte-Carlo calculations \cite{bulgacQMC,zhenyaPRL,zhenyaNJP,Juillet,ChangAFMC,BulgacCrossover}. They can also be convenient for analytics, as used in \cite{MoraCastin,LudoYvanBoite, MoraCastin2D} and in this work.
Here particles live on a lattice, i. e. the coordinates are integer multiples of the lattice spacing $b$. The Hamiltonian reads
\be
H=H_0+g_0 \,W
\ee
where
\bea
H_0&=&\sum_{i=1}^N \left[ -\frac{\hbar^2}{2m}\Delta_{\rr_i} +U(\rr_i) \right]
\label{eq:def_H0}
\\
W&=&\sum_{i<j} \delta_{\rr_i,\rr_j} b^{-d}
\label{eq:def_W}
\eea
in first quantization, i.e.
\bea
H_0&=&\sum_\sigma\int_D \frac{d^dk}{(2\pi)^d}\,\epsk c_\sigma^\dagger(\kk)c_\sigma(\kk) + \sum_{\rr,\sigma} b^d U(\rr) (\psi_\sigma^\dagger \psi_\sigma)(\rr) 
\\
W&=&\sum_\rr b^d (\psi^\dagger_\up\psi^\dagger_\down\psi_\down\psi_\up)(\rr)
\label{eq:defW_2e_quant}
\eea
in second quantization.
Here $d$ is the space dimension, 
$\epsilon_\kk$ is the dispersion relation
and $ c_\sigma^\dagger(\kk)$ is creates a particle
in the plane wave state $|\kk\rangle$ defined by
$\langle \rr | \kk \rangle = e^{i \kk \cdot \rr}$
for any $\kk$ belonging to the first Brillouin zone
$D=\left(-\frac{\pi}{b},\frac{\pi}{b}\right]^d$.
Accordingly the operator $\Delta$ in (\ref{eq:def_H0}) is the discrete representation of the Laplacian defined by
$-\frac{\hbar^2}{2m}\langle \rr | \Delta_\rr | \kk \rangle \equiv \epsilon_\kk \langle \rr | \kk \rangle$.
The simplest choice for the dispersion relation is $\ds\epsk=\frac{\hbar^2k^2}{2m}$~\cite{MoraCastin,Juillet,LudoYvanBoite, MoraCastin2D,ChangAFMC}. Another choice, used in~\cite{zhenyaPRL,zhenyaNJP}, is the dispersion relation of the Hubbard model: $\ds\epsk=\frac{\hbar^2}{m b^2}\sum_{i=1}^d\left[1-\cos(k_i b)\right]$. More generally, what follows applies to any $\epsk$ such that $\ds\epsk\underset{b\to0}{\rightarrow}\frac{\hbar^2k^2}{2m}$ sufficiently rapidly and $\epsilon_{-\kk}=\epsk$.

A key quantity is the zero-energy scattering state $\phi(\rr)$, defined by the two-body Schr\"odinger equation (with the center of mass at rest)
\be
\left(-\frac{\hbar^2}{m}\,\Delta_\rr +g_0 \frac{\delta_{\rr,\vn}}{b^d}\right) \phi(\rr) = 0
\ee
and by the normalization conditions
\bea
\phi(\rr)&\underset{r\gg b}{\simeq}& \frac{1}{r}-\frac{1}{a} \ \ \ {\rm in}\ 3D
\label{eq:normalisation_phi_tilde_3D}
\\
\phi(\rr)&\underset{r\gg b}{\simeq}& \ln(r/a) \ \ \ {\rm in}\ 2D.
\label{eq:normalisation_phi_tilde_2D}
\eea

A straightforward two-body analysis, detailed in App.~\ref{app:2body}, yields the relation between the scattering length and the bare coupling constant $g_0$:
\bea
\frac{1}{g_0}=\frac{m}{4\pi\hbar^2 a}-\int_D \frac{d^3 k}{(2\pi)^3} \frac{1}{2\epsk} & & \ \ \ {\rm in}\ 3D
\label{eq:g0_3D}
\\
\frac{1}{g_0}= \lim_{q\to 0} 
-\frac{m}{2\pi\hbar^2}\ln(a q e^\gamma /2)
+\int_D \frac{d^2 k}{(2\pi)^2} \mathcal{P} \frac{1}{2(\epsilon_{\mathbf{q}} - \epsk)}
& & \ \ \ {\rm in}\ 2D
\label{eq:g0_2D}
\eea
where $\gamma=0.577216\ldots$ is Euler's constant and $\mathcal{P}$ is the principal value.
Other useful relations derived in App.~\ref{app:2body} are
\bea
\phi(\vn)&=&-\frac{4\pi\hbar^2}{m g_0}\ \ \ {\rm in}\ 3D
\label{eq:phi0_vs_g0}
\\
\phi(\vn)&=&\frac{2\pi\hbar^2}{m g_0}\ \ \ {\rm in}\ 2D
\label{eq:phi0_vs_g0_2D}
\eea
and
\bea
|\phi(\vn)|^2&=&\frac{4\pi\hbar^2}{m}\frac{d(-1/a)}{dg_0}\ \ \ {\rm in}\ 3D
\label{eq:phi_tilde_3D}
\\
|\phi(\vn)|^2&=&\frac{2\pi\hbar^2}{m}\frac{d(\ln a)}{dg_0}\ \ \ {\rm in}\ 2D
\label{eq:phi_tilde_2D}.
\eea

In the zero-range limit ($b\to 0$ with $g_0$ adjusted in such a way that $a$ remains constant), the spectrum of the lattice model is expected to converge to the one of the zero-range model~\cite{LudoYvanBoite,zhenyaPRL}, and any eigenfunction $\psi(\rr_1,\dots,\rr_N)$ of the lattice model tends to the corresponding eigenfunction of the zero-range model, 
provided all interparticle distances remain much larger than $b$.
Let us denote by $1/\ktyp$ the typical length-scale on which the zero-range model's wavefunction varies:
e.g. for the lowest eigenstates,
it
is on the order of the mean interparticle distance, or
on the order of $a$ in the regime where $a$ is small and positive and dimers are formed.
The zero-range limit is then reached if $\ktyp b\ll1$.

For lattice models, it will prove convenient to define the regular part  $A$ by
\be
\psi(\rr_1,\ldots,\rr_i=\RR_{ij},\ldots,\rr_j=\RR_{ij},\ldots,\rr_N)=\phi(\vn)\,A_{ij}(\RR_{ij},(\rr_k)_{k\neq i,j}).
\label{eq:def_A_reseau}
\ee
In the zero-range regime $k_{\rm typ} b\ll1$, we expect that when the distance $r_{ij}$ between two particles of opposite spin is $\ll 1/\ktyp$ while all the other interparticle distances are much larger than $b$ and than $r_{ij}$, the many-body wavefunction is proportional to $\phi(r_{ij})$, with a proportionality constant given by~(\ref{eq:def_A_reseau}):
\be
\psi(\rr_1,\ldots,\rr_N)\simeq\phi(\rr_j-\rr_i)\,A_{ij}(\RR_{ij},(\rr_k)_{k\neq i,j}) 
\label{eq:psi_courte_dist}
\ee
where $\RR_{ij}=(\rr_i+\rr_j)/2$.
If moreover $r_{ij}\gg b$, $\phi$ can be replaced by its asymptotic form (\ref{eq:normalisation_phi_tilde_3D},\ref{eq:normalisation_phi_tilde_2D}); since
the contact conditions (\ref{eq:CL_3D}), (\ref{eq:CL_2D}) of the zero-range model must be recovered, we see that the lattice model's regular part tends to the zero-range model's regular part in the zero-range limit.

\subsubsection{Finite-range continuous-space model}

Such models are used in numerical few-body
correlated Gaussian  and many-body fixed-node Monte-Carlo  calculations (see e. g. \cite{Pandha,Giorgini,BlumeUnivPRL,StecherLong,BlumeRelations, RevueTrentoFermions} and refs. therein).
They are also relevant to neutron matter \cite{GezerlisCarlson}.
The Hamiltonian reads
\be
H=H_0+\sum_{i=1}^{N_\up}\sum_{j=N_\up+1}^N\,V(r_{ij}),
\ee
$H_0$ being defined by (\ref{eq:def_H0}) where $\Delta_{\rr_i}$ now stands for the usual Laplacian,
and $V(r)$ is an
interaction potential between particles of opposite spin, which vanishes for $r>b$ or at least decays quickly enough for $r\gg b$.
The two-body zero-energy scattering state $\phi(r)$ is again defined by the Schr\"odinger equation $-(\hbar^2/m)\Delta_\rr\phi+V(r)\phi=0$ and the boundary condition (\ref{eq:normalisation_phi_tilde_3D},\ref{eq:normalisation_phi_tilde_2D}).
The
zero-range regime
is again reached for $k_{\rm typ} b\ll1$ with $k_{\rm typ}$ the typical relative wavevector~\footnote{
For purely attractive interaction potentials such as the square-well potential, above a critical particle number, 
the ground state is a collapsed state and
the zero-range regime can only be reached for certain excited states.}. Equation (\ref{eq:psi_courte_dist}) again holds in the zero-range regime, where $A$ now simply stands for the zero-range model's regular part.

\subsection{Relations in the zero-range limit}
\label{sec:ZR}
\subsubsection{First order derivative of the energy with respect to the scattering length}
\label{sec:dEda}

We now derive relations for the zero-range model. For some of the derivations we will use a lattice model and take the zero-range limit in the end.

\noindent{\underline{\it Three dimensions:}}
\\
Let us consider a wavefunction $\psi_1$ satisfying the contact condition~(\ref{eq:CL_3D}) for  a scattering length $a_1$. We denote by $A^{(1)}_{ij}$ the regular part of $\psi_1$ appearing in the contact condition~(\ref{eq:CL_3D}). Similarly, $\psi_2$ satisfies the contact condition for a scattering length $a_2$ and a regular part $A^{(2)}_{ij}$. 
Then, as shown in Appendix~\ref{app:lemme}, the following lemma holds:
\be
\langle \psi_1, H \psi_2 \rangle - \langle H \psi_1, \psi_2 \rangle = 
\frac{4\pi\hbar^2}{m}\left(\frac{1}{a_1}-\frac{1}{a_2}\right) \ ( A^{(1)},A^{(2)} )
\label{eq:lemme_3D}
\ee
where the scalar product between regular parts is defined by
\be
( A^{(1)},A^{(2)} )\equiv \sum_{i<j} \int \Big( \prod_{k\neq i,j} d^d r_k \Big) \int d^d R_{ij}
A^{(1)*}_{ij}(\mathbf{R}_{ij}, (\mathbf{r}_k)_{k\neq i,j})
A^{(2)}_{ij}(\mathbf{R}_{ij}, (\mathbf{r}_k)_{k\neq i,j}).
\label{eq:def_AA}
\ee
We then apply (\ref{eq:lemme_3D}) to the case where $\psi_1$ and $\psi_2$ are $N$-body eigenstates of energy $E_1$ and $E_2$. The left hand side of (\ref{eq:lemme_3D}) then reduces to $(E_2-E_1) \langle \psi_1 | \psi_2 \rangle$. Taking the limit $a_2\to a_1$ gives the final result
\be
\frac{dE}{d(-1/a)} = \frac{4\pi\hbar^2}{m} (A,A)
\label{eq:thm_dE_3D}
\ee
for any eigenstate. This result is contained in the work of Tan~\cite{TanEnergetics,TanLargeMomentum}\footnote{Our derivation is similar to the one given in the two-body case and sketched in the many-body case in Section~3 of~\cite{TanLargeMomentum}.}.
Note that, here and in what follows, we have assumed that the wavefunction is normalized: $\la\psi|\psi\ra=1$.

\noindent{\underline{\it Two dimensions:}}
\\
The $2D$ version of the lemma~(\ref{eq:lemme_3D}) is
\be
\langle \psi_1, H \psi_2 \rangle - \langle H \psi_1, \psi_2 \rangle = 
\frac{2\pi\hbar^2}{m}\ln\left(a_2/a_1\right) \ ( A^{(1)},A^{(2)} ),
\label{eq:lemme_2D}
\ee
as shown in Appendix~\ref{app:lemme}.
As in $3D$, we deduce from the lemma the final result
\be
\frac{dE}{d(\ln a)} = \frac{2\pi\hbar^2}{m} (A,A).
\label{eq:thm_dE_2D}
\ee

\subsubsection{Large-momentum tail of the momentum distribution}
\label{sec:C_nk}

The momentum distribution is defined in second quantization by
\be
n_\sigma(\kk) = \la \hat{c}^\dagger_\si(\kk) \hat{c}_\si(\kk) \ra
\ee
where $\hat{c}_\si(\kk)$ annihilates a particle of spin $\si$ in the plane-wave state $|\kk\ra$ defined by $\la\rr|\kk\ra=e^{i \kk\cdot\rr}$.
This corresponds to the normalization
\be
\int \frac{d^d k}{(2\pi)^d}\,n_\sigma(\kk) = N_\sigma.
\label{eq:def_nk_fermions}
\ee
In first quantization,
\be
n_\sigma(\kk)=\sum_{i:\sigma} \int \Big( \prod_{l\neq i} d^d r_l \Big)
\left| \int d^d r_i e^{-i \kk\cdot\rr_i} \psi(\rr_1,\ldots,\rr_N)\right|^2
\label{eq:nk_1e_quant}
\ee
where the sum is taken over all particles of spin $\sigma$, i.e. $i$ runs from $1,$ to $N_\uparrow$ for $\sigma=\uparrow$ and from $N_\uparrow+1$ to $N$ for $\sigma=\downarrow$.

\noindent{\underline{\it Three dimensions:}}
\\
The key point is that in the large-$k$ limit, the Fourier transform with respect to $\rr_i$ is dominated by the contribution of the short-distance divergence coming from the contact condition~(\ref{eq:CL_3D}):
\be
\int d^3 r_i\, e^{-i \kk\cdot\rr_i} \psi(\rr_1,\ldots,\rr_N)
\underset{k\to\infty}{\simeq}\int d^3 r_i\, e^{-i \kk\cdot\rr_i}
\sum_{j,j\neq i} \frac{1}{r_{ij}} A_{ij}(\rr_j,(\rr_k)_{k\neq i,j}).
\label{eq:FT_sing_3D}
\ee
From $\Delta(1/r)=-4\pi\delta(\rr)$, we have the identity
\be
\int d^3 r \,e^{-i \kk\cdot \rr}\frac{1}{r}=\frac{4\pi}{k^2},
\label{eq:TF_de_1/r}
\ee
so that
\be
\int d^3 r_i \,e^{-i \kk\cdot\rr_i} \psi(\rr_1,\ldots,\rr_N)
\underset{k\to\infty}{\simeq} \frac{4\pi}{k^2}
\sum_{j,j\neq i} e^{-i \kk\cdot\rr_j} A_{ij}(\rr_j,(\rr_l)_{l\neq i,j}).
\ee
Inserting this into (\ref{eq:nk_1e_quant}) and expanding the modulus squared, the cross terms vanish in the large-$k$ limit, so that
\be
C =(4\pi)^2 (A,A)
\label{eq:C_vs_AA}
\ee
where $C\equiv \lim_{k\to \infty} k^4 n_\sigma(\kk)$.
This can be rewritten using (\ref{eq:thm_dE_3D}) as:
\be
C = \frac{4\pi m}{\hbar^2}\frac{dE}{d(-1/a)},
\label{eq:thm_nk_3D}
\ee 
in agreement with Tan~\cite{TanLargeMomentum}.

\noindent{\underline{\it Two dimensions:}}
\\
The $2D$ contact condition~(\ref{eq:CL_2D}) now gives
\be
\int d^2 r_i \, e^{-i \kk\cdot\rr_i} \psi(\rr_1,\ldots,\rr_N)
\underset{k\to\infty}{\simeq}\int d^2 r_i, e^{-i \kk\cdot\rr_i}
\sum_{j,j\neq i} \ln (r_{ij}) A_{ij}(\rr_j,(\rr_l)_{l\neq i,j}).
\label{eq:FT_sing_2D}
\ee
From $\Delta(\ln r)=2\pi\delta(\rr)$, we have the identity
\be
\int d^2 r \,e^{-i \kk\cdot \rr}\ln r=-\frac{2\pi}{k^2},
\ee
so that
\be
\int d^2 r_i \,e^{-i \kk\cdot\rr_i} \psi(\rr_1,\ldots,\rr_N)
\underset{k\to\infty}{\simeq} -\frac{2\pi}{k^2}
\sum_{j,j\neq i} e^{-i \kk\cdot\rr_j} A_{ij}(\rr_j,(\rr_l)_{l\neq i,j}).
\ee
As in $3D$ this leads to
\be
C=(2\pi)^2 (A,A)
\label{eq:C_vs_AA_2D}
\ee
where $C\equiv\lim_{k\to\infty} k^4 n_\sigma(\kk)$,
and thus from (\ref{eq:thm_dE_3D}):
\be
C=\frac{2\pi m}{\hbar^2}\frac{dE}{d(\ln a)}.
\label{eq:thm_nk_2D}
\ee

\subsubsection{Short-distance asymptotic behavior of the pair distribution function}
\label{sec:g2}

The pair distribution function, giving the probability density of finding a spin-$\uparrow$ particle at point $\RR+\rr/2$ and a spin-$\downarrow$ particle at point $\RR-\rr/2$, reads~\footnote{
In second quantization, (\ref{eq:def_g2_fermions}) corresponds to $g_{\uparrow\downarrow}^{(2)}\left(\RR+\rr/2,\RR-\rr/2\right)=\langle
\hat{\psi}^\dagger_\uparrow(\RR+\rr/2)
\hat{\psi}^\dagger_\downarrow(\RR-\rr/2)
\hat{\psi}_\downarrow(\RR-\rr/2)
\hat{\psi}_\uparrow(\RR+\rr/2)
\rangle$.
}
\bea
g_{\uparrow\downarrow}^{(2)}\left(\RR+\frac{\rr}{2},\RR-\frac{\rr}{2}\right)
&=&
\int d^d r_1\ldots d^d r_N 
\left|
\psi(\rr_1,\ldots,\rr_N)
\right|^2
\sum_{i=1}^{N_\uparrow}
\sum_{j=N_\uparrow+1}^{N} 
\delta\left(\RR+\frac{\rr}{2}-\rr_i\right)
\delta\left(\RR-\frac{\rr}{2}-\rr_j\right)
\label{eq:def_g2_fermions}
\\
&=&
\sum_{i=1}^{N_\uparrow}
\sum_{j=N_\uparrow+1}^{N} 
\int \Big(  \prod_{k\neq i,j} d^d r_k \Big) \left| \psi\left(\rr_1,\ldots,\rr_i=\RR+\frac{\rr}{2},\ldots,\rr_j=\RR-\frac{\rr}{2},\ldots,\rr_N\right) \right|^2.
\label{eq:def_g2_psi}
\eea
In what follows we consider the spatially integrated pair distribution function
\be
G^{(2)}_{\uparrow\downarrow}(\rr)\equiv
\int d^3R \,
 g_{\uparrow \downarrow}^{(2)} \left(\mathbf{R}+\frac{\mathbf{r}}{2},
\mathbf{R}-\frac{\mathbf{r}}{2}\right).
\ee
\noindent{\underline{\it Three dimensions:}}
\\
Replacing the wavefunction in~(\ref{eq:def_g2_psi}) by its asymptotic behavior given by the contact condition~(\ref{eq:CL_3D}) immediately yields:
\be
G^{(2)}_{\uparrow\downarrow}(\rr)
\underset{r\to 0}{\sim}
(A,A) \frac{1}{r^2}.
\ee
Expressing $(A,A)$ in terms of $C$ trough~(\ref{eq:C_vs_AA}) finally gives:
\be
G^{(2)}_{\uparrow\downarrow}(\rr)
\underset{r\to 0}{\sim}
\frac{C}{(4\pi)^2} \frac{1}{r^2}.
\label{eq:thm_g2_3D}
\ee
In a measurement of all particle positions, the total number of pairs of particles of opposite spin which are separated by a distance smaller than $s$ is
\be
N_{\rm pair}(s)=\int_{r<s} d^d r\, G^{(2)}_{\uparrow\downarrow}(\rr)
\ee
so that from (\ref{eq:thm_g2_3D})
\be
N_{\rm pair}(s)\underset{s\to 0}{\sim} \frac{C}{4\pi}
s,
\label{eq:Npair_3D}
\ee
as obtained in~\cite{TanEnergetics,TanLargeMomentum}.

\noindent{\underline{\it Two dimensions:}}
\\
The contact condition~(\ref{eq:CL_2D}) similarly leads to
\be
G^{(2)}_{\uparrow\downarrow}(\rr)
\underset{r\to0}{\sim}
\frac{C}{(2\pi)^2}
\ln^2 r.
\label{eq:thm_g2_2D}
\ee
After integration over the region $r<s$ this gives
\be
N_{\rm pair}(s)\underset{s\to 0}{\sim} \frac{C}{4\pi}
s^2
\ln^2 s.
\label{eq:Npair_2D}
\ee

\subsubsection{Expression of the energy in terms of the momentum distribution}
\label{sec:energy_thm}
\noindent{\underline{\it Three dimensions:}}
\\
As shown by Tan~\cite{TanEnergetics}, the total energy of any eigenstate has a simple expression in terms of the momentum distribution:
\be
 E - E_{\rm trap}  = \frac{\hbar^2 C}{4\pi m a}  
 +\sum_{\sigma} \int \frac{d^3\!k}{(2\pi)^3}  \frac{\hbar^2 k^2}{2m} 
\left[n_\sigma(\kk) - \frac{C}{k^4}\right]
\label{eq:energy_thm_3D}
\ee
or equivalently
\be
 E - E_{\rm trap}  = \lim_{\Lambda\to\infty}\left[
 \frac{\hbar^2 C}{4\pi m}\left(\frac{1}{a}-\frac{2\Lambda}{\pi}\right) + \sum_\sigma\int_{k<\Lambda} \frac{d^3k}{(2\pi)^3}  \frac{\hbar^2 k^2}{2m} 
n_\sigma(\kk) \right]
\label{eq:energy_thm_3D_Lambda}
\ee
where
\be
C\equiv\lim_{k\to\infty} k^4\,n_\sigma(\kk),
\ee
and
\be
E_{\rm trap}\equiv \left< \sum_{i=1}^N U(\rr_i)\right>
\ee
is the trapping potential energy.
A simple rederivation of this result is obtained using the lattice model (defined in Sec.~\ref{sec:models:lattice}):
As shown in Section~\ref{sec:energy_thm_latt}, one easily obtains 
Eq.~(\ref{eq:energy_thm_3D_latt}), which
yields (\ref{eq:energy_thm_3D}) in the zero-range limit since $\ds D\rightarrow\mathbb{R}^3$ and $\epsilon_\kk\to\hbar^2 k^2/(2m)$ for $b\to0$.

\noindent{\underline{\it Two dimensions:}}
\\
The $2D$ version of (\ref{eq:energy_thm_3D_Lambda}) is
\be
 E - E_{\rm trap}  = \lim_{\Lambda\to\infty}\left[
 -\frac{\hbar^2 C}{2\pi m} \ln \left(\frac{a \Lambda e^\gamma}{2}\right) 
+
 \sum_{\sigma} \int_{k<\Lambda} \frac{d^2\!k}{(2\pi)^2}  \frac{\hbar^2 k^2}{2m} 
n_\sigma(\kk) 
\right]
\label{eq:energy_thm_2D}
\ee
as was shown (for a homogeneous system) in \cite{CombescotC}. This can easily be rewritten in the following forms, which resemble~(\ref{eq:energy_thm_3D}):
\be
 E - E_{\rm trap}  = -\frac{\hbar^2 C}{2\pi m }  \ln\left(\frac{a q e^\gamma}{2}\right)
 +\sum_{\sigma} \int \frac{d^2\!k}{(2\pi)^2}  \frac{\hbar^2 k^2}{2m} 
\left[n_\sigma(\kk) - \frac{C}{k^4}\theta(k-q)\right]\ \ {\rm for\ any}\ q>0,
\label{eq:energy_thm_2D_heaviside}
\ee
where the Heaviside function $\theta$ ensures that the integral converges at small $k$,
or equivalently
\be
 E - E_{\rm trap}  = -\frac{\hbar^2 C}{2\pi m} \ln \left(\frac{a q e^\gamma}{2}\right) 
+\sum_{\sigma} \int \frac{d^2\!k}{(2\pi)^2}  \frac{\hbar^2 k^2}{2m} 
\left[n_\sigma(\kk) - \frac{C}{k^2(k^2+q^2)}\right]\ \ {\rm for\ any}\ q>0.
\label{eq:energy_thm_2D_yvan}
\ee
To derive this we again use the lattice model. 
We note that, if the limit $q\to0$ is replaced by the limit $b\to0$ taken for fixed $a$, Eq.~(\ref{eq:g0_2D}) remains true (see App.~\ref{app:2body}); repeating the reasoning of Section~\ref{sec:energy_thm_latt} then shows that (\ref{eq:energy_thm_2D_latt}) remains true; taking the limit $b\to0$ finally gives
\be
 E - E_{\rm trap}  = -\frac{\hbar^2 C}{2\pi m} \ln \left(\frac{a q e^\gamma}{2}\right) 
+\sum_{\sigma} \int \frac{d^2\!k}{(2\pi)^2}  \frac{\hbar^2 k^2}{2m} 
\left[n_\sigma(\kk) - \frac{C}{k^2}\mathcal{P}\frac{1}{k^2-q^2}\right]
\label{eq:energy_thm_2D_PP}
\ee
where $q>0$ is arbitrary;
this can be rewritten as (\ref{eq:energy_thm_2D}).

\subsubsection{Short-distance expansion of the one-body density matrix}
\label{sec:g1}

The one-body density matrix is defined as
\be
g_{\sigma \sigma}^{(1)} \left(\mathbf{R}-\frac{\mathbf{r}}{2},
\mathbf{R}+\frac{\mathbf{r}}{2}\right)=\left< \hat{\psi}_\sigma^\dagger
\left(\mathbf{R}-\frac{\mathbf{r}}{2}\right)
\hat{\psi}_\sigma\left(\mathbf{R}+\frac{\mathbf{r}}{2}\right) \right>
\ee
where $\hat{\psi}_\sigma(\rr)$ annihilates a particle of spin $\sigma$ at point $\rr$.
Let us define a spatially integrated one-body density matrix
\be
G^{(1)}_{\si\si}(\rr)\equiv\int d^dR \,
 g_{\sigma \sigma}^{(1)} \left(\mathbf{R}-\frac{\mathbf{r}}{2},
\mathbf{R}+\frac{\mathbf{r}}{2}\right).
\ee
This is related to the momentum distribution by Fourier transformation:
\be  G^{(1)}_{\si\si}(\rr) = 
\int \frac{d^d k}{(2\pi)^d}\,e^{i\kk\cdot\rr} n_\si(\kk).
\label{eq:G1_vs_nk}
\ee

\noindent \underline{{\it Three dimensions:}}
\\
As shown below,
\be
 G^{(1)}_{\si\si}(\rr) \underset{r\to 0}{=}N_\si - \frac{C}{8\pi}r+O(r^2),
 \label{eq:thm_g1_a}
\ee
and moreover the expansion can pushed to second order if one sums over spin and averages over three orthogonal directions of $\rr$:
\be
\frac{1}{3} \sum_{i=1}^3 \sum_\si  G^{(1)}_{\si\si}(r {\bf u_i}) 
\underset{r\to 0}{=}
N-\frac{C}{4\pi}r
-\frac{m}{3\hbar^2}\left(
E-E_{\rm trap} - \frac{\hbar^2 C}{4\pi m a}
\right) r^2 + o(r^2)
\label{eq:thm_g1_b}
\ee
where the ${\bf u_i}$'s are three orthogonal unit vectors.
This last relation, as well as its $2D$ version (\ref{eq:thm_g1_b_2D}),  also hold if one averages over all directions of $\rr$ uniformly on the unit sphere or unit circle.
They generalize the result obtained in $1D$ in~\cite{Olshanii_nk}, but the derivation is different from the $1D$ case~\footnote{In $2D$ and $3D$, our result does not follow from the well-known fact that, for a finite-range interaction potential in continuous space, $-\frac{\hbar^2}{2m}\sum_\si \Delta G^{(1)}_{\si\si}(\rr=\vn)$ equals the kinetic energy; indeed, the Laplacian does not commute with the zero-range limit in that case.}.

To derive (\ref{eq:thm_g1_a},\ref{eq:thm_g1_b}) we rewrite (\ref{eq:G1_vs_nk}) as
\be
G^{(1)}_{\si\si}(\rr) = N_\si + 
\int \frac{d^3 k}{(2\pi)^3}\,\left(e^{i\kk\cdot\rr} -1\right)\frac{C}{k^4}
+ \int \frac{d^3 k}{(2\pi)^3}\,\left(e^{i\kk\cdot\rr} -1\right)\left(n_\si(\kk)-\frac{C}{k^4}\right).
\ee
The first integral equals $-(C/8\pi) r$. In the second integral, we use
\be
e^{i\kk\cdot\rr}-1\underset{r\to 0}{=}i\kk\cdot\rr-\frac{(\kk\cdot\rr)^2}{2}+o(r^2).
\label{eq:expand_exp}
\ee
The first term of this expansion gives a contribution to the integral  proportional to the total momentum of the gas, which vanishes since the eigenfunctions are real.
The second term is $O(r^2)$, which gives~(\ref{eq:thm_g1_a}).
Eq.~(\ref{eq:thm_g1_b}) follows from the fact that the contribution of the second term, after averaging over the directions of $\rr$, 
is given by the integral of $k^2 [n_\si(\kk)-C/k^4]$, which
is related to the total energy by~(\ref{eq:energy_thm_3D}).

\noindent \underline{{\it Two dimensions:}}
\\
As shown below,
\be
 G^{(1)}_{\si\si}(\rr) \underset{r\to 0}{=}N_\si + \frac{C}{8\pi}r^2\ln r+O(r^2),
 \label{eq:thm_g1_a_2D}
\ee
and for any pair of orthogonal unit vectors $({\bf u_1},{\bf u_2})$
\be
\frac{1}{2} \sum_{i=1}^2 \sum_\si G^{(1)}_{\si\si}(r {\bf u_i}) 
\underset{r\to 0}{=}
N
+\frac{C}{4\pi}r^2
\left[\ln\left(\frac{r}{a}\right)+\frac{\mathcal{F}}{32}\right]
-\frac{m}{2\hbar^2}\left(
E-E_{\rm trap}
\right) r^2 + o(r^2)
\label{eq:thm_g1_b_2D}
\ee
where
\be
\mathcal{F}\equiv\  _2F_3\left(1,1;2,3,3;-\frac{1}{4}\right)
=-16\sum_{i=1}^\infty \frac{1}{i\left[(i+1)!\right]^2}\left(-\frac{1}{4}\right)^i=0.98625471\ldots.
\label{eq:hypergeo}
\ee

To derive~(\ref{eq:thm_g1_a_2D},\ref{eq:thm_g1_b_2D}) we rewrite (\ref{eq:G1_vs_nk}) as
\be
G^{(1)}_{\si\si}(\rr) = N_\si + I(\rr)+J(\rr)
\ee
with
\be
I(\rr)=
\int \frac{d^2 k}{(2\pi)^2}\,\left(e^{i\kk\cdot\rr} -1\right)\frac{C}{k^4}\theta(k-q)
\ee
and
\be
J(\rr)= \int \frac{d^2 k}{(2\pi)^2}\,\left(e^{i\kk\cdot\rr} -1\right)\left(n_\si(\kk)-\frac{C}{k^4}\theta(k-q)\right)
\ee
where $q>0$ is arbitrary and the Heaviside function $\theta$ ensures that the integrals converge.

To evaluate $I(\rr)$ we use standard manipulations to get
\be
I(\rr)=\frac{C}{8\pi}r^2\left[\ln(q r) + 4 \mathcal{I}\right]+O(r^4)
\ee
where
\be
\mathcal{I}=\int_1^\infty dx \frac{J_0(x)-1}{x^3},
\ee
$J_0$ being a Bessel function. Evaluating this integral with Maple gives
\be
\mathcal{I}=\frac{\gamma-1-\ln 2}{4}+\frac{\mathcal{F}}{128}
\ee
where $\mathcal{F}$ is the hypergeometric function defined in~(\ref{eq:hypergeo}).

Finally we evaluate $J(\rr)$ using the same procedure as in $3D$:  expanding the exponential [see~(\ref{eq:expand_exp})] yields an integral which can be related to the total energy thanks to~(\ref{eq:energy_thm_2D_heaviside}).

\subsubsection{Second order derivative of the energy with respect to the scattering length}
We denote by $|\psi_n\ra$ an orthonormal basis of $N$-body eigenstates which vary smoothly with $1/a$, and  by $E_n$ the corresponding eigenenergies.
We will show that
\be
\frac{1}{2} \frac{d^2E_n}{d(-1/a)^2}
= \left(\frac{4\pi\hbar^2}{m}\right)^2 \sum_{n',E_{n'}\neq E_n} 
\frac{|(A^{(n')},A^{(n)})|^2}{E_n-E_{n'}} \ \ \ {\rm in}\ 3D
\label{eq:d^2E_3D}
\ee
\be
\frac{1}{2} \frac{d^2E_n}{d(\ln a)^2}
= \left(\frac{2\pi\hbar^2}{m}\right)^2 \sum_{n',E_{n'}\neq E_n} 
\frac{|(A^{(n')},A^{(n)})|^2}{E_n-E_{n'}}\ \ \ {\rm in}\ 2D
\label{eq:d^2E_2D}
\ee
where the sum is taken on all values of $n'$ such that $E_{n'}\neq E_n$.
This implies that for the ground state energy $E_0$,
\bea
\frac{d^2E_0}{d(-1/a)^2} &<& 0\ \ \ {\rm in}\ 3D
\label{eq:d^2E0_3D}
\\
\frac{d^2E_0}{d(\ln a)^2} &<& 0\ \ \ {\rm in}\ 2D.
\label{eq:d^2E0_2D}
\eea
Eq.(\ref{eq:d^2E0_3D}) was intuitively expected~\cite{LeticiaSoutenance}: Eq.~(\ref{eq:Npair_3D}) shows that $dE_0/d(-1/a)$ is proportional to the probability of finding two particles very close to each other, and it is natural that this probability decreases when one goes from the BEC limit ($-1/a\to-\infty$) to the BCS limit ($-1/a\to+\infty$), i.e. when the interactions become less attractive~\footnote{In the lattice model in $3D$, the coupling constant $g_0$ is always negative in the zero-range limit $|a|\gg b$, and is an increasing function of $-1/a$, as can be seen from (\ref{eq:g0_3D}).}.
Eq.(\ref{eq:d^2E0_2D})
also agrees with intuition~\footnote{Eq.~(\ref{eq:Npair_2D}) shows that $dE_0/d(\ln a)$ is proportional to the probability of finding two particles very close to each other, and it is natural that this probability decreases when one goes from the BEC limit ($\ln a\to-\infty$) to the BCS limit ($\ln a\to+\infty$), i.e. when the interactions become less attractive [in the lattice model in $2D$, the coupling constant $g_0$ is always negative in the zero-range limit $a\gg b$, and is an increasing function of $\ln a$, as can be seen from (\ref{eq:g0_2D})].}.

For the derivation,  it is convenient to use the lattice model
(defined in Sec.~\ref{sec:models:lattice}): As shown in Sec.\ref{sec:d^2E_reseau} one easily obtains (\ref{eq:d^2E_dg0}), from which the result is deduced as follows.
$|\phi(\vn)|^2$ is eliminated using (\ref{eq:phi_tilde_3D},\ref{eq:phi_tilde_2D}). Then,
in $3D$, one uses
\be
\frac{d^2E_n}{d(-1/a)^2}=\frac{d^2E_n}{dg_0^{\phantom{0}2}} \left(\frac{dg_0}{d(-1/a)}\right)^2+\frac{dE_n}{dg_0}\frac{d^2 g_0}{d(-1/a)^2}
\label{eq:deriv_2_fois}
\ee
where the second term equals $2g_0\,dE_n/d(-1/a)\,m/(4\pi\hbar^2)$ and thus vanishes in the zero-range limit.
Similarly, in $2D$ one uses the fact that
\be
\frac{d^2E_n}{d(\ln a)^2}=\frac{d^2E_n}{dg_0^{\phantom{0}2}} \left(\frac{dg_0}{d(\ln a)}\right)^2
\ee
in the zero-range limit.

\subsubsection{Time derivative of the energy}

We now consider the case where the scattering length $a(t)$ and the trapping potential $U(\rr,t)$ are varied with time. The time-dependent version of the zero-range model (see e.g.~\cite{CRAS}) is given by Schr\"odinger's equation
\be
i\hbar \frac{\partial}{\partial t} \psi(\rr_1,\ldots,\rr_N;t) = 
H(t)\,  \psi(\rr_1,\ldots,\rr_N;t)
\ee
when all particle positions are distinct, with
\be
H(t)=
\sum_{i=1}^N \left[
-\frac{\hbar^2}{2 m}\Delta_{\rr_i} + U(\rr_i,t)
\right],
\ee
and by the contact condition~(\ref{eq:CL_3D}) in~$3D$ or~(\ref{eq:CL_2D}) in~$2D$ for the scattering length $a=a(t)$.
One then has the relations
\bea
\frac{dE}{dt} &=& \frac{\hbar^2 C}{4\pi m} \frac{d(-1/a)}{dt}+
\la\psi(t)|\sum_{i=1}^N\partial_t U(\rr_i,t)|\psi(t)\ra
\ \ \ {\rm in} 3D
\label{eq:thm_dEdt_3D}
\\
\frac{dE}{dt} &=& \frac{\hbar^2 C}{2\pi m} \frac{d(\ln a)}{dt}+
\la\psi(t)|\sum_{i=1}^N\partial_t U(\rr_i,t)|\psi(t)\ra
\ \ \ {\rm in} 2D,
\label{eq:thm_dEdt_2D}
\eea
where $E(t)=\langle \psi(t) | H(t)|\psi(t)\rangle$ is the total energy and
$E_{\rm trap}(t)=\langle \psi(t) | \sum_{i=1}^N U(\rr_i,t) |\psi(t)\rangle$ is the trapping potential energy.
The relation (\ref{eq:thm_dEdt_3D}) was first obtained by Tan~\cite{TanLargeMomentum}.
 A very simple derivation of these relations using the lattice model is given in Section \ref{sec:dEdt_reseau}. Here we give a derivation within the zero-range model.

\noindent{\underline{\it Three dimensions:}}
\\
 We first note that the
generalization of the
 lemma (\ref{eq:lemme_3D})  to the case of two Hamiltonians $H_1$ and $H_2$ with corresponding trapping potentials $U_1(\rr)$ and $U_2(\rr)$ reads:
\be
\langle \psi_1, H_2 \psi_2 \rangle - \langle H_1 \psi_1, \psi_2 \rangle = 
\frac{4\pi\hbar^2}{m}\left(\frac{1}{a_1}-\frac{1}{a_2}\right) \ ( A^{(1)},A^{(2)}) + 
\langle \psi_1 | \sum_{i=1}^N \left[U_2(\rr_i,t)-U_1(\rr_i,t)\right] |\psi_2 \rangle.
\label{eq:lemme_modif_3D}
\ee
Applying this relation for $|\psi_1\rangle=|\psi(t)\rangle$ and $|\psi_2\rangle=|\psi(t+\delta t)\rangle$ [and correspondingly $a_1=a(t)$,
$a_2=a(t+\delta t)$ and $H_1=H(t)$, $H_2=H(t+\delta t)$] gives:
\bea
\la \psi(t), H(t+\delta t) \psi(t+\delta t)\ra -
\la H(t)\psi(t),\psi(t+\delta t)\ra& =& \frac{4\pi\hbar^2}{m}\left(\frac{1}{a(t)}-\frac{1}{a(t+\delta t)}\right) (A(t),A(t+\delta t))
\nonumber
\\ &&+\la\psi(t)| \sum_{i=1}^N U(\rr_i,t+\delta t)-U(\rr_i,t)|\psi(t+\delta t)\ra.
\label{eq:interm_dt}
\eea
 Dividing by $\delta t$, taking the limit $\delta t\to0$,
and using the expression~(\ref{eq:C_vs_AA}) of $(A,A)$ in terms of $C$ ,
the right-hand-side of (\ref{eq:interm_dt})  reduces to the right-hand-side of (\ref{eq:thm_dEdt_3D}).
The left-hand-side of (\ref{eq:interm_dt}) can be rewritten using Schr\"odinger's equation as $i\hbar d\la\psi(t)|\psi(t+\delta t)\ra/dt$.
Using Schr\"odinger's equation again to Taylor expand $|\psi(t+\delta t)\ra$ in this last expression finally gives
the result (\ref{eq:thm_dEdt_3D}).

\noindent{\underline{\it Two dimensions:}}
\\
The relation (\ref{eq:thm_dEdt_2D})
is derived similarly from the lemma
\be
\langle \psi_1, H_2 \psi_2 \rangle - \langle H_1 \psi_1, \psi_2 \rangle 
=
\frac{2\pi\hbar^2}{m}\ln(a_2/a_1) ( A^{(1)},A^{(2)}) + 
\langle \psi_1 | \sum_{i=1}^N \left[U_2(\rr_i,t)-U_1(\rr_i,t)\right] |\psi_2 \rangle.
\label{eq:lemme_modif_2D}
\ee
\subsection{Relations for the lattice model}\label{sec:latt}
In this Section, as well as in Section~\ref{sec:V(r)}, it will prove convenient to {\it define} $C$ by
\bea
C&\equiv&\frac{4\pi m}{\hbar^2}\frac{dE}{d(-1/a)}\ \ {\rm in}\ 3D
\label{eq:defC_b_finie_3D}
\\
C&\equiv&\frac{2\pi m}{\hbar^2}\frac{dE}{d(\ln a)}\ \ {\rm in}\ 2D.
\label{eq:defC_b_finie_2D}
\eea 
This new definition of $C$ coincides with the identities (\ref{eq:C_vs_AA}) and (\ref{eq:C_vs_AA_2D}) of Section~\ref{sec:ZR} in the zero-range limit, as follows from (\ref{eq:thm_dE_3D},\ref{eq:thm_dE_2D}).

We will use the following lemma: 
For any wavefunctions $\psi$, $\psi'$,
\be
\la\psi'|W|\psi\ra = |\phi({\bf 0})|^2\ ( A',A)
\label{eq:lemme_W}
\ee
where $A$ and $A'$ are the regular parts related to $\psi$ and $\psi'$ through (\ref{eq:def_A_reseau}), and the scalar product between regular parts is naturally defined as the discrete version of (\ref{eq:def_AA}):
\be
( A',A )\equiv \sum_{i<j} 
\sum_{(\rr_k)_{k\neq i,j}} \sum_{\RR_{ij}} b^{(N-1)d}
A'^*_{ij}(\mathbf{R}_{ij}, (\mathbf{r}_k)_{k\neq i,j})
A_{ij}(\mathbf{R}_{ij}, (\mathbf{r}_k)_{k\neq i,j}).
\ee
The lemma simply follows from
\be
\la\psi'|W|\psi\ra=\sum_{i<j} 
\sum_{(\rr_k)_{k\neq i,j}} b^{(N-2)d}
(\psi'^*\psi)(\rr_1,\ldots,\rr_i=\rr_j,\ldots,\rr_j,\ldots,\rr_N).
\label{eq:note_W}
\ee

\subsubsection{First order derivative of the energy with respect to the scattering length}\label{subsec:dE_latt}
The Hellmann-Feynman theorem gives
\be
\frac{dE}{dg_0} = \langle \psi | \frac{dH}{dg_0} |\psi \rangle =
\langle \psi |W| \psi \rangle.
\ee
But lemma (\ref{eq:lemme_W}) with $\psi'=\psi$ writes
\be
\langle \psi |W| \psi \rangle
= |\phi({\bf 0})|^2\ ( A,A).
\label{eq:W_AA}
\ee
Using the expressions (\ref{eq:phi_tilde_3D},\ref{eq:phi_tilde_2D}) of $|\phi(\vn)|^2$, we conclude that
\bea
\frac{dE}{d(-1/a)}&=&\frac{4\pi\hbar^2}{m}(A,A)\ \ {\rm in}\ 3D
\label{eq:dEda_latt_3D}
\\
\frac{dE}{d(\ln a)}&=&\frac{2\pi\hbar^2}{m}(A,A)\ \ {\rm in}\ 2D.
\label{eq:dEda_latt_2D}
\eea

\subsubsection{Interaction energy}
The left-hand-side of (\ref{eq:W_AA}) is obviously equal to the mean interaction energy $E_{\rm int}$ divided by $g_0$;
in the right-hand-side of (\ref{eq:W_AA}), $(A,A)$ can be expressed in terms of $C$ using  (\ref{eq:dEda_latt_3D},\ref{eq:dEda_latt_2D}) and the definition (\ref{eq:defC_b_finie_3D},\ref{eq:defC_b_finie_2D}) of $C$:
\bea
(A,A)&=&\frac{C}{(4\pi)^2}\ \ \ {\rm in}\ 3D
\label{eq:AA_vs_C_rezo_3D}
\\
(A,A)&=&\frac{C}{(2\pi)^2}\ \ \ {\rm in}\ 2D.
\label{eq:AA_vs_C_rezo_2D}
\eea
This gives
\bea
\frac{E_{\rm int}}{g_0}
&=&\frac{C}{(4\pi)^2}|\phi(\vn)|^2
\ \ {\rm in}\ 3D
\label{eq:Eint_g2_C}
\\
\frac{E_{\rm int}}{g_0}
&=&\frac{C}{(2\pi)^2}|\phi(\vn)|^2
\ \ {\rm in}\ 2D.
\label{eq:Eint_g2_C_2D}
\eea
Eliminating $\phi(\vn)$ thanks to (\ref{eq:phi0_vs_g0},\ref{eq:phi0_vs_g0_2D}) finally gives
\be
E_{\rm int}=C \left(\frac{\hbar^2}{m}\right)^2 \frac{1}{g_0}
\label{eq:Eint}
\ee
both in $3D$ and $2D$.

\subsubsection{Relation between energy, momentum distribution and $C$}\label{sec:energy_thm_latt}
Here we show that
\bea
 E - E_{\rm trap}  &=& \frac{\hbar^2 C}{4\pi m a}  
 +\sum_{\sigma} \int_D \frac{d^3\!k}{(2\pi)^3}\  \epsk \left[n_\sigma(\kk) - C\left(\frac{\hbar^2}{2m\epsk}\right)^2\right]\ \ \ \ \ \ \ \ {\rm in}\ 3D
\label{eq:energy_thm_3D_latt}
\\
 E - E_{\rm trap}  &= & \lim_{q\to 0} -\frac{\hbar^2 C}{2\pi m} \ln \left(\frac{a q e^\gamma}{2}\right) 
+\sum_{\sigma} \int_D \frac{d^2\!k}{(2\pi)^2}\,  \epsk \left[n_\sigma(\kk) - C\frac{\hbar^2}{2m\epsk}\mathcal{P}\frac{\hbar^2}{2m(\epsk-\epsq)}\right]\ \ \ {\rm in}\ 2D.
\label{eq:energy_thm_2D_latt}
\eea
To derive this we start from the expression (\ref{eq:Eint}) of the interaction energy and eliminate $1/g_0$ thanks to (\ref{eq:g0_3D},\ref{eq:g0_2D}).
The desired quantity $E-E_{\rm trap}=E_{\rm int}+E_{\rm kin}$ is then obtained from
\be
E_{\rm kin}=\sum_\sigma\int_D\frac{d^d k}{(2\pi)^d}\,\epsk\, n_\sigma(\kk).
\ee

\subsubsection{Second order derivative of the energy with respect to the coupling constant}\label{sec:d^2E_reseau}
We denote by $|\psi_n\ra$ an orthonormal basis of $N$-body eigenstates which vary smoothly with $g_0$, and  by $E_n$ the corresponding eigenenergies.
We apply second order perturbation theory to determine how an eigenenergy varies for an infinitesimal change of $g_0$. This gives:
\be
\frac{1}{2}\frac{d^2 E_n}{dg_0^{\phantom{0}2}}=\sum_{n', E_{n'}\neq E_n} \frac{\left|\la\psi_{n'}|W|\psi_n\ra\right|^2}{E_n-E_{n'}},
\label{eq:d^2E/dg0^2}
\ee
where the sum is taken over all values of $n'$ such that $E_{n'}\neq E_n$.
Lemma (\ref{eq:lemme_W}) then yields:
\be
\frac{1}{2}\frac{d^2 E_n}{dg_0^{\phantom{0}2}}=|\phi(\vn)|^4\sum_{n', E_{n'}\neq E_n} \frac{|(A^{(n')},A^{(n)})|^2}{E_n-E_{n'}}.
\label{eq:d^2E_dg0}
\ee

\subsubsection{Time derivative of the energy}\label{sec:dEdt_reseau}

Equations (\ref{eq:thm_dEdt_3D},\ref{eq:thm_dEdt_2D}) remain exact for the lattice model. Indeed, the Hellmann-Feynman theorem gives
\be
\frac{dE}{dt}=\la\frac{dH}{dt}\ra=\frac{dg_0}{dt}\la W\ra+\la\sum_{i=1}^N \partial_t U(\rr_i,t)\ra.
\ee
The result then follows by using the lemma (\ref{eq:lemme_W}), the expressions (\ref{eq:phi_tilde_3D},\ref{eq:phi_tilde_2D}) of $|\phi(\vn)|^2$, and the expressions (\ref{eq:AA_vs_C_rezo_3D},\ref{eq:AA_vs_C_rezo_2D}) of $(A,A)$ in terms of $C$.

\subsubsection{On-site pair distribution function}
We define the spatially integrated pair distribution function
\be
G^{(2)}_{\uparrow\downarrow}(\rr)\equiv
\sum_{\RR} b^d
 g_{\uparrow \downarrow}^{(2)} \left(\mathbf{R}+\frac{\mathbf{r}}{2},
\mathbf{R}-\frac{\mathbf{r}}{2}\right).
\ee
Using (\ref{eq:Eint_g2_C},\ref{eq:Eint_g2_C_2D})
and expressing the interaction energy in terms of $g^{(2)}_{\uparrow\downarrow}$ thanks to the second-quantized form (\ref{eq:defW_2e_quant}) yields:
\bea
G^{(2)}_{\uparrow\downarrow}(\vn)
&=&\frac{C}{(4\pi)^2}|\phi(\vn)|^2
\label{eq:g2_latt}
\ \ \ {\rm in}\ 3D
\\
G^{(2)}_{\uparrow\downarrow}(\vn)
&=&\frac{C}{(2\pi)^2}|\phi(\vn)|^2
\ \ {\rm in}\ 2D.
\eea

\subsubsection{Pair distribution function at short distances}
\label{subsec:G2_short_dist_latt}
The last result can be generalized to finite but small $r$, 
assuming that we are in the zero-range regime $\ktyp b\ll1$ (introduced at the end of Sec.~\ref{sec:models:lattice}):
\bea
G^{(2)}_{\uparrow\downarrow}(\rr)&\underset{r\ll1/\ktyp}{\simeq}&\frac{C}{(4\pi)^2}|\phi(\rr)|^2
\ \ \ {\rm in }\ 3D
\label{eq:g2_rezo_3D}
\\
G^{(2)}_{\uparrow\downarrow}(\rr)&\underset{r\ll1/\ktyp}{\simeq}&\frac{C}{(2\pi)^2}|\phi(\rr)|^2
\ \ \ {\rm in }\ 2D.
\label{eq:g2_rezo_2D}
\eea
Indeed, the expression (\ref{eq:def_g2_psi}) of $g^{(2)}_{\uparrow\downarrow}$ 
in terms of the wavefunction is valid for the lattice model with the obvious replacement of the integrals by sums, so that
\be
G^{(2)}_{\uparrow\downarrow}(\rr)=\sum_\RR b^d \sum_{i=1}^{N_\up}\sum_{j=N_\up+1}^N
\sum_{(\rr_k)_{k\neq i,j}} b^{(N-2)d}
\left| \psi\left(\rr_1,\ldots,\rr_i=\RR+\frac{\rr}{2},\ldots,\rr_j=\RR-\frac{\rr}{2},\ldots,\rr_N\right) \right|^2.
\ee
For $r\ll1/\ktyp$, we can replace $\psi$ by the short-distance expression (\ref{eq:psi_courte_dist}),
assuming that the multiple sum is dominated by the configurations where all the distances $|\rr_k-\RR|$ and $r_{k k'}$ are much larger than $b$:
\be
G^{(2)}_{\uparrow\downarrow}(\rr)\simeq (A,A)\ |\phi(\rr)|^2.
\label{eq:G2_AA}
\ee
Expressing $(A,A)$ in terms of $C$ thanks to (\ref{eq:AA_vs_C_rezo_3D},\ref{eq:AA_vs_C_rezo_2D}) gives (\ref{eq:g2_rezo_3D},\ref{eq:g2_rezo_2D}).

\subsubsection{Momentum distribution at large momenta}\label{subsec:nk_latt}

Assuming again that we are in the zero-range regime $\ktyp b\ll1$, we will show that
\be
n_\sigma(\kk)\underset{k\gg\ktyp}{\simeq} C \left(\frac{\hbar^2}{2m\epsk}\right)^2
\label{eq:nk_latt}
\ee
both in $3D$ and in $2D$. We start from
\be
n_\sigma(\kk)=\sum_{i:\sigma} \sum_{(\rr_l)_{l\neq i}} b^{d(N-1)}
\left|\sum_{\rr_i} b^d e^{-i\kk\cdot\rr_i}\psi(\rr_1,\dots,\rr_N)
\right|^2.
\label{eq:nk_psi_latt}
\ee
We are interested in the limit $k\gg\ktyp$.
Since $\psi(\rr_1,\ldots,\rr_N)$ is a function of $\rr_i$ which varies on the scale of $1/\ktyp$, except when $\rr_i$ is close to another particle $\rr_j$ where it varies on the scale of $b$,
we can replace $\psi$ by its short-distance form (\ref{eq:psi_courte_dist}):
\be
\sum_{\rr_i} b^d e^{-i\kk\cdot\rr_i}\psi(\rr_1,\ldots,\rr_N)\simeq
\tilde{\phi}(\kk) \sum_{j,j\neq i} e^{-i\kk\cdot\rr_j} A_{ij}(\rr_j,(\rr_l)_{l\neq i,j}),
\label{eq:TF_approx}
\ee
where $\tilde{\phi}(\kk)=\la\rr|\phi\ra=\sum_\rr b^d e^{-i\kk\cdot\rr}\phi(\rr)$.
Here we excluded the configurations where more than two particles are at distances $\lesssim b$, which are expected to have a negligible contribution to (\ref{eq:nk_psi_latt}).
Inserting (\ref{eq:TF_approx}) into (\ref{eq:nk_psi_latt}),
expanding the modulus squared, and neglecting the cross-product terms in the limit $k\gg\ktyp$, we obtain
\be
n_\sigma(\kk)\simeq|\tilde{\phi}(\kk)|^2 (A,A).
\label{eq:nk_AA_latt}
\ee
Finally, $\tilde{\phi}(\kk)$ is easily computed for the lattice model: for $k\neq0$, the two-body Schr\"odinger equation (\ref{eq:schro_2corps}) directly gives
$\tilde{\phi}(\kk)=-g_0\phi(\vn)/(2\epsk)$, and $\phi(\vn)$ is given by (\ref{eq:phi0_vs_g0},\ref{eq:phi0_vs_g0_2D}), which yields (\ref{eq:nk_latt}).

\subsection{Relations for a finite-range interaction in continuous space}
\label{sec:V(r)}
We recall that in this Section, $C$ is again defined by (\ref{eq:defC_b_finie_3D},\ref{eq:defC_b_finie_2D}).

\subsubsection{Interaction energy}
As for the lattice model, we find that the interaction energy is proportional to $C$:
\bea
E_{\rm int}&=&\frac{C}{(4\pi)^2}\int d^3 r\, V(r) |\phi(\rr)|^2\ \ \ {\rm in}\ 3D
\label{eq:Eint_V(r)_3D}
\\
E_{\rm int}&=&\frac{C}{(2\pi)^2}\int d^2 r\, V(r) |\phi(\rr)|^2\ \ \ {\rm in}\ 2D.
\label{eq:Eint_V(r)_2D}
\eea
It was shown in~\cite{ZhangLeggettUniv} that this relation is asymptotically valid in the zero-range limit in $3D$. 
Here we show that it remains exact for any finite value of the range and we generalize it to $2D$.

For the derivation, we set
\be
V(r)=g_0 W(r)
\ee
where $g_0$ is a dimensionless coupling constant which allows to tune $a$.
The Hellmann-Feynman theorem then gives $E_{\rm int}=g_0 dE/dg_0$. The result then follows by writing $dE/dg_0=dE/d(-1/a)\cdot d(-1/a)/dg_0$ in $3D$ and
$dE/dg_0=dE/d(\ln a)\cdot d(\ln a)/dg_0$ in $2D$,
and by using the definition (\ref{eq:defC_b_finie_3D},\ref{eq:defC_b_finie_2D}) of $C$
as well as the following lemmas:
\bea
g_0\frac{d(-1/a)}{d g_0}&=& \frac{m}{4\pi\hbar^2}\int d^3 r\, V(r) |\phi(r)|^2
\ \ \ {\rm in}\ 3D
\label{eq:lemme_g0_vs_a_3D}
\\
g_0\frac{d(\ln a)}{d g_0}&=& \frac{m}{2\pi\hbar^2}\int d^2 r\, V(r) |\phi(r)|^2
\ \ \ {\rm in}\ 2D.
\label{eq:lemme_g0_vs_a_2D}
\eea
To derive these lemmas, we
consider two values of the scattering length $a_i,\ i=1,2$, and the corresponding scattering states $\phi_i$ and coupling constants $g_{0,i}$. The corresponding two-particle relative-motion Hamiltonians are $H_i=-(\hbar^2/m)\,\Delta_\rr + g_{0,i} W(r)$. Since $H_i \phi_i=0$, we have
\be
\lim_{R\to\infty} \int_{r<R} d^d r \left( \phi_1 H_2 \phi_2
- \phi_2 H_1 \phi_1 \right) = 0.
\ee
The contribution of the kinetic energies can be computed from Ostrogradsky's theorem and the large-distance form of $\phi$~\footnote{In $2D$ we assume, to facilitate the derivation, that $V(r)=0$ for $r>b$, but the result is expected to hold for any $V(r)$ which vanishes quickly enough at infinity.}.
\setcounter{fnnumberter}{\thefootnote}
The contribution of the potential energies is proportional to $g_{0,2}-g_{0,1}$. Taking the limit $a_2\to a_1$ gives the results (\ref{eq:lemme_g0_vs_a_3D},\ref{eq:lemme_g0_vs_a_2D}).
Lemma (\ref{eq:lemme_g0_vs_a_3D}) was also used in~\cite{ZhangLeggettUniv} and the above derivation is essentially identical to the one of~\cite{ZhangLeggettUniv}.
For this $3D$ lemma, there also exists an alternative derivation based on  the two-body problem in a large box~\footnote{We consider two particles of opposite spin in a cubic box of side $L$ with periodic boundary conditions, and we work in the limit where $L$ is much larger than $|a|$ and $b$. In this limit, there exists a ``weakly interacting'' eigenstate $\psi$ whose energy is given by the ``mean-field'' shift $E=g/L^3$. The Hellmann-Feynman theorem gives $g_0\,dE/dg_0=E_{\rm int}[\psi]$.
But the wavefunction $\psi(\rr_1,\rr_2)\simeq\Phi(r_{12})/L^3$ where $\Phi$ is the zero-energy scattering state normalized by $\Phi\to1$ at infinity. Thus $E_{\rm int}=\int d^3 r\,V(r) |\Phi(r)|^2/L^3$. The desired Eq.(\ref{eq:lemme_g0_vs_a_3D}) then follows, since $\Phi=-a \phi$.}.

\subsubsection{Relation between energy and momentum distribution}
\noindent
\underline{\it Three dimensions:}
\be
E-E_{\rm trap} =
\frac{\hbar^2 C}{4\pi m a} + \sum_\sigma
\int \frac{d^3 k}{(2\pi)^3} \frac{\hbar^2 k^2}{2m}\left[ n_\sigma(\kk)-\frac{C}{(4\pi)^2}|\tilde{\phi}'(\kk)|^2\right]
\label{eq:energy_thm_3D_V(r)}
\ee
where $\tilde{\phi}'(k)=\tilde{\phi}(k)+a^{-1}(2\pi)^3\delta(\kk)=\int d^3r\,e^{-i\kk\cdot\rr}\phi'(r)$ with
\be
\phi'(r)=\phi(r)+\frac{1}{a}.
\ee
This is simply obtained by adding the kinetic energy to (\ref{eq:Eint_V(r)_3D})
and by using the lemma:
\be
\int d^3 r\, V(r) |\phi(r)|^2 = \frac{4\pi\hbar^2}{m a}-\int \frac{d^3 k}{(2\pi)^3}\frac{\hbar^2k^2}{m}|\tilde{\phi}'(k)|^2.
\label{eq:lemme_phi'(k)_3D}
\ee
To derive this lemma, we start from Schr\"odinger's equation $-(\hbar^2/m)\Delta\phi+V(r)\phi=0$, which implies
\be
\int d^3 r\, V(r) |\phi(r)|^2=\frac{\hbar^2}{m}\int d^3 r \, \phi\Delta\phi.
\label{eq:phiDeltaphi}
\ee
On the other hand, applying Ostrogradsky's theorem 
 over the sphere of radius $R$, using the asymptotic expression (\ref{eq:normalisation_phi_tilde_3D}) of $\phi$
 and taking the limit $R\to\infty$ yields
\be
\int d^3 r \, \phi\Delta\phi = \frac{4\pi}{a}-\int d^3 r\, (\mathbf{\nabla} \phi)^2.
\ee
We then replace $\nabla \phi$ by $\nabla\phi'$. Applying the Parseval-Plancherel relation to $\partial_i \phi$,
and using the fact that $\phi'(r)$ vanishes at infinity, we get:
\be
\int d^3 r\,(\nabla\phi')^2 = \int \frac{d^3 k}{(2\pi)^3}\,k^2 |\tilde{\phi}'(k)|^2
\ee
The desired result (\ref{eq:lemme_phi'(k)_3D}) follows.

\noindent
\underline{\it Two dimensions:}
\be
E-E_{\rm trap} =\lim_{R\to\infty}\left\{\frac{\hbar^2 C}{2\pi m}\ln\left(\frac{R}{a}\right)
+\sum_\sigma \int \frac{d^2k}{(2\pi)^2}\,\frac{\hbar^2k^2}{2m}\left[n_\sigma(\kk)
-\frac{C}{(2\pi)^2}|\tilde{\phi}_R'(k)|^2\right]\right\}
\label{eq:energy_thm_2D_V(r)}
\ee
where $\tilde{\phi}_R'(k)=\int d^3r\,e^{-i\kk\cdot\rr}\phi'_R(r)$ with
\be
\phi_R'(r)=\left[\phi(r)-\ln(R/a)\right]\,\theta(R-r).
\ee
This follows from (\ref{eq:Eint_V(r)_2D}) and from the lemma:
\be
\int d^2r\,V(r)|\phi(r)|^2=\lim_{R\to\infty}\left\{\frac{2\pi\hbar^2}{m}\ln\left(\frac{R}{a}\right)
-\int\frac{d^2k}{(2\pi)^2}\,\frac{\hbar^2k^2}{m}|\tilde{\phi}'_R(k)|^2\right\}.
\label{eq:lemme_phi'(k)_2D}
\ee
The derivation of this lemma again starts with (\ref{eq:phiDeltaphi}). Ostrogradsky's theorem then gives~[\thefnnumberter]
\be
\int d^2r\,\phi\Delta\phi=\lim_{R\to\infty}\left\{2\pi\ln\left(\frac{R}{a}\right)-\int_{r<R} d^2r\,(\mathbf{\nabla}\phi)^2\right\}.
\ee
We can then replace $\int_{r<R}d^2r\,(\nabla \phi)^2$ by $\int d^2r\,(\nabla\phi'_R)^2$, since $\phi'_R(r)$ is continuous at $r=R$~[\thefnnumberter] so that $\nabla\phi'_R$ does not contain any delta distribution. The Parseval-Plancherel relation can be applied to $\partial_i \phi'_R$, since this function is square-integrable. Then, using the fact that $\phi'_R(r)$ vanishes at infinity, we get
\be
\int d^2r\,(\nabla\phi'_R)^2 = \int \frac{d^2k}{(2\pi)^2}\,k^2|\tilde{\phi}'_R(k)|^2,
\ee
and the lemma (\ref{eq:lemme_phi'(k)_2D}) follows.

\subsubsection{Pair distribution function at short distances}
\label{subsec:g2_V(r)}
In the zero-range limit $k_{\rm typ} b \ll 1$, the short-distance behavior of the pair distribution function is given by the same expressions (\ref{eq:g2_rezo_3D},\ref{eq:g2_rezo_2D}) as for the lattice model.
Indeed, Eq.(\ref{eq:G2_AA}) is derived in the same way as for the lattice model;
one can then use the
zero-range model's expressions (\ref{eq:C_vs_AA},\ref{eq:C_vs_AA_2D})
of $(A,A)$ in terms of $C$, since the finite range model's quantities $C$ and $A$ tend to the zero-range model's ones in the zero-range limit.

\subsubsection{Momentum distribution at large momenta}\label{subsec:nk_V(r)}
In the zero-range regime $\ktyp b\ll1$ the momentum distribution at large momenta $k\gg\ktyp$ is given by
\bea
n_\sigma(\kk)&\simeq&\frac{C}{(4\pi)^2}|\tilde{\phi}(\kk)|^2
\ \ \ {\rm in}\ 3D
\label{eq:nk_V(r)_3D}
\\
n_\sigma(\kk)&\simeq&\frac{C}{(2\pi)^2}|\tilde{\phi}(\kk)|^2
\ \ \ {\rm in}\ 2D.
\eea
Indeed, Eq.(\ref{eq:nk_AA_latt}) is derived as for the lattice model,
and $(A,A)$ can be expressed in terms of $C$ as in Subsec.\ref{subsec:g2_V(r)}.

\subsection{Derivative of the energy with respect to the effective range}\label{sec:re}

We now show that, in $3D$, the leading order finite-range correction to the zero-range model's spectrum is given 
by the model-independent expression
\be
 \left(\frac{\partial E}{\partial r_e}\right)_{\!a}=
2\pi\sum_{i<j}\int d^3R\int\Big(\prod_{k\neq i,j}d^3r_k\Big)
A_{ij}(\RR,(\rr_k)_{k\neq i,j})
\left[E+\frac{\hbar^2}{4m}\Delta_{\RR}+\frac{\hbar^2}{2m}\sum_{k\neq i,j}\Delta_{\rr_k} -\sum_{l=1}^N U(\rr_l)\right]
A_{ij}(\RR,(\rr_k)_{k\neq i,j})
\label{eq:dEdre}
\ee
where $r_e$ is the effective range of the interaction potential,
the derivative is taken in $r_e=0$, the function $A$ is assumed to be real without loss
of generality, and in the sum over $l$ we have set $\rr_i=\rr_j=\RR$.
To obtain this result we use a modified version of the zero-range model, 
where the boundary condition (\ref{eq:CL_3D}) is replaced by
\be
 \psi(\rr_1,\ldots,\rr_N)\underset{r_{ij}\to0}{=}
\left( \frac{1}{r_{ij}}-\frac{1}{a}+\frac{m}{2\hbar^2}\mathcal{E} r_e
 \right) \, A_{ij}\left(
\mathbf{R}_{ij}, (\mathbf{r}_k)_{k\neq i,j}
\right)
+O(r_{ij}),
\label{eq:CL_re}
\ee
where
\be
\mathcal{E}=E-2 U(\RR_{ij})-\left(\sum_{k\neq i,j} U(\rr_k)\right)
+\frac{1}{A_{ij}\left(
\mathbf{R}_{ij}, (\mathbf{r}_k)_{k\neq i,j}
\right)}
\left[
\frac{\hbar^2}{4m}\Delta_{\RR}+\frac{\hbar^2}{2m}\sum_{k\neq i,j}\Delta_{\rr_k}\right]
A_{ij}\left(
\mathbf{R}_{ij}, (\mathbf{r}_k)_{k\neq i,j}
\right).
\label{eq:E=}
\ee
Equations (\ref{eq:CL_re},\ref{eq:E=}) generalize the ones already used for 3 bosons in free space in \cite{Efimov93,PetrovBosons}
(the predictions of \cite{Efimov93} and \cite{PetrovBosons} have been confirmed using different approaches, see \cite{PlatterRangeCorrections} and Refs. therein, and \cite{MoraBosons,LudoMattia} respectively).
Such a model was also used in the two-body case, see e.g. \cite{Greene,Tiesinga,Naidon}, and the modified scalar product that makes it hermitian
was constructed in \cite{LudovicOndeL}.

For the derivation of (\ref{eq:dEdre}), we consider an eigenstate $\psi_1$ of the zero-range model, satisfying the boundary condition (\ref{eq:CL_3D}) with a scattering length $a$ and a regular part $A^{(1)}$, and the corresponding finite-range eigenstate $\psi_2$ satisfying (\ref{eq:CL_re},\ref{eq:E=}) with the same scattering length $a$ and a regular part $A^{(2)}$.
As in App.~\ref{app:lemme} we get (\ref{eq:ostro}), as well as 
(\ref{eq:int_surface_3D}) with $1/a_1-1/a_2$ replaced by $m\mathcal{E}r_e/(2\hbar^2)$.
This yields (\ref{eq:dEdre}).

\subsection{Generalization to statistical mixtures and to thermodynamic equilibrium in the canonical ensemble} \label{subsec:finiteT}

The above results, summarized in Tables~\ref{tab:fermions}, \ref{tab:latt} and \ref{tab:V(r)}, hold for any eigenstate (apart from lines 8-9 of Tab. \ref{tab:fermions} and line 11 of Tab. \ref{tab:latt}). 
Thus they can be generalized 
straightforwardly to statistical mixtures of eigenstates.
The relation for the time derivative of $E$ (Tab. \ref{tab:fermions} line 9) holds for any time-evolving pure state, and thus also for any time-evolving statistical mixture.

We turn to the case of thermal
equilibrium in the canonical ensemble. We shall use the notation
\be
\lambda\equiv\left\{
\begin{array}{lr}
-1/a & {\rm in}\ 3D
\\
\frac{1}{2}\ln a & {\rm in}\ 2D.
\end{array}
\right.
\label{eq:def_lambda}
\ee

\paragraph{First order derivative of $E$.}

The thermal average in the canonical ensemble $\overline{dE/d\lambda}$ 
can be rewritten in the following  more familiar way,
as detailed in Appendix \ref{app:adiab}:
\be
\overline{\left(\frac{dE}{d\lambda}\right)}
=
\left(
\frac{dF}{d\lambda}
\right)_{\!T}
=
\left(
\frac{dU}{d\lambda}
\right)_{\!S}
\label{eq:relation_T}
\ee
where $\overline{(\dots)}$ is the canonical thermal average,
$F$ is the free energy, $U=\bar{E}$ is the mean energy and $S$ is the entropy.
Taking the thermal average of (\ref{eq:thm_dE_3D},\ref{eq:thm_dE_2D}) thus gives
\be
\left(
\frac{dF}{d\lambda}
\right)_{\!T}
=
\left(
\frac{dU}{d\lambda}
\right)_{\!S} 
 = 
 \frac{4\pi\hbar^2}{m} \ \overline{(A,A)}
 \ee
 The other results (\ref{eq:thm_g2_3D},\ref{eq:thm_g2_2D},\ref{eq:thm_nk_3D},\ref{eq:thm_nk_2D},\ref{eq:energy_thm_3D},\ref{eq:energy_thm_2D})
 are generalized to finite temperatures in the same way.

\paragraph{Second order derivative of $E$.}

Taking a thermal average of the expression (\ref{eq:d^2E_3D},\ref{eq:d^2E_2D}) we get after a simple manipulation:
\be
\frac{1}{2} \overline{\left( \frac{d^2 E}{d\lambda^2}\right)} = \left( \frac{4\pi\hbar^2}{m}\right)^2
 \frac{1}{2 Z}\sum_{n,n';E_n\neq E_{n'}}\frac{e^{-\beta E_n}-e^{-\beta E_{n'}}}{E_n-E_{n'}}|(A^{(n')},A^{(n)})|^2
\ee
where $Z=\sum_n \exp(-\beta E_n)$.
This implies
\be
 \overline{\left( \frac{d^2 E}{d\lambda^2}\right)}<0.
 \label{eq:d2Ebar<0}
 \ee
Moreover one can check that
\be
\left(\frac{d^2F}{d\lambda^2}\right)_T - \overline{\left( \frac{d^2 E}{d\lambda^2}\right)} = -\beta\left[\,
\overline{\left( \frac{dE}{d\lambda}\right)^{\phantom{,}2}}
- \overline{\left( \frac{dE}{d\lambda}\right)}^{\phantom{,}2} \right]<0,
\ee
which implies
\be
\left(\frac{d^2F}{d\lambda^2}\right)_T < 0.
\label{eq:thm_F}
\ee
In usual cold atom experiments, however, there is no thermal reservoir imposing
a fixed temperature to the gas, one rather can achieve
adiabatic transformations by a slow variation of the scattering
length of the gas 
\cite{SalomonMolecules,HuletMolecules,GrimmCrossover,CarrCastin,WernerAF}. 
One also more directly accesses
the mean energy $U$ of the gas rather than its
free energy, even if the entropy is also measurable \cite{thomas_entropie_PRL}. The second order derivative of $U$ with
respect to $\lambda$ for a fixed entropy is thus the relevant
quantity to consider \cite{LandauLifschitzPhysStat}. 
As shown in the appendix \ref{app:adiab}
one has in the canonical ensemble:
\be
\label{eq:d2us}
\left(\frac{d^2U}{d\lambda^2}\right)_S
=
\overline{\left(\frac{d^2E}{d\lambda^2}\right)}
+\frac
{
\left[\mbox{Cov}\!\left(E,\frac{dE}{d\lambda}\right)\right]^2
-\mbox{Var}(E)\mbox{Var}\!\left(\frac{dE}{d\lambda}\right)
}
{k_B T\,\mbox{Var}(E)}
.
\ee
where $\mbox{Var}(X)$ and $\mbox{Cov}(X,Y)$ stand for the variance
of the quantity $X$ and the covariance of the quantities $X$ and $Y$
in the canonical ensemble, respectively.
From the Cauchy-Schwarz inequality 
$[\mbox{Cov}(X,Y)]^2\leq \mbox{Var}(X)\mbox{Var}(Y)$,
and from the inequality (\ref{eq:d2Ebar<0}), we thus conclude that
\be
\left(\frac{d^2U}{d\lambda^2}\right)_S < 0.
\label{eq:thm_d^2E_S}
\ee

To be complete, we also consider the process where 
$\lambda$ is varied so slowly that there is adiabaticity in the many-body
quantum mechanical sense:
The adiabatic theorem of quantum mechanics~\cite{AdiabThmKato} 
implies that in the limit where $\lambda$ is changed infinitely slowly, 
the occupation probabilities of each eigenspace of the many-body Hamiltonian 
do not change with time, 
even in presence of level crossings~\cite{AdiabThmAvron}. 
We note that this may require macroscopically long evolution times
for a large system.
For an initial equilibrium state in the
canonical ensemble, the mean energy then varies with $\lambda$ as
\be
E^{\rm quant}_{\rm adiab}(\lambda)=
\sum_n \frac{e^{-\beta_0 E_n(\lambda_0)}}{Z_0}\,E_n(\lambda)
\label{eq:Ead1}
\ee
where the subscript $0$ refers to the initial state.
Taking the second order derivative of (\ref{eq:Ead1}) with respect
to $\lambda$ in $\lambda=\lambda_0$ gives
\be
\frac{d^2 E_{\rm adiab}^{\rm quant}}{d\lambda^2}
= \overline{\left(\frac{d^2E}{d\lambda^2}\right)}
<0.
\label{eq:d^2E_quant}
\ee

Finally, we compare the result of isentropic transformation~(\ref{eq:d2us}) to the one of the adiabatic transformation in the quantum sense~(\ref{eq:d^2E_quant}). They differ by the second term in the right hand side of~(\ref{eq:d2us}). A priori this term is extensive, and thus not negligible compared to the first term. We have explicitly checked this expectation for the Bogoliubov model Hamiltonian of a weakly interacting Bose gas.
The discrepancy between (\ref{eq:d2us}) and (\ref{eq:d^2E_quant}) indicates that the limit of infinitely slow transformation does not commute with the thermodynamic limit.
More explicitly, we see that if $\lambda$ is varied so slowly that the quantum adiabaticity (\ref{eq:Ead1}) is achieved, one cannot assume any more that the system follows a sequence of thermal equilibrium states with a constant entropy;
in practice this may require evolution times which grow exponentially with the system size.


\section{Spinless bosons} \label{sec:bosons}

The wavefunction $\psi(\rr_1,\ldots,\rr_N)$ is now completely symmetric.

Our results for bosons are shown in Table~\ref{tab:bosons}. 
An obvious difference with the fermionic case is that there are no more spin indices in the pair distribution function $g^{(2)}$ and in the momentum distribution $n(\kk)$. Accordingly, Eqs.~(\ref{eq:def_g2_fermions},\ref{eq:def_nk_fermions}) are replaced by~\footnote{In second quantization, (\ref{eq:def_g2_bosons}) corresponds to $g^{(2)}\left(\RR+\rr/2,\RR-\rr/2\right)=\langle
\hat{\psi}^\dagger(\RR+\rr/2)
\hat{\psi}^\dagger(\RR-\rr/2)
\hat{\psi}(\RR-\rr/2)
\hat{\psi}(\RR+\rr/2)
\rangle$.
}
\be
g^{(2)}\left(\RR+\frac{\rr}{2},\RR-\frac{\rr}{2}\right)
=
\int d\rr_1\ldots d\rr_N 
\left|
\psi(\rr_1,\ldots,\rr_N)
\right|^2
\sum_{i\neq j}
\delta\left(\RR+\frac{\rr}{2}-\rr_i\right)
\delta\left(\RR-\frac{\rr}{2}-\rr_j\right)
\label{eq:def_g2_bosons}
\ee
and
\be
\int \frac{d^d k}{(2\pi)^d} n(\kk) = N.
\ee

An important difference with the fermionic case is that in $3D$,
 the Efimov effect occurs~\cite{Efimov}, and the zero-range model is defined not only by the contact condition~(\ref{eq:CL_3D}) and the Schr\"odinger equation~(\ref{eq:schroZR}), but also by a boundary condition in the limit where three particles approach each other: There exists a function $B$, hereafter called three-body regular part, such that
\be
\psi(\rr_1,\ldots,\rr_N)\underset{R\to0}{\sim}\frac{1}{R^2}\sin\left[|s_0|\ln\frac{R}{R_t}\right] \varPhi(\Oom)\, B(\CC,\rr_4,\ldots,\rr_N).
\label{eq:danilov}
\ee
where $R_t$ is the so-called three-body parameter and is an additional parameter of the zero-range model;
$\CC=(\rr_1+\rr_2+\rr_3)/3$ is the center of mass of particles $1$,$2$ and $3$;
$R$ and $\Oom$ are the hyperradius and the hyperangles associated with particles $1$,$2$ and $3$
~\footnote{See Chapter 3 in \cite{WernerThese}.}
\setcounter{fnnumberthesechaptroissectrois}{\thefootnote}
We recall the definition of $R$ and $\Oom$: The Jacobi coordinates are defined by $\rr=\rr_2-\rr_1$ and $\rhob=(2\rr_3-\rr_1-\rr_2)/\sqrt{3}$; then, $R\equiv\sqrt{(r^2+\rho^2)/2}$, and $\Oom\equiv(\alpha,\hat{r},\hat{\rho})$ with $\alpha\equiv {\rm arctan}(r/\rho)$, $\hat{r}\equiv\rr/r$ and $\hat{\rho}\equiv\rhob/\rho$.
$s_0=i\cdot1.00624\ldots$ is Efimov's transcendental number, it is the imaginary solution of (\ref{eq:s}).
$\varPhi(\Oom)$ is the normalized hyperangular part of an Efimov states' wavefunction: $\varPhi(\Oom)=\phi_{s_0}(\Oom)/\sqrt{(\phi_{s_0}|\phi_{s_0})}$ [\thefnnumberthesechaptroissectrois]. We recall that, in the present case (bosons with zero total angular momentum), $\phi_{s_0}(\Oom)\equiv(1+Q)\sin\left[s_0\left(\frac{\pi}{2}-\alpha\right)\right]/[\sqrt{4\pi}\sin(2\alpha)]$ where $Q=P_{13}+P_{23}$ and $P_{ij}$ exchanges particles $i$ and $j$.
The hyperangular scalar product is defined by $(\phi|\phi)=\int d\Oom\,|\phi(\Oom)|^2$ with $\int d\Oom\equiv2\int_0^{\pi/2}d\alpha\sin^2(2\alpha)\int d\hat{r}\int d\hat{\rho}$, where $d\hat{r}$ and $d\hat{\rho}$ are the differential solid angles. Its value is given in App.~\ref{app:Efi_psi}.

For $N=3$ it is well established that this model is self-adjoint and that it is the zero-range limit of finite-range models, 
see e.g.~\cite{WernerThese} and references therein.
A recent numerical study of several finite range models predicted for $N=4$ the existence of tetramers of energies weakly depending on the model for
a fixed three-body parameter
\cite{Stecher4corps}, an existence confirmed experimentally~\cite{Grimm4body}. This numerical study, together with other ones~\cite{Hammer4corps1,Hammer4corps2}
claim that there is no need to introduce a four-body parameter in the zero-range limit,
implying that the here considered zero-range model is self-adjoint for $N=4$.
Here we consider an arbitrary value of $N$ such that the model is self-adjoint.

The main relations for zero-range interactions are displayed in Table~\ref{tab:bosons}.
Moreover we note that the relations for finite-range interactions, given in Tables \ref{tab:latt} and \ref{tab:V(r)} for fermions, can be easily generalized to the bosonic case.

\subsection{Relations which are analogous to the fermionic case}
The derivations of all relations of Table~\ref{tab:bosons}  are completely analogous to the fermionic case, except for lines 4 and 6 of the left
column (3D case).

The first result in Table~\ref{tab:bosons}  was first obtained in~\cite{WernerThese} in the case $N=3$.
A simple way to derive it for any $N$ is to use the lattice model
and to apply the reasoning of Sec.\ref{subsec:dE_latt}. The key point is that in the limit of a lattice spacing $b$ much smaller than $|a|$, the three-body parameter corresponding to the lattice model is equal to a numerical constant times $b$~\footnote{The value of this constant is irrelevant for what follows. It could be calculated e.g. by equating the energies of the weakly bound Efimov trimers of the lattice model with the ones of the zero-range model. This was done in~\cite{Werner3corpsPRL,WernerThese}, not for the lattice model, but for a Gaussian separable potential model.}. Thus, varying the coupling constant $g_0$ while keeping $b$ fixed
is equivalent to 
varying $a$ while keeping $R_t$ fixed:
\be
\frac{dE}{dg_0}=\left(\frac{dE}{d(-1/a)}\right)_{\!R_t}\ \frac{d(-1/a)}{dg_0}
\label{eq:dEdg0_Rt}
\ee
where the lattice model's $(dE/d(-1/a))_{R_t}$ tends to the zero-range model's one 
if one takes the zero-range limit while keeping $R_t$ fixed~\footnote{The zero-range limit for a fixed $R_t$ can be taken by repeatedly dividing $b$ by the discrete scaling factor $\exp(\pi/|s_0|)$
where $s_0$ is defined in~(\ref{eq:def_s0}) and by adjusting $g_0$ so that $a$ remains fixed. In this limit the ground state energy tends to $-\infty$ as follows from the Thomas effect, but 
the restriction of the spectrum to any fixed energy window converges (see e.g.~\cite{WernerThese}).}.
\setcounter{fnnumberbis}{\thefootnote}
The same reasoning explains why the second derivative in the last line of Table~\ref{tab:bosons} 
also has to be taken for a fixed $R_t$.

\subsection{Derivative of the energy with respect to the three-body parameter}

The Efimov effect also gives rise to the following new relation between the derivative of the energy with respect to the three-body parameter and the three-body regular part:
\be
\left(\frac{\partial E}{\partial \ln R_t}\right)_a=\frac{\hbar^2}{m}\frac{\sqrt{3}}{32}|s_0|^2 N(N-1)(N-2)\int d\CC\int d\rr_4\ldots d\rr_N\,|B(\CC,\rr_4,\ldots,\rr_N)|^2.
\label{eq:dEdRt}
\ee
This is similar to the relation 
(\ref{eq:thm_dE_3D})
between the derivative with respect to the scattering length and the (two-body) regular part
\footnote{We note that it was already speculated in \cite{Braaten} that, in presence of the Efimov effect, ``a three-body analog of the contact'' may ``play an important role''.}.
We will first derive this relation using the zero-range model in the case $N=3$, and then using a lattice model for any $N$.

\subsubsection{Derivation using the zero-range model for three particles}
We consider two wavefunctions $\psi_1$, $\psi_2$, satisfying the two-body boundary condition (\ref{eq:CL_3D}) with the same scattering length $a$, and satisfying the three-body boundary condition with different three-body parameters $R_{t 1}$, $R_{t 2}$. The corresponding three-body regular parts are denoted by $B_1$, $B_2$.
We show in the App.~\ref{app:3b} that
\be
\la \psi_1, H \psi_2\ra-\la H\psi_1,\psi_2\ra=\frac{\hbar^2}{m}\frac{3\sqrt{3}|s_0|}{16}\sin\left[|s_0|\ln\frac{R_{t 2}}{R_{t 1}}\right]\,\int d\CC\,B^*_1(\CC)B_2(\CC),
\label{eq:lemme_dEdRt}
\ee
which yields (\ref{eq:dEdRt}) by choosing $\psi_i$ as an eigenstate of energy $E_i$ and taking the limit $R_{t 2}\to R_{t 1}$.
We note that $\psi_1$ and $\psi_2$ do not satisfy lemma (\ref{eq:lemme_3D}) because they are too singular for $R\to0$.

\subsubsection{Derivation using a lattice model}

We now rederive (\ref{eq:dEdRt}) using a lattice model which is analogous to the model defined in Sec.\ref{sec:models:lattice}, except that the Hamiltonian now contains a three-body interaction term:
\be
H=\int_D \frac{d^3k}{(2\pi)^3}\,\epsk c^\dagger(\kk)c(\kk) + \sum_{\rr} b^3 U(\rr) (\psi^\dagger \psi)(\rr) +g_0 \sum_\rr b^3 (\psi^\dagger\psi^\dagger\psi\psi)(\rr)
+h_0 \sum_\rr b^3 (\psi^\dagger\psi^\dagger\psi^\dagger\psi\psi\psi)(\rr).
\ee
The scattering length is related to $g_0$ as in Sec.\ref{sec:models:lattice}.
We define the zero-energy three-body scattering state $\phi_0(\rr_1,\rr_2,\rr_3)$
as the solution of $H|\phi_0\ra=0$ for $a=\infty$, with the boundary condition
\be
\phi_0(\rr_1,\rr_2,\rr_3)\sim \frac{1}{R^2}\sin\left[|s_0|\ln\frac{R}{R_t}\right] \varPhi(\Oom)
\ee
in the limit where all interparticle distances tend to infinity.
This defines the three-body parameter $R_t(b,h_0)$ for the lattice model.
In order to derive (\ref{eq:dEdRt}) for any desired values of $a$ and $R_t$, we first choose $b$ in such a way that $R_t(b,h_0=0)$ is equal to the desired $R_t$ divided by $e^{n\pi/|s_0|}$, the zero-range limit $n\to\infty$ being taken in the end. We then choose $g_0$ to reproduce the desired value of $a$. We then change $h_0$ to a small non-zero value, keeping fixed $b$ and $g_0$ (and thus also $a$). The Hellman-Feynman theorem writes:
\be
\frac{\partial E}{\partial h_0} = \sum_{\rr} b^3\,\la (\psi^\dagger\psi^\dagger\psi^\dagger\psi\psi\psi)(\rr)\ra=N(N-1)(N-2)\sum_{\rr_4,\ldots,\rr_N}b^{3(N-3)}|\psi(\rr,\rr,\rr,\rr_4,\ldots,\rr_N)|^2.
\ee
For the lattice model we define the three-body regular part $B$ through:
\be
\psi(\rr,\rr,\rr,\rr_4,\ldots,\rr_N)=\phi_0(\vn,\vn,\vn)\,B(\rr,\rr_4,\ldots,\rr_N);
\ee
in the zero-range limit, we expect that this lattice model's regular part tends to the regular part of the zero-range model defined in (\ref{eq:danilov}), as was discussed for the two-body regular part $A$ in Sec. \ref{sec:models:lattice}.
We thus have, in the zero-range limit:
\be
\left( \frac{\partial E}{\partial(\ln R_t)}\right)_a=N(N-1)(N-2)
|\phi_0(\vn,\vn,\vn)|^2 \left( \frac{\partial h_0}{\partial(\ln R_t)}\right)_{\!b}
\int d^3 r\, d^3 r_4\ldots d^3 r_N\,|B(\rr,\rr_4,\ldots,\rr_N)|^2.
\label{eq:proto_dEdRt}
\ee
It remains to evaluate the expression between brackets: This is achieved by applying (\ref{eq:proto_dEdRt}) to the case of an Efimov trimer in free space, where the regular part can be deduced from the expression for the normalized wavefunction given in App.~\ref{app:Efi_psi}.

\subsection{Momentum distribution for an Efimov trimer}

For an Efimov trimer state at rest, for an infinite scattering length,
we show in Appendix~\ref{app:efimov}  that the atomic momentum distribution has the asymptotic expansion
\be
n(k)\underset{k\to\infty}{=} \frac{C}{k^4}+
\frac{D}{k^5} \cos\left[2|s_0|\ln(k\sqrt{3}/\kappa_0)+\varphi\right]+\ldots
\label{eq:nk_efi}
\ee
where
$s_0=i\cdot1.00624\ldots$
solves
\be
s\cos(s\pi/2)-8/\sqrt{3}\sin(s\pi/6)=0,
\label{eq:def_s0}
\ee
the energy of the trimer $E_{\rm trim}=-\hbar^2 \kappa_0^2/m$ depends on the three-body parameter
$R_t$ as specified in (\ref{eq:kappa0}), and the quantities $C$, $D$ and $\varphi$ are derived in the appendix
\ref{app:efimov}.
The crucial point is that $D\neq 0$: The momentum distribution has a slowly decaying oscillatory subleading tail. 

\begin{figure}[htb]
\includegraphics[width=0.445\linewidth,clip=]{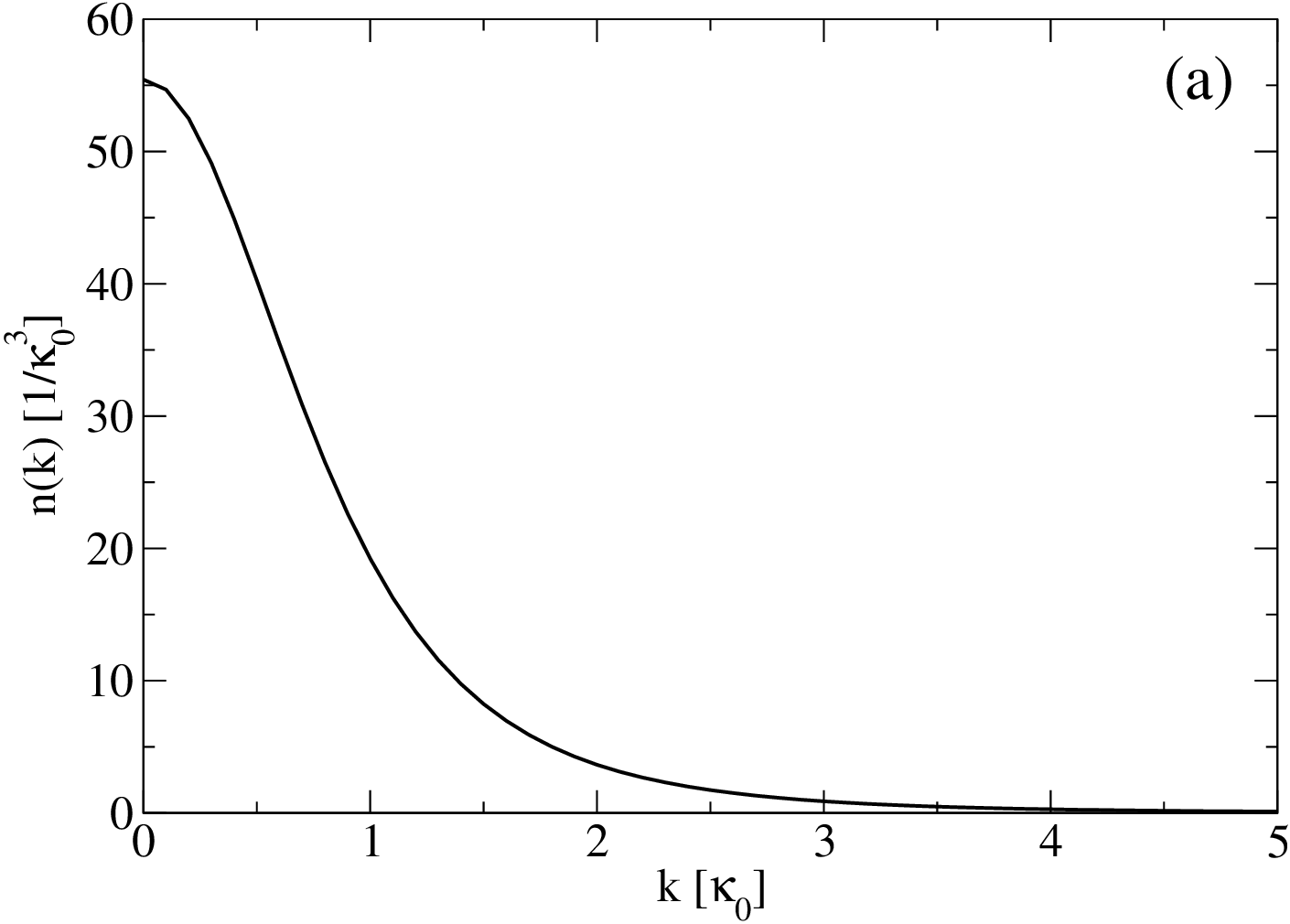}
\hfill
\includegraphics[width=0.46\linewidth,clip=]{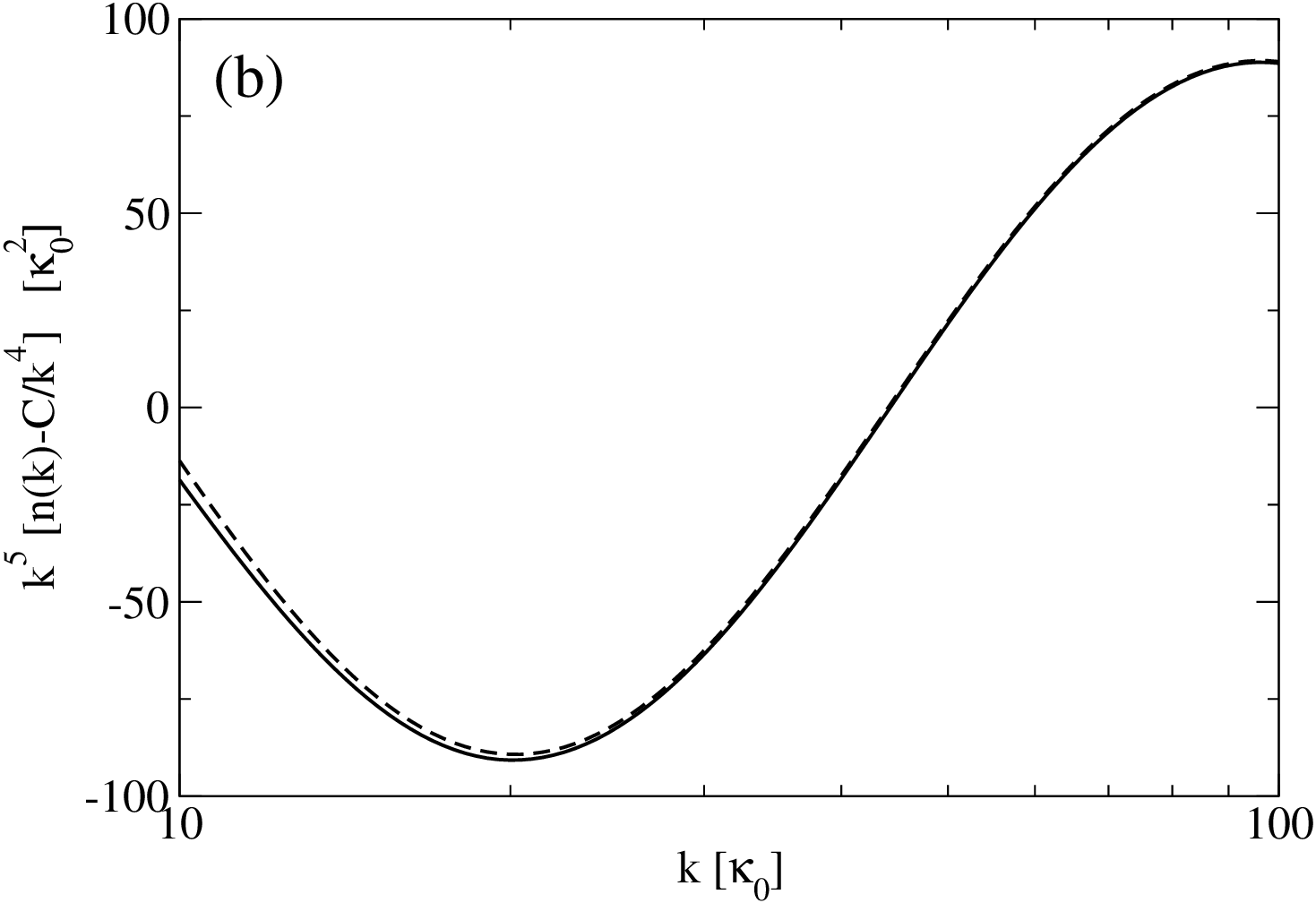}
\caption{For a free space Efimov trimer at rest
composed of three bosonic atoms of mass $m$
interacting {\sl via} a zero range, 
infinite scattering length potential, atomic momentum distribution $n(k)$ 
as a function of $k$. (a) Numerical calculation from the expressions 
derived in the appendix \ref{app:efimov}. (b) Numerical calculation
(solid line) and asymptotic behavior (\ref{eq:nk_efi}) (dashed line),
with the horizontal axis in log scale.
The unit of momentum is $\kappa_0$, such that the trimer energy
is $-\hbar^2/m\kappa_0^2$.
\label{fig:nk}}
\end{figure}

The calculations performed 
in appendix \ref{app:efimov} also allow a straightforward
numerical calculation of the atomic momentum distribution 
for an
Efimov trimer, both for
low values of $k$, see Fig.\ref{fig:nk}a, and for high values of $k$,
see Fig.\ref{fig:nk}b showing how $n(k)$ approaches the asymptotic
behavior (\ref{eq:nk_efi}).
We have also derived in appendix \ref{app:efimov}
the exact value in $\kk=\mathbf{0}$:
\be
\label{eq:n_ori}
n(\kk=\mathbf{0}) =  \frac{55.43379775608\ldots}{\kappa_0^3}.
\ee

\subsection{Breakdown of the energy-momentum relation in the zero-range model}

\subsubsection{A non-converging integral}
As a consequence of (\ref{eq:nk_efi}), the integral $\int d^3 k\,k^2 [n(k)-C/k^4]$ is not well-defined: 
After the change of variables $x=\ln k$, the integrand behaves for $x\to\infty$ as a linear superposition
of $e^{i |s_0| x}$ and $e^{-i |s_0| x}$, that is 
as a periodic function of $x$ oscillating around zero. 
This was overlooked in~\cite{CombescotC}.

\subsubsection{Failure of a naive regularisation}\label{sec:failure_naive}

At first sight, however, this does not look too serious: one often argues, when one faces
the integral of such an oscillating function of zero mean, that the oscillations 
at infinity simply average to zero.
More precisely, let us define the cut-off dependent energy of the Efimov trimer (here $1/a=0$):
\be
E(\Lambda) = \int_{k<\Lambda} \frac{d^3\!k}{(2\pi)^3}  \frac{\hbar^2 k^2}{2m}
\left[n(k)-\frac{C}{k^4}\right].
\label{eq:elam}
\ee
For $\Lambda\to\infty$, $E(\Lambda)$ is asymptotically an oscillating function of the logarithm of $\Lambda$,
oscillating around a mean value $\bar{E}$.
The naive expectation would be that the trimer energy $E_{\rm trim}$ equals $\bar{E}$.
This naive expectation is equivalent to the usual trick used to regularize oscillating integrals, consisting
here in introducing a convergence factor $e^{-\eta \ln(k/\kappa_0)}$ in the integral
without momentum cut-off and then taking the limit $\eta\to 0^+$:
\be
\lim_{\eta\to 0^+} \int_{\mathbb{R}^3} \frac{d^3\!k}{(2\pi)^3}  \frac{\hbar^2 k^2}{2m}
\left[n(k)-\frac{C}{k^4}\right] e^{-\eta \ln(k/\kappa_0)} = \bar{E}.
\label{eq:espoir}
\ee
We have decided to really test this naive regularisation.
As we now show, remarkably, it actually fails to give the correct energy of
the Efimov trimer.

\paragraph{Numerical evidence for $E\neq \bar{E}$.}

\begin{figure}[htb]
\includegraphics[width=0.8\linewidth,clip=]{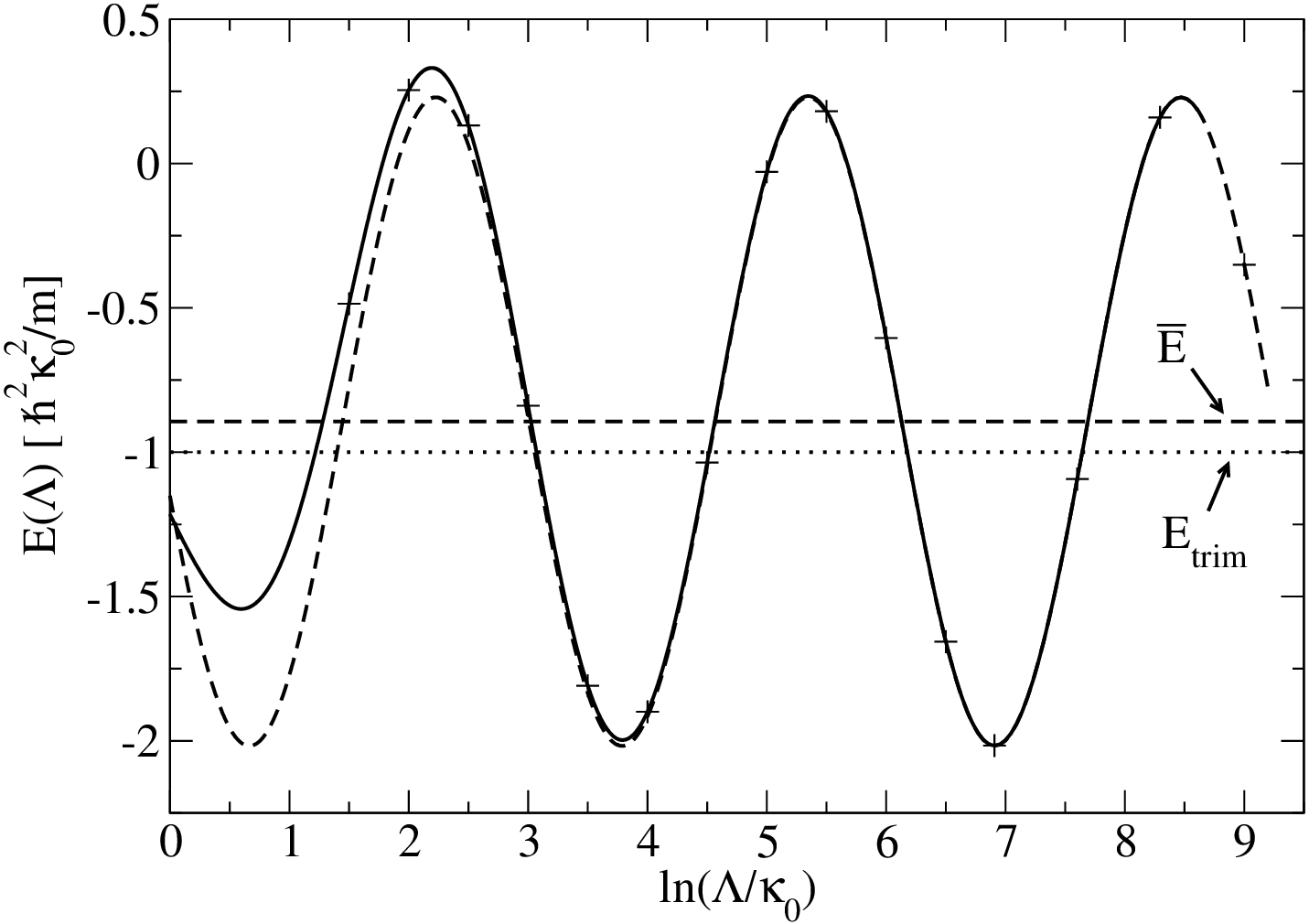}
\caption{Cut-off dependent energy $E(\Lambda)$ as defined in (\ref{eq:elam}) for a free space infinite scattering length Efimov trimer
with a zero range interaction, as a function of the logarithm of the momentum cut-off $\Lambda$.
Solid line: numerical result obtained {\sl via} a calculation of the momentum distribution
$n(k)$. Symbols $+$: direct numerical calculation of $E(\Lambda)$ as exposed
in the Appendix \ref{app:brutale}. Dashed sinusoidal line: asymptotic oscillatory behavior of $E(\Lambda)$ 
for large $\Lambda$, obtained in omitting $O(1/\Lambda)$ in (\ref{eq:elamasympt}). 
Dashed horizontal line: mean value $\bar{E}$ around which $E(\Lambda)$ oscillates
at large $\Lambda$. The values of $\bar{E}$ obtained analytically 
(\ref{eq:ebar_analy}) and numerically (\ref{eq:nest}) are indistinguishable at the scale
of the figure, and clearly deviate from the dotted line giving the true energy
$E_{\rm trim}$
of the trimer, exemplifying the failure of a at a first sight convincing application
of an energy-momentum relation for bosons in three dimensions. 
The unit of momentum $\kappa_0$ is such that the true trimer energy
is $E_{\rm trim}=-\hbar^2 \kappa_0^2/m$.
\label{fig:elam}}
\end{figure}

We first performed a numerical calculation of $E(\Lambda)$, using the results of 
Appendix~\ref{app:efimov} to perform a very accurate
numerical calculation of $n(k)$. The result is shown as a solid line in Fig.\ref{fig:elam}.
We also developed a more direct technique allowing a numerical calculation
of $E(\Lambda)$ without the knowledge of $n(k)$, see Appendix \ref{app:brutale}:
The corresponding results are represented as $+$ symbols in Fig.\ref{fig:elam}
and are in perfect agreement with the solid line.
As expected, $E(\Lambda)$ is asymptotically an oscillating function of the logarithm of $\Lambda$,
oscillating around a mean value $\bar{E}$.
This may be formalized as follows. We introduce an arbitrary, non-zero value
$k_{\rm min}$ of the momentum, and we define
\bea
\label{eq:defdn1}
\delta n(k) &\equiv& n(k) -\frac{C}{k^4} \ \ \ \mbox{for}\ k<k_{\rm min} \\
\delta n(k) &\equiv& n(k) - \left\{\frac{C}{k^4}+ \frac{D}{k^5} \cos\left[2|s_0|\ln(k\sqrt{3}/\kappa_0)+\varphi\right]\right\} 
\ \ \ \mbox{for}\ k>k_{\rm min} .
\label{eq:defdn2}
\eea
The introduction of $k_{\rm min}$ ensures that the integral of $k^2 \delta n(k)$ over
$\kk$ converges around
$\kk=\mathbf{0}$. The subtraction of the asymptotic behavior of $n(k)$ up to order $O(1/k^5)$
for $k>k_{\rm min}$ ensures that the integral of $k^2 \delta n(k)$ 
over $\mathbb{R}^3$ converges at infinity.
As a consequence we get for $\Lambda > k_{\rm min}$ the splitting
\be
E(\Lambda) = \int_{k<\Lambda} \frac{d^3k}{(2\pi)^3} \frac{\hbar^2 k^2}{2m} \delta n(k)
+\int_{k_{\rm min}<k<\Lambda} \frac{d^3k}{(2\pi)^3} \frac{\hbar^2 k^2}{2m} \frac{D}{k^5} \cos\left[2|s_0|\ln(k\sqrt{3}/\kappa_0)+\varphi\right].
\ee
With the change of variable $x=\ln(k\sqrt{3}/\kappa_0)$, one can calculate the second integral explicitly.
Since the first integral converges in the limit $\Lambda\to +\infty$ we obtain 
\be
E(\Lambda) = \bar{E} + \frac{\hbar^2 D}{8\pi^2 m |s_0|} \sin[2|s_0| \ln(\Lambda\sqrt{3}/\kappa_0)+\varphi] + O(1/\Lambda),
\label{eq:elamasympt}
\ee
with
\be
\bar{E} = -\frac{\hbar^2 D}{8\pi^2 m |s_0|} \sin[2|s_0| \ln(k_{\rm min}\sqrt{3}/\kappa_0)+\varphi] + \int_0^{+\infty} dk\,
\frac{\hbar^2 k^4}{4\pi^2 m}\delta n(k).
\label{eq:qveb}
\ee
From this last equation and the numerical calculations of $n(k)$ first up to $k=1000\kappa_0$ and
then up to $k\simeq 5500\kappa_0$, 
we get two slightly different values of $\bar{E}$ which give
an estimate with an error bar \footnote{The following tricks are used in the numerical 
calculation of the integral appearing in (\ref{eq:qveb}). 
The integral is split in $\int_0^{k_{\rm max}} + \int_{k_{\rm max}}^{+\infty}$.
A linear scale is used to discretized $k/\kappa_0$ in between 0
and 10. For $k$ in between 10 and $k_{\rm max}$ ($k_{\rm max}$ is either 1000 or 5500), $\ln(k/\kappa_0)$
is discretized on a linear scale. The integral from $k_{\rm max}$ to infinity
is estimated with the formula $\delta n(k)\simeq -36/(k/\kappa_0)^6$, an approximate asymptotic
expression that was tested numerically over the range $100<k/\kappa_0<1000$.
A simple test of this overall procedure is to check the normalization of $n(k)$. It is found that the
numerical calculation gives the correct normalization factor within a 
$\simeq 10^{-6}$ relative error.
We note that calculating $\delta n(k)$ with a one percent error for $k=5500$
requires a calculation of $n(k)$ with a relative error $\simeq 2\times 10^{-10}$.
}:
\be
\bar{E} \simeq 0.89397(3) E_{\rm trim}.
\label{eq:nest}
\ee
The key point is that $\bar{E}$ significantly differs from $E_{\rm trim}$:
the naive regularisation does not give the correct value of the trimer energy!

\paragraph{Physical explanation and expression of $\bar{E}$.}
We now show that an analytical expression for $\bar{E}$ may be obtained, which confirms
the numerical conclusion and has the crucial advantage
of explaining the physics behind the discrepancy $\bar{E}\neq E_{\rm trim}$.

The first step is to realize what happens in a finite range interaction model, when
one takes the zero range limit.
E.g.\ in the lattice model, for a non-zero value
of the lattice spacing $b$, one readily realizes that an exact energy-momentum
relation holds for an arbitrary number of bosons even in three dimensions.
Repeating the reasoning of Sec.~\ref{sec:energy_thm} while
keeping finite the lattice spacing $b$, we get
\be
E'-E'_{\rm trap}=\frac{\hbar^2 C'}{8\pi m a}
+\int_{[-\pi/b;\pi/b]^3} \frac{d^3 k}{(2\pi)^3}\frac{\hbar^2 k^2}{2m}
\left[n'(\kk)-\frac{C'}{k^4}\right]
\label{eq:energy_thm_reseau}
\ee
where
\be
C'\equiv\frac{8\pi m}{\hbar^2}\frac{dE'}{d(-1/a)},
\ee
and the prime denotes quantities calculated within the lattice model.
If one then takes the zero-range limit by repeatedly 
dividing $b$ by the discrete scaling factor~[\thefnnumberbis], so as to ensure
a well-defined value for the three-body parameter $R_t$,
the lattice model's quantities $E'$, $E'_{\rm trap}$, $C'$ and $n'(\kk)$ 
converge to the zero-range model's quantities $E$, $E_{\rm trap}$, $C$ and $n(\kk)$;
however one cannot simply remove the primes and replace the integration domain by $\mathbb{R}^3$ in (\ref{eq:energy_thm_reseau}), 
because this would lead to an ill-defined integral, as we have seen.
This paradox is due to the fact that the finite range interaction $b$ in the lattice model
has two effects, that conspire to ensure that the energy-momentum relation (\ref{eq:energy_thm_reseau})
is correct: (i) it introduces a momentum cut-off of order $1/b$, and 
(ii) it changes the large momentum tails of the 
momentum distribution
in a way that tends to zero when $b\to 0$ for fixed $k$, and that still
can have a non-zero impact on the energy-momentum relation since the integral
of $k^2/k^5$ is UV divergent in 3D.
The previous failure of the energy-momentum relation for the zero range model
is thus due to the fact that we have introduced a cut-off (\ref{eq:elam})
or a regularisation (\ref{eq:espoir}) 
in the integral over $\kk$ {\sl without} subsequently modifying in an appropriate
way the coefficient of the $1/k^5$ part of $n(k)$.

In a second step, we introduce a {\sl consistent} way of performing a UV regularisation,
that will both show up in the integral over $\kk$ in the energy-momentum relation 
and affect the tail of the momentum distribution.
To be explicit, we turn back to the example of $N=3$ bosons forming an Efimov
trimer in free space, of energy $E_{\rm trim}=-\hbar^2 \kappa_0^2/m$,
for an infinite scattering length.
In the zero range model, when the positions $\rr_1$ and
$\rr_2$ of the particles $1$ and $2$ symmetrically converge 
to a common point $\RR_{12}$ in space, the three-body wavefunction diverges
as $B(2|\rr_3-\RR_{12}|/\sqrt{3})/(-4\pi r_{12})$ where the function $B$ is known,
see Appendix \ref{app:efimov}. The one body momentum distribution has 
an explicit integral expression in terms of the Fourier transform $\tilde{B}(k)$ 
of this function $B$, as shown in that Appendix. The idea is then to introduce
a smooth regularisation simply replacing $\tilde{B}(k)$ with
\footnote{The argument of the exponential
in (\ref{eq:replace}) is a regular function of $k$ around $k=0$, contrarily 
to the more primitive choice $e^{-2\eta\ln(k/\kappa_0)}$, and is quite
natural considering (\ref{eq:Bt}) and (\ref{eq:sacdv}).}
\be
\tilde{B}_{\eta} (k) \equiv \tilde{B}(k)  \,
e^{-\eta\ln\left[\sqrt{1+k^2/\kappa_0^2}+k/\kappa_0\right]}
\label{eq:replace}
\ee
and eventually take the limit $\eta\to 0^+$. 
 Performing the replacement 
(\ref{eq:replace}) in the expression (\ref{eq:decomp},\ref{eq:nI},\ref{eq:nII},\ref{eq:nIII},\ref{eq:nIV}) of the momentum distribution 
leads to a modified momentum distribution $n_{\eta}(k)$,
with an asymptotic behavior $n_{\eta}^{\rm asymp}(k)$ for $k\to +\infty$
that is modified as compared to (\ref{eq:nk_efi}): 
$n_\eta(k) \underset{k\to+\infty}{=} n_{\eta}^{\rm asymp}(k) + O(1/k^6)$ with
\be
n^{\rm asymp}_\eta(k) = \frac{C_\eta}{k^4}+
\frac{e^{-2\eta\ln(k\sqrt{3}/\kappa_0)}}{k^5} \left\{\bar{D}_\eta +
D_\eta \cos\left[2|s_0|\ln(k\sqrt{3}/\kappa_0)+\varphi_\eta\right]\right\}.
\label{eq:nketaasymp}
\ee
In the limit $\eta\to 0^+$ one has to recover (\ref{eq:nk_efi})
so that $C_\eta\to C$, $D_\eta\to D$, $\varphi_\eta\to \varphi$ and $\bar{D}_\eta\to 0$.
The expressions of $C_\eta$ and $\bar{D}_\eta$ are given in the Appendix \ref{app:douce}
and confirm these requirements. What shall play a crucial role in what follows
is that, however, $\bar{D}_\eta$ is not zero for $\eta>0$, it vanishes linearly with $\eta$
in $\eta=0$.
For this {\sl consistent} regularisation, the energy-momentum relation holds in the limit of
vanishing $\eta$ for the Efimov trimer, as we prove in Appendix \ref{app:pour_les_sceptiques}
\be
E_{\rm trim} = \lim_{\eta\to 0^+} E_{\eta}\ \ \ \mbox{with}\ \ \ \ 
E_\eta\equiv \int_{\mathbb{R}^3} \frac{d^3k}{(2\pi)^3} 
\frac{\hbar^2 k^2}{2m} \left[n_\eta(k) - \frac{C_\eta}{k^4}\right].
\label{eq:emrfsr}
\ee
The definitions (\ref{eq:defdn1}),(\ref{eq:defdn2}) are then modified as
\bea
\delta n_\eta(k) &\equiv& n_\eta(k) -\frac{C}{k^4} \ \ \ \mbox{for}\ k<k_{\rm min} \\
\delta n_\eta(k) &\equiv& n_\eta(k) - n_\eta^{\rm asymp}(k)
\ \ \ \mbox{for}\ k>k_{\rm min} .
\eea
This results in the splitting
\be
E_\eta = \int_{\mathbb{R}^3} \frac{d^3k}{(2\pi)^3} 
\, \frac{\hbar^2 k^2}{2m} \delta n_\eta(k)
+ \frac{\hbar^2}{4\pi^2 m} 
\int_{x_{\rm min}}^{+\infty}
dx\, e^{-2\eta x} [\bar{D}_\eta + D_\eta \cos(2|s_0|x+\varphi_\eta)]
\label{eq:eta_exp}
\ee
where the change of variable $x=\ln(k\sqrt{3}/\kappa_0)$ was used
so that $x_{\rm min}=\ln(k_{\rm min}\sqrt{3}/\kappa_0)$.
For $\eta\to 0^+$, we can replace in the right hand side of (\ref{eq:eta_exp}) 
$\delta n_\eta(k)$ with $\delta n(k)$ since
the first integral converges absolutely, but we cannot exchange the limit
and the integration in the second integral.
After explicit calculation of this second integral, we take
$\eta\to 0^+$ and  we recognize $\bar{E}$
from (\ref{eq:qveb}) so that
\be
E_{\rm trim} = \bar{E} + \frac{\hbar^2}{8\pi^2 m} \lim_{\eta\to0^+}
\frac{\bar{D}_\eta}{\eta}.
\label{eq:tec}
\ee
As detailed in the Appendix \ref{app:douce}, $\bar{D}_\eta$ and the above limit
may be expressed as single integrals, which allows to evaluate $\bar{E}$ with a good
precision:
\be
\bar{E} = 0.8939667780883\ldots E_{\rm trim}
\label{eq:ebar_analy}
\ee
This confirms the numerical estimate (\ref{eq:nest}).

\section{Arbitrary mixture}\label{sec:melange}

In this Section we consider a mixture of bosonic and/or fermionic atoms with an arbitrary number of spin components. The $N$ particles are thus divided into groups, each group corresponding to a given chemical species and to a given spin state. We label these groups by an integer $\sigma\in\{1,\ldots,n\}$.
Assuming that there are no spin-changing collisions, the number $N_\sigma$ of atoms in each group is fixed, and one can consider that particle $i$ belongs to the group $\sigma$ if $i\in I_\sigma$, where the $I_\sigma$'s are a fixed partition of
$\{1,\ldots,N\}$ which can be chosen arbitrarily. For example, 
a possible choice is $I_1=\{1,\ldots,N_1\}$; $I_2=\{N_1+1,\ldots,N_1+N_2\}$; etc.
 The wavefunction $\psi(\rr_1,\ldots,\rr_N)$ is then symmetric (resp. antisymmetric) with respect to the exchange of two particles  belonging to the same group $I_\sigma$ of bosonic (resp. fermionic) particles.
Each atom has a mass $m_i$ and is subject to a trapping potential $U_i(\rr_i)$, and the scattering length between atoms $i$ and $j$ is $a_{ij}$.  We set $m_i=m_\sigma$ and $a_{ij}=a_{\sigma \sigma'}$
for $i\in I_\sigma$ and $j\in I_{\sigma'}$. The reduced masses are $\mu_{\si\sip}=m_\si m_\sip/(m_\si+m_\sip)$.
We shall denote by $P_{\sigma \sigma'}$ the set of all pairs of particles with one particle in group $\sigma$ and the other one in group $\sigma'$, each pair being counted only once:
\be
P_{\sigma \sigma'}\equiv\left\{ (i,j)\in (I_\sigma\times I_{\sigma'})\cup (I_{\sigma'}\times I_\sigma) \ /\ i<j \right\}.
\ee

The definition of the zero-range model is modified as follows: In the contact conditions~(\ref{eq:CL_3D},\ref{eq:CL_2D})
the scattering length $a$ is replaced by $a_{ij}$,
and the limit $r_{ij}\to0$ is taken for a fixed center of mass position $\mathbf{R}_{ij}=(m_i\rr_i+m_j\rr_j)/(m_i+m_j)$;
moreover Schr\"odinger's equation becomes
\be
\sum_{i=1}^N \left[
-\frac{\hbar^2}{2 m_i}\Delta_{\rr_i} + U_i(\rr_i)
\right] \psi = E\,\psi.
\ee

Our results are summarized in Table~\ref{tab:melange}, where we introduced the notation
\be
( A^{(1)},A^{(2)})_{\si\sip}\equiv \sum_{(i,j)\in P_{\si \sip}} \int \Big( \prod_{k\neq i,j} d^d r_k \Big) \int d^d R_{ij}
A^{(1)}_{ij}(\mathbf{R}_{ij}, (\mathbf{r}_k)_{k\neq i,j})^*
A^{(2)}_{ij}(\mathbf{R}_{ij}, (\mathbf{r}_k)_{k\neq i,j}).
\ee
We note that since $a_{\si\sip}=a_{\sip\si}$ there are only $n(n+1)/2$ independent scattering length, and the 
 partial derivatives with respect to one of these independent scattering lengths are taken while keeping fixed the other independent scattering lengths.
 
 In $3D$ the Efimov effect can occur, e.g. if the mixture contains a bosonic group, or at least three fermionic groups, or two fermionic groups with a mass ratio strictly larger than a critical value $13.6\dots$~\cite{Efimov73}. In this case, as in the previous Section, the derivatives with respect to any scattering length have to be taken for fixed three-body parameter(s), and the relation between $E$ and the momentum distribution (line~4 of Table~\ref{tab:melange}) breaks down~\footnote{This relation also breaks down for a Fermi-Fermi mixture with a mass ratio equal to the critical value $13.6\dots$ above which the Efimov effect occurs. Indeed, the momentum distribution then has a subleading contribution $\delta n_\si(k)\propto 1/k^5$~\cite{TanPrivate}, leading to a divergent integral in this relation. For a mass ratio strictly smaller than the critical ratio the integral converges, because $\delta n_\si(k)\propto 1/k^{5+2s}$ where $s>0$ is the scaling exponent of the three-body wavefunction, $\psi(\lambda\rr_1,\lambda\rr_2,\lambda\rr_3)\propto\lambda^{s-2}$ for $\lambda\to0$~\cite{TanLargeMomentum}.}.
 This relation was first obtained in~\cite{CombescotC} in $3D$, and in $2D$ for Fermi-Fermi mixtures. Here we express the first sum in a more explicit way in terms of the  partial derivative of the energy, and we point out the breakdown of this relation in presence of the Efimov effect.
  

The derivations are analogous to the ones of Sections~\ref{sec:fermions} and~\ref{sec:bosons}. 
The lemmas~(\ref{eq:lemme_3D},\ref{eq:lemme_2D}) are replaced by
\bea
\la\psi_1,H\psi_2\ra-\la H\psi_1,\psi_2\ra=\left\{
\begin{array}{lr}
\ds \frac{2\pi\hbar^2}{\mu_{\si\sip}}\left(\frac{1}{a_1}-\frac{1}{a_2}\right)(A^{(1)},A^{(2)})_{\si\sip} & {\rm in}\ 3D
\\
\ds\frac{\pi\hbar^2}{\mu_{\si\sip}}\ln(a_2/a_1)(A^{(1)},A^{(2)})_{\si\sip} & {\rm in}\ 2D.
\end{array}
\right.
\eea
The pair distribution function is now defined by
\be
g^{(2)}_{\si\sip}(\mathbf{u},\mathbf{v})
=
\int d\rr_1\ldots d\rr_N 
\left|
\psi(\rr_1,\ldots,\rr_N)
\right|^2
\sum_{ i\in I_\si , j\in I_\sip , i\neq j}
\delta\left(\mathbf{u}-\rr_i\right)
\delta\left(\mathbf{v}-\rr_j\right).
\ee
The Hamiltonian of the lattice model used in some of the derivations now reads
\be
H=H_0+\sum_{\sigma\leq\sigma'}g_{0,\sigma \sigma'} \,W_{\sigma \sigma'}
\ee
where
\be
H_0=\sum_{i=1}^N \left[ -\frac{\hbar^2}{2m_i}\Delta_{\rr_i} +U_i(\rr_i) \right]
\ee
with $\langle \rr | \Delta_\rr | \kk \rangle \equiv -k^2 \langle \rr | \kk \rangle$
and
\be
W_{\sigma \sigma'}=\sum_{(i,j)\in P_{\sigma \sigma'}} \delta_{\rr_i,\rr_j} b^{-d}.
\ee
In the formulas of Sec.~\ref{sec:fermions} and App.~\ref{app:2body}
coming from the two-body scattering problem, one has to replace $g_0$ by $g_{0,\sigma\sigma'}$, $a$ by $a_{\si\sip}$ and $m$ by $2\mu_{\si\sip}$. Denoting the corresponding scattering state by $\phi_{\si\si'}(\rr)$,
the lemma~(\ref{eq:lemme_W}) becomes
\be
\la\psi_{n'}|W_{\si\sip}|\psi_n\ra = |\phi_{\si\sip}({\bf 0})|^2\ ( A^{(n')},A^{(n)})_{\si\sip}.
\ee

\section{Applications}\label{sec:appl}
In this Section, we apply some of the above relations, first to the three-body problem and then to the many-body problem. We consider the unitary limit $a=\infty$ in three dimensions.

\subsection{Three-body problem: corrections to exactly solvable cases and comparison with numerics}\label{sec:appl_3body}

Here we use the known analytical expressions for the three-body wavefunctions to compute the corrections to the spectrum to first order in the inverse scattering length $1/a$ and in the effective range $r_e$.

\subsubsection{Universal eigenstates in a trap}
The problem of three identical spinless bosons~\cite{Werner3corpsPRL,Pethick3corps} 
or two-component fermions (say $N_\up=2$ and $N_\down=1$)~\cite{Werner3corpsPRL,TanScaling} 
is exactly solvable in the unitary limit in an isotropic harmonic trap $U(\rr)=1/2\,m\omega^2 r^2$.
Here we restrict to
zero total angular momentum, with a center of mass in its ground state,
so that the normalization constants of the wavefunctions are also known analytically \cite{WernerThese}. Moreover we restrict to universal eigenstates~\footnote{For Efimovian eigenstates, computing the derivative of the energy with respect to the effective range would require to use a regularisation procedure similar to the one employed in free space in \cite{Efimov93,PlatterRangeCorrections}. However the derivative with respect to $1/a$ can be computed \cite{WernerThese}.}. The spectrum is then given by
\be
E=E_{\rm cm}+(s+1+2q)\hbar\omega
\ee
where $E_{\rm cm}$ is the energy of the center of mass motion,
$s$ belongs to the infinite set of real positive solutions of
\be
-s \cos\left(s\frac{\pi}{2}\right) + \eta\frac{4}{\sqrt{3}}\sin\left(s\frac{\pi}{6}\right)=0
\label{eq:s}
\ee
with $\eta=+2$ for bosons and $-1$ for fermions,
and
$q$ is a non-negative integer quantum number describing the degree of excitation of an exactly decoupled bosonic breathing mode
\cite{CRAS,WernerSym}.
We restrict for simplicity to states with $q=0$.

\paragraph{Derivative of the energy with respect to $1/a$.}

Injecting the expression of the regular part $A$ of the normalized wavefunction \cite{WernerThese} into the relation (\ref{eq:thm_dE_3D}) or its bosonic version (Table~\ref{tab:bosons}, line 1) we obtain
\begin{equation}
\frac{\partial E}{\partial(-1/a)}\Big|_{a=\infty} = 
\frac{\Gamma(s+1/2)\sqrt{2}s\sin\left(s\frac{\pi}{2}\right)}{\Gamma(s+1)
\left[
-\cos\left(s\frac{ \pi}{2 } \right) + s\frac{ \pi}{2 } \sin\left(s\frac{ \pi}{2 } \right)
+\eta\frac{2\pi }{3\sqrt{3} } \cos\left(s\frac{ \pi}{6 } \right)
\right]}
\sqrt{\frac{\hbar^3\omega}{m}}.
\label{eq:dEda3corps}
\end{equation}
For the lowest fermionic state, this gives
$(\partial E/\partial(1/a))_{a=\infty}\simeq -1.1980\sqrt{\hbar^3\omega/m}$,
in agreement with the value  $-1.19(2)$ which we extracted from the numerical solution of a finite-range model
presented in Fig.~4a of \cite{StecherLong},
where the error bar comes from our simple way of extracting the derivative from the numerical data of \cite{StecherLong}.

\paragraph{Derivative of the energy with respect to the effective range.}

Using relation (\ref{eq:dEdre}), which holds not only for fermions but also for bosonic universal states, we obtain
\be
\left(\frac{\partial E}{\partial r_e}\right)_a=
\frac{\Gamma(s-1/2) s (s^2-1/2) \sin(s\pi/2)}{\Gamma(s+1) 2\sqrt{2}\left[
-\cos(s\pi/2)+s\pi/2\cdot\sin(s\pi/2)+\eta \,2\pi/(3\sqrt{3})\cdot\cos(s\pi/6)
\right]}
 \sqrt{\hbar m \omega^3}.
 \label{eq:dEdre_3}
\ee
For bosons, this result was derived previously using the method of \cite{Efimov93} and found to agree with the numerical solution of a finite-range separable potential model for the lowest state \cite{WernerThese}.
For fermions, (\ref{eq:dEdre_3}) agrees with the numerical data from Fig. 3 of \cite{StecherLong} to $\sim0.3\%$ for the two lowest states and $5\%$ for the third lowest state
\footnote{Here we used the value of the effective range $r_e=1.435\,r_0$~\cite{ThogersenThese}
for the Gaussian interaction potential $V(r)=-V_0 e^{-r^2/r_0^2}$
with $V_0$ equal to the value where the first two-body bound state appears.};
(\ref{eq:dEdre_3}) also agrees to $3\%$
with the numerical data
from p.~21 of \cite{WernerThese} for the lowest state of a finite-range separable potential model. 
All these deviations are compatible with the estimated numerical accuracy.

\subsubsection{Derivative of the energy of an Efimov trimer with respect to $1/a$.}

The same relation (Table~\ref{tab:bosons}, line 1) can be applied to Efimov trimers in free space. Using the expression of the normalized three-body wavefunction (see App.~\ref{app:Efi_psi}) we get
\bea
 \left(\frac{\partial E}{\partial(-1/a)}\right)_{R_t} &=&
\left(-\frac{\hbar^2}{m}\,  E\right)^{1/2} \,
\frac{\pi \sinh(|s_0|\pi/2) \tanh(|s_0|\pi)}{\cosh(|s_0|\pi/2)+\frac{\pi |s_0|}{2}
\sinh(|s_0|\pi/2) - \frac{4\pi}{3\sqrt{3}}\cosh(|s_0|\pi/6)}
\label{eq:Cefi} \\
&=& 2.11267159347\ldots \times \left(-\frac{\hbar^2}{m}\,  E\right)^{1/2}.
\eea
This confirms and refines the value of the coefficient $2.11$ estimated numerically in \cite{RevueBraaten}.
We note that this numerical coefficient is simply $\Delta'(-\pi/2)/|s_0|$, where $\Delta(\xi)$
is Efimov's universal function that was estimated numerically in
\cite{RevueBraaten} and computed very
precisely in \cite{MohrThese}. Our analytical calculation gives the exact value
of the derivative 
\be
\Delta'(-\pi/2)= 2.125850069373\ldots,
\ee
to be compared with the numerical estimate $\Delta'(-\pi/2)\simeq 2.12$ in \cite{RevueBraaten}.

The expression (\ref{eq:Cefi}) can also be obtained
from the tail of the momentum distribution:
The expression of the coefficient $C$ follows from Eq.(\ref{eq:Cexact}),
and (\ref{eq:Cefi}) is recovered
using the relation on the second line of the left column of
Table~\ref{tab:bosons}.

\subsection{Unitary Fermi gas: comparison with fixed-node Monte-Carlo}
\label{sec:FNMC}

\begin{figure}
   \begin{minipage}[c]{.46\linewidth}
      \includegraphics[width=\linewidth]{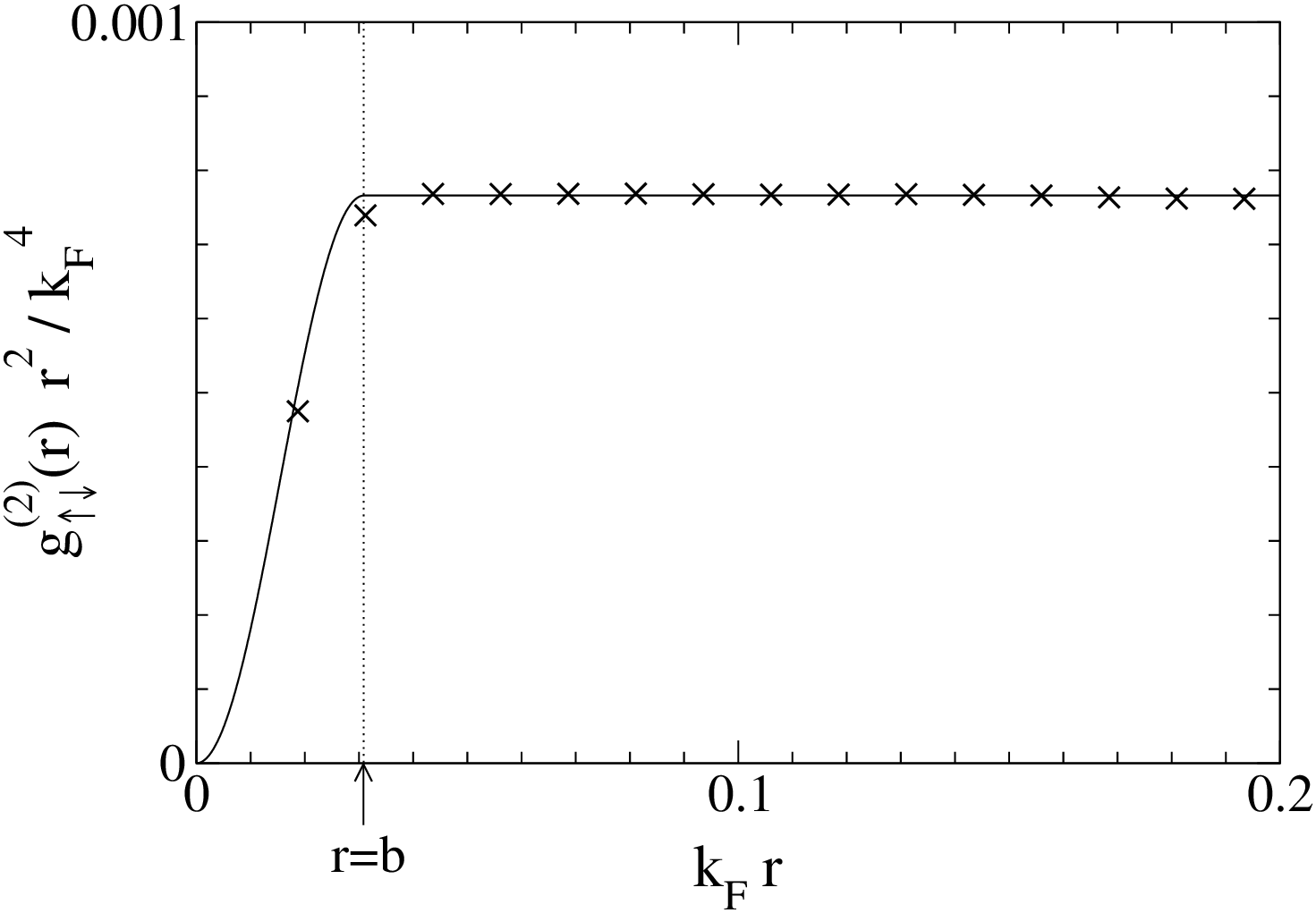}
       \caption{Pair distribution function
       $g_{\uparrow\downarrow}^{(2)}(r)=\langle
\hat{\psi}^\dagger_\uparrow(\rr)
\hat{\psi}^\dagger_\downarrow(\vn)
\hat{\psi}_\downarrow(\vn)
\hat{\psi}_\uparrow(\rr)
\rangle$ of the homogeneous unitary gas at zero temperature. Crosses: fixed-node Monte-Carlo results from Ref.~\cite{LoboGiorgini_g2}. Line: analytic expression~(\ref{eq:g2_pour_MC}), where the value $\zeta=0.95$ was taken to fit the Monte-Carlo results.
The arrow indicates the range $b$ of the square-well interaction potential.
       \label{fig:g2}}
   \end{minipage}
\hfill
  \begin{minipage}[c]{.46\linewidth}
      \includegraphics[width=\linewidth]{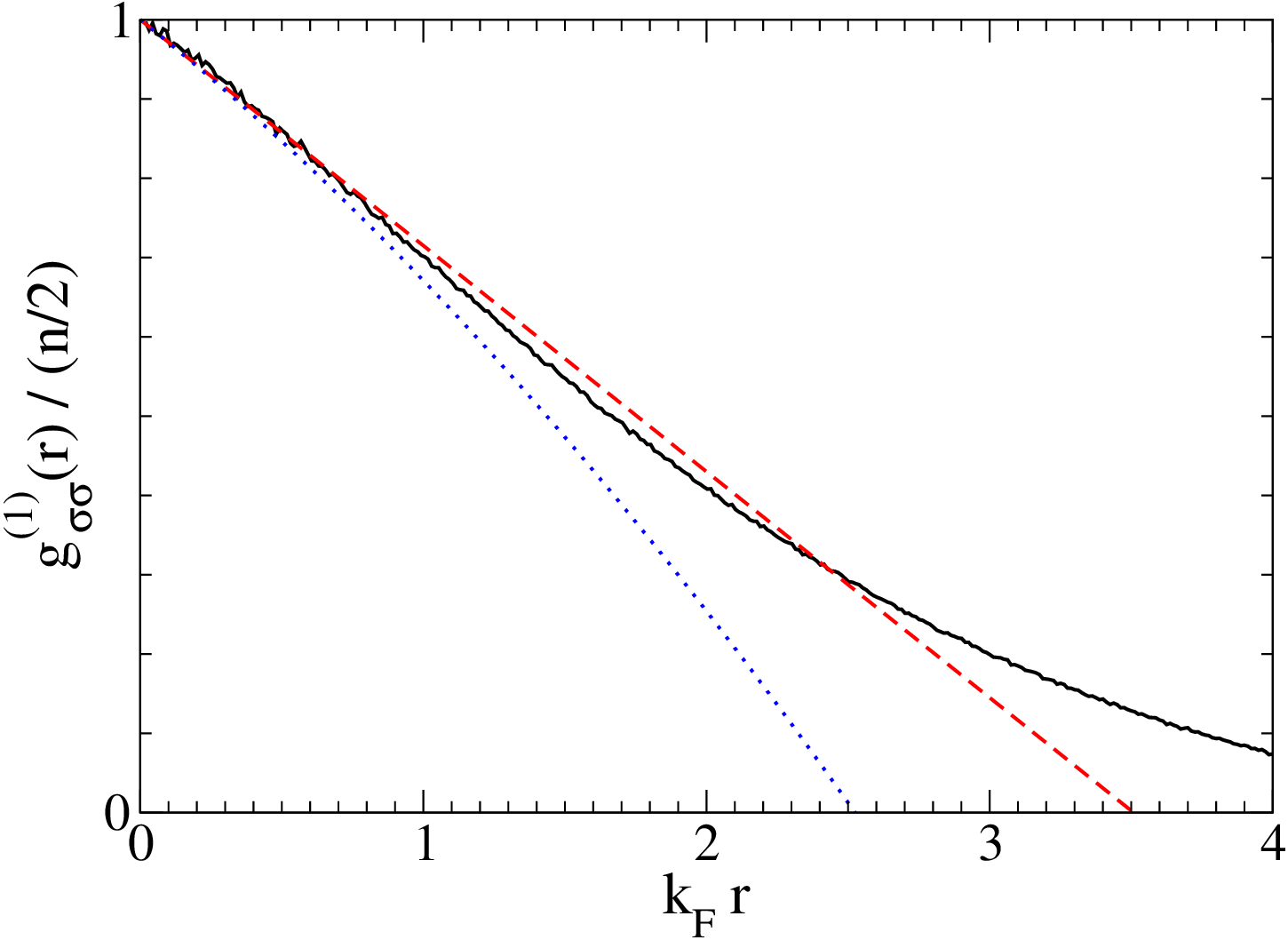}
      \caption{One-body density matrix $g^{(1)}_{\si\si}(r)=\la\hat{\psi}^\dagger_\si(\rr)\hat{\psi}_\si(\vn)\ra$ of the homogeneous unitary gas at zero temperature: comparison between the fixed-node Monte-Carlo results from Ref.~\cite{Giorgini_nk}
      (continuous curve)
      and the
      analytic expression~(\ref{eq:g1_pour_MC}) for the
       small-$k_F r$ expansion of $g^{(1)}_{\sigma\sigma}$ up to first order (dashed straight line, red online) and second order (dotted parabola, blue online) where we took the value $\zeta=0.95$ extracted from the Monte-Carlo data for $g^{(2)}_{\uparrow\downarrow}$, see Fig.~\ref{fig:g2}.
      \label{fig:g1} }
   \end{minipage} 
\end{figure}

For the homogeneous unitary gas (i.e. the two-component Fermi gas in $3D$ with $a=\infty$) at zero temperature, we can compare our analytical expressions for the short-distance behavior of the one-body density matrix $g^{(1)}_{\sigma\sigma}$ and the pair distribution function $g^{(2)}_{\uparrow\downarrow}$ to the fixed-node Monte-Carlo results published by the Trento group in~\cite{Giorgini,Giorgini_nk,LoboGiorgini_g2}. In this case, $g^{(1)}_{\si\si}({\bf R}-\rr/2,{\bf R}+\rr/2)$ and
$g^{(2)}_{\up\down}({\bf R}-\rr/2,{\bf R}+\rr/2)$ depend only on $r$ and not on $\si$, ${\bf R}$ and  the direction of $\rr$.
Expanding the energy to first order in $1/(k_F a)$ around the unitary limit yields:
\be
E=E_0\left(\xi - \frac{\zeta}{k_F a}+\dots\right)
\label{eq:eq_d_etat}
\ee
where $E_0$ is the ground state energy of the ideal gas, $\xi$ and $\zeta$ are universal dimensionless numbers, and the Fermi wavevector is related to the density through $k_F=(3\pi^2 n)^{1/3}$. 
Expressing $C$ in terms of $\zeta$ thanks to (\ref{eq:thm_nk_3D},\ref{eq:eq_d_etat}) and inserting this into~(\ref{eq:thm_g1_b}) gives
\be
g^{(1)}_{\si\si}(r)\simeq\frac{n}{2}\left[ 1 - \frac{3\zeta}{10} k_F r - \frac{\xi}{10} (k_F r)^2 + \dots\right].
\label{eq:g1_pour_MC}
\ee
For a finite interaction range $b$, this expression is valid for $b\ll r \ll k_F^{-1}$ 
\footnote{For a finite-range potential one has $g^{(1)}_{\si\si}(r)=n/2-r^2 m E_{\rm kin}/(3\hbar^2 \mathcal{V})+\dots$ where $\mathcal{V}$ is the volume; the kinetic energy diverges in the zero-range limit as $E_{\rm kin}\sim -E_{\rm int}$, thus $E_{\rm kin}\sim-C/(4\pi)^2 \int d^3r\,V(r)|\phi(r)|^2$ from (\ref{eq:Eint}), so that $E_{\rm kin}\sim C\pi\hbar^2/(32 m b)$ for the square-well interaction. This behavior of $g^{(1)}(r)$ only holds at very short distance $r\ll b$ and is below the resolution of the Monte-Carlo data.}.
Equation~(\ref{eq:g2_rezo_3D}) yields
\be
g^{(2)}_{\up\down}(r)\underset{k_F r\ll1}{\simeq}\frac{\zeta}{40\pi^3}k_F^4 |\phi(r)|^2.
\label{eq:g2_pour_MC}
\ee
The interaction potential used in the Monte-Carlo simulations~\cite{Giorgini,Giorgini_nk,LoboGiorgini_g2} is a square-well
\be
V(r)=\left\{\begin{array}{lr} -\left(\frac{\pi}{2}\right)^2 \frac{\hbar^2}{m b^2} & {\rm for}\  r<b
\\ 0 & {\rm for}\ r>b
\end{array}\right.
\ee
for which the zero-energy scattering state is
\be
\phi(r)=\left\{\begin{array}{lr} \sin\left(\frac{\pi r}{2 b}\right)/r & {\rm for}\  r<b
\\ 1/r & {\rm for}\ r>b
\end{array}\right.
\ee
and the range $b$ was taken such that $n b^3=10^{-6}$ i.e. $k_F b=0.0309367\dots$. Thus we can assume that we are in the zero-range limit $k_F b\ll1$, so that (\ref{eq:g1_pour_MC},\ref{eq:g2_pour_MC}) are applicable.

Figure \ref{fig:g2} shows that the expression (\ref{eq:g2_pour_MC}) for $g^{(2)}_{\uparrow\downarrow}$ fits well the Monte-Carlo data of \cite{LoboGiorgini_g2} if one adjusts the value of $\zeta$ to $0.95$. This value is close to the value $\zeta\simeq1.0$ extracted from (\ref{eq:eq_d_etat}) and the $E(1/a)$-data of~\cite{Giorgini}.

Using $\zeta=0.95$ we can compare the expression (\ref{eq:g1_pour_MC}) for $g^{(1)}_{\sigma\sigma}$ with  Monte-Carlo data of~\cite{Giorgini_nk}  without adjustable parameters.
Figure \ref{fig:g1} shows that the first order derivatives agree, while the second order derivatives are compatible within the statistical noise. This provides an interesting check of the numerical results, even though any wavefunction satisfying the contact condition~(\ref{eq:CL_3D}) would lead to $g^{(1)}_{\sigma\sigma}$ and $g^{(2)}_{\uparrow\downarrow}$ functions satisfying (\ref{eq:thm_g2_3D},\ref{eq:thm_g1_a}) with values of $C$ compatible with each other. 

\subsection{Dependence of the unitary gas critical temperature on the interaction range}

The key property underlying (\ref{eq:dEdre}) is that the leading order change of each eigenstate energy of a spin $1/2$ Fermi gas due to a small but non-zero
interaction range is linear in the effective range $r_e$ of the interaction potential, which a coefficient which is model
independent. As a consequence, the leading order change of the thermodynamical potentials, and even of the critical temperature $T_c$
of the Fermi gas, are also expected to be linear in $r_e$, with model independent coefficients.

\begin{figure}[htb]
\includegraphics[width=0.8\linewidth,clip=]{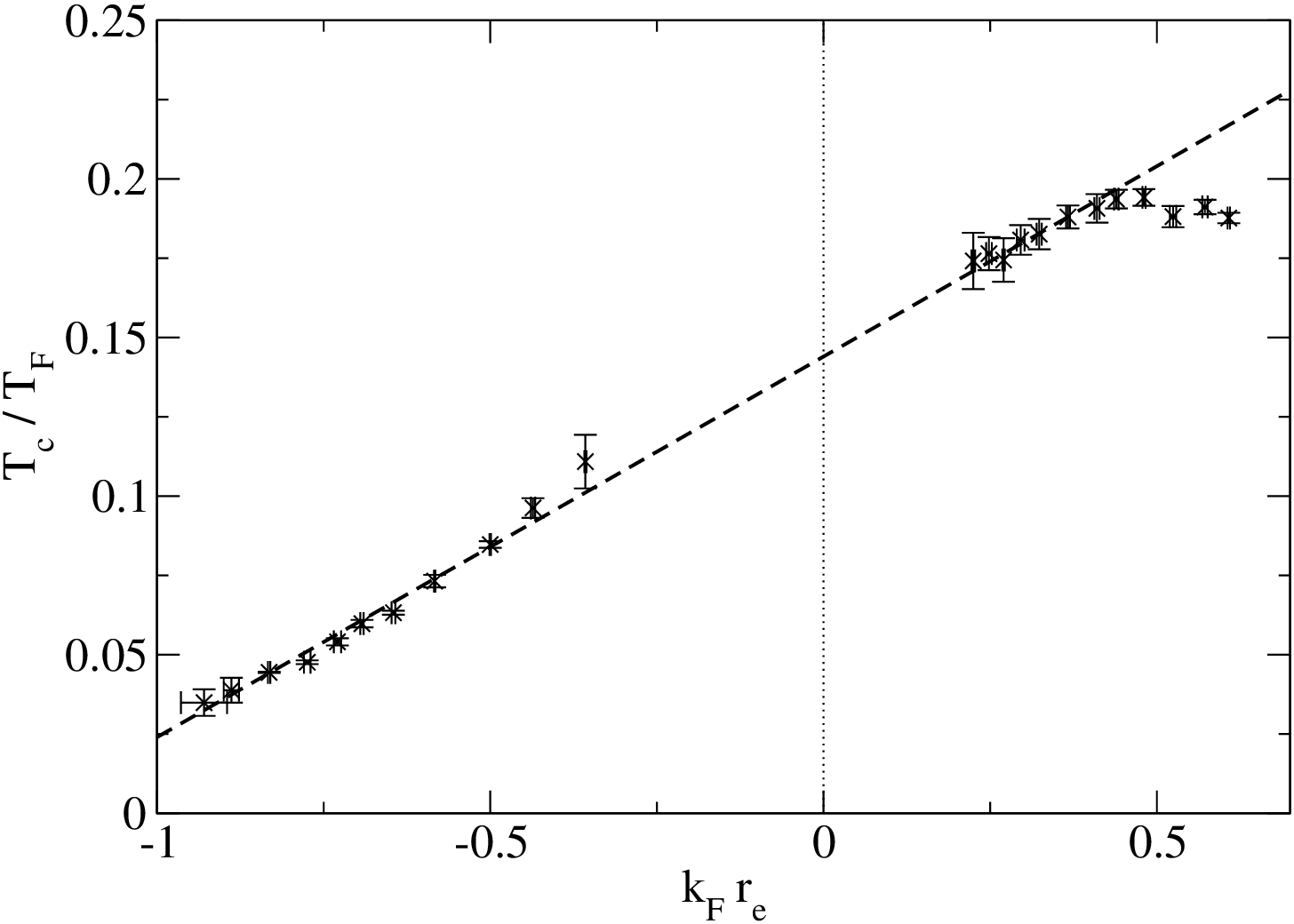}
\caption{Critical temperature $T_c$ of the unitary Fermi gas as a function of the effective range $r_e$ of the interaction potential,
as given by the Quantum Monte Carlo results of \cite{zhenyaPRL} for the Hubbard model (symbols with $r_e <0$) and of 
\cite{zhenyas_crossover} for a continuous space model (symbols with $r_e>0$). The dashed line corresponds to a linear
fit of the data over the interval $k_F r_e \in [-0.8,0.45]$. Here $k_B T_F =\hbar^2 k_F^2/(2m)$ is the Fermi energy of the ideal gas
with the same density as the unitary gas. 
\label{fig:Tc}}
\end{figure}

This expectation can be tested with the Quantum Monte Carlo data of \cite{zhenyaPRL} and \cite{zhenyas_crossover}, where the critical temperature of the unitary gas
was calculated for two different, finite range interaction models \footnote{Strictly speaking, (\ref{eq:dEdre}) was derived for a parabolic kinetic
energy dispersion relation, whereas the dispersion relation in the Quantum Monte Carlo Hubbard lattice model deviates from a parabola at large $k$.
The deviations however scale as $k^4 r_e^2$ at low $k$, so are expected to give rise to a higher order, $O(r_e^2)$ deviation to $T_c$.}. 
In Fig.\ref{fig:Tc}, we plot the Monte Carlo data
as a function of $k_F r_e$, where $r_e$ is the effective range of the corresponding model.  Our expectation is that all the data lie
on the same straight line for low enough $k_F |r_e|$, which is indeed essentially the case (considering the size of the error bars).
We then conclude that the shift of $T_c$ due to the interaction range is of order
\be
\frac{\delta T_c}{T_F} \simeq 0.12 k_F r_e,
\ee
at low $r_e$ and in a model independent way.
This results in a relative shift at the percent level for typical experiments on lithium, 
where $k_F r_e \approx 0.01$.

\section{Conclusion} \label{sec:conclusion}

We derived relations between various observables for $N$ particles of arbitrary masses and statistics in an external potential
with zero-range or short-range interactions, in continuous space or on a lattice, in two or three dimensions.

Some of our results generalize the ones of
\cite{Olshanii_nk, TanLargeMomentum, TanEnergetics, ZhangLeggettUniv, CombescotC}:
The
large-momentum behavior of the momentum distribution,
the short-distance behavior of the pair correlation function and of the one-body density matrix, the derivative of the energy with respect to the scattering length or to time, the norm of the regular part of the wavefunction (defined through the behavior of the wavefunction when two particles approach each other),
and, in the case of finite-range interactions, the interaction energy, 
are all related to the same quantity $C$; 
and
the difference between the total energy and the trapping potential energy is
related to $C$ and to a functional of the momentum distribution (which is also equal to 
 the second order term in the short-distance expansion of the one-body density matrix).
For
Efimov states
with zero-range interactions, 
we found that this last relation breaks down, because the large-momentum tail of the momentum distribution contains a subleading oscillatory term.

We also obtained entirely new relations:
The second order derivative of the energy with respect to the inverse scattering length (or to the logarithm of the scattering length in two dimensions) is related to the regular part of the wavefunctions, and is negative at fixed entropy;
the derivative of the energy of a universal state with respect to the effective range of the interaction potential is also related to the regular part;
and the derivative of the energy of an efimovian state
with respect to the three-body parameter is related to the 
a three-body analog of the regular part.
Applications were presented in three dimensions for an infinite scattering length: the derivative of the energy with respect to the inverse scattering length was computed analytically and found to agree with numerics for Efimov trimers; the same was done for universal three-body states in a harmonic trap, not only for the derivative of the energy with respect to the inverse scattering length but also with respect to the effective range; existing fixed-node Monte-Carlo data for the unitary Fermi gas were checked to satisfy exact relations. Also the derivative of the critical temperature of the unitary gas with respect to the effective range, expected from
our results to be model-independent, is estimated from the Quantum Monte Carlo results of \cite{zhenyaPRL}.

The relations obtained here may be used in various other contexts. For example, 
the result (\ref{eq:thm_d^2E_S}) on the sign of the second order derivative of $E$ at constant entropy is relevant to adiabatic ramp experiments~\cite{CarrCastin,GrimmCrossover,JinPotentialEnergy,thomas_entropie_PRL,thomas_entropie_JLTP},
and the relation (\ref{eq:g2_latt}) allows to directly compute $C$ using diagrammatic Monte-Carlo~\cite{VanHouckePrepa}. $C$ is directly related to the closed-channel fraction in a two-channel model \cite{BraatenLong,WernerTarruellCastin}, which allows to extract it \cite{WernerTarruellCastin} from experimental photoassociation measurements \cite{HuletClosedChannel}. $C$ also plays an important role in the theory of radiofrequency spectra \cite{ZwergerRF,BaymRF,ZhangLeggettUniv,StrinatiRF,RanderiaRF,ZwergerRFLong} and in finite-$a$ virial theorems \cite{TanViriel,Braaten,WernerViriel}.

We can think of several generalizations of
the relations presented here.
All relations can be rederived in the case of periodic boundary conditions.
The relations for finite-range models, obtained here for two-component fermions, can be generalized to arbitrary mixtures.
The techniques used here can be applied to the
 one-dimensional case to generalize the relations of \cite{Olshanii_nk}.
For two-channel or multi-channel models one may derive relations other than the ones of \cite{BraatenLong,WernerTarruellCastin,ZhangLeggettUniv}.
In presence of the Efimov effect, the derivative of the energy with respect to the three-body parameter
can easily be related to the short-distance behavior of the third order density correlation function
thanks to (\ref{eq:dEdRt},\ref{eq:danilov});
moreover
the asymptotic behavior (\ref{eq:nk_efi}) of the momentum distribution is expected to hold for any state satisfying the three-body boundary condition (\ref{eq:danilov}), with
a coefficient $D$ of the subleading tail in 
(\ref{eq:nk_efi}) related to
$\partial E/\partial(\ln R_t)$ through a simple proportionality factor.
Indeed, the singularities of the wavefunction at short interparticle distances are generally related to large momentum tails.

\acknowledgements
We thank S.~Tan, M. Olshanii, E.~Burovski, N.~Prokof'ev and R.~Ignat  for useful discussions, as well as S. Giorgini, J.~von~Stecher and M. Th{\o}gersen for sending numerical data from~\cite{Giorgini_nk,LoboGiorgini_g2,StecherLong,ThogersenThese}.
F.W. is supported by NSF under Grant No.~PHY-0653183.
Y.C. is a member of IFRAF. 

\section*{Note}
While completing this work, we became aware of unpublished notes by Tan where
Eqs.~(\ref{eq:thm_nk_2D},\ref{eq:thm_dEdt_2D}) were obtained independently using the formalism of~\cite{TanSimple}~\footnote{Tan also derived Equation~(\ref{eq:energy_thm_2D_heaviside}) independently from~\cite{CombescotC} using
 Ref.~\cite{TanSimple}'s Equation~(17).}.
 
\appendix

\section{Two-body scattering for the lattice model}

\label{app:2body}
For the lattice model defined in Sec.~\ref{sec:models:lattice},
we recall that $\phi(\rr)$ denotes the
zero-energy two-body scattering state with the normalization~(\ref{eq:normalisation_phi_tilde_3D},\ref{eq:normalisation_phi_tilde_2D}).
In this Appendix we derive the relation (\ref{eq:g0_3D},\ref{eq:g0_2D}) between the coupling constant $g_0$ and the scattering length, as well as the expressions (\ref{eq:phi0_vs_g0},\ref{eq:phi0_vs_g0_2D},\ref{eq:phi_tilde_3D},\ref{eq:phi_tilde_2D}) of $\phi(\vn)$.
Some of the calculation resemble the ones in~\cite{MoraCastin, YvanHouchesLowDShort}.

We consider a low-energy scattering state $\Phi_\qq(\rr)$ of wavevector $q\ll b^{-1}$ and energy $E=2\epsq\simeq\hbar^2q^2/m$, i.e. the solution of the two-body Schr\"odinger equation (with the center of mass at rest):
\be
(H_0+V)|\Phi_\qq\ra=E |\Phi_\qq\ra
\label{eq:schro_2corps}
\ee
where $H_0=\int_D d^dk/(2\pi)^d\,2\epsk|\kk\ra\la\kk|$ and $V=g_0|\rr=\vn\ra\la\rr=\vn|$, with the asymptotic behavior
\bea
\Phi_\qq(\rr)&\underset{r\to\infty}{=}&e^{i\qq\rr}+f_\qq \frac{e^{iqr}}{r}+\ldots\ \ \ \ {\rm in}\ 3D
\label{eq:phiq_asymp_3D}
\\
\Phi_\qq(\rr)&\underset{r\to\infty}{=}&e^{i\qq\rr}-f_\qq \sqrt{\frac{i}{8\pi q r}}e^{iqr}+\ldots\ \ \ \ {\rm in}\ 2D.
\label{eq:phiq_asymp_2D}
\eea
Here $f_\qq$ is the scattering amplitude (in $2D$ the present definition is from~\cite{ShlyapHoucheslowDShort}, and $\sqrt{i}\equiv e^{i\pi/4}$), which in the present case is independent of the direction of $\rr$ as we will see.
We then have the well-known expression
\be
|\Phi_\qq\ra=(1+G V)|\qq\ra
\label{eq:phiq_GV}
\ee
where $G=(E+i0^+-H)^{-1}$, which, since
\be
G=G_0+G_0 V G,
\label{eq:G_G0}
\ee
is equivalent to
\be
|\Phi_\qq\ra=(1+G_0 T)|\qq\ra
\label{eq:phiq_G0T}
\ee
where
\be
T=V+VGV.
\label{eq:defT}
\ee
Indeed, (\ref{eq:phiq_GV}) clearly solves (\ref{eq:schro_2corps}), and one can check [using the fact that $\la \rr |G_0|\rr=\vn\ra$ behaves for $r\to\infty$ as
$-m/(4\pi\hbar^2)\,e^{iqr}/r$ in $3D$ and
$-m/\hbar^2\sqrt{i/(8\pi q r)}e^{iqr}$ in $2D$] that
(\ref{eq:phiq_G0T}) satisfies (\ref{eq:phiq_asymp_3D},\ref{eq:phiq_asymp_2D}) with
\bea
f_\qq&=&-\frac{m}{4\pi\hbar^2}b^3\la\rr=\vn|T|\qq\ra\ \ \ \ {\rm in}\ 3D
\label{eq:f_T_3D}
\\
f_\qq&=&\frac{m}{\hbar^2}b^2\la\rr=\vn|T|\qq\ra\ \ \ \ {\rm in}\ 2D.
\label{eq:f_T_2D}
\eea
Using (\ref{eq:defT}) and (\ref{eq:G_G0}) one gets
\be
\la\rr=\vn|T|\qq\ra=b^{-d}\left[\frac{1}{g_0}-\int_D \frac{d^d k}{(2\pi)^d}\,\frac{1}{E+i0^+-2\epsk}\right]^{-1}.
\ee
In $3D$ the scattering length in defined by $\ds f_\qq\underset{q\to0}{\rightarrow}-a$, which gives the relation (\ref{eq:g0_3D}) between $a$ and $g_0$. In $2D$,
\be
f_\qq\underset{q\to0}{=}-\frac{2\pi}{\ln(qae^\gamma/2)-i\pi/2+o(1)}
\label{eq:fq_2D_lowE}
\ee
where $a$ is by definition the $2D$ scattering length.
Identifying the inverse of the right-hand-sides of Eqs.(\ref{eq:f_T_2D}) and (\ref{eq:fq_2D_lowE}) and taking the real part gives the desired (\ref{eq:g0_2D}). We note that Eqs.~(\ref{eq:fq_2D_lowE},\ref{eq:g0_2D}) remain true if $q\to0$ is replaced by the limit $b\to0$ taken for fixed $a$.

To derive (\ref{eq:phi0_vs_g0},\ref{eq:phi0_vs_g0_2D}) we start from
\be
V|\Phi\ra = T(E+i 0^+) |\qq\ra
\ee
which directly follows from (\ref{eq:phiq_GV}).
Applying $\la\rr=\vn|$ on the left and using (\ref{eq:f_T_3D},\ref{eq:f_T_2D}) yields
\bea
g_0 \Phi_\qq(\vn)&=&-\frac{4\pi\hbar^2}{m}f_\qq\ \ \ {\rm in}\ 3D
\\
g_0 \Phi_\qq(\vn)&=&\frac{\hbar^2}{m}f_\qq\ \ \ {\rm in}\ 2D.
\label{eq:phiq0_f_2D}
\eea
In $3D$, we simply have $\ds\phi=-a^{-1}\underset{q\to0}{\lim}\Phi_\qq$, and the result (\ref{eq:phi0_vs_g0}) follows. 
In $2D$, the situation is a bit more tricky because $\underset{q\to0}{\lim}\Phi_\qq(\vn)=0$. We thus start with $q>0$, and we will take the limit $q\to0$ later on. 
At finite $q$, we define
$\phi_\qq(\rr)$ as being proportional to $\Phi_\qq(\rr)$, and normalized by imposing the same condition 
(\ref{eq:normalisation_phi_tilde_2D}) than at zero energy, but only for $b\ll r\ll q^{-1}$.
Inserting (\ref{eq:fq_2D_lowE}) into (\ref{eq:phiq0_f_2D}) gives an expression for $\Phi_\qq(\vn)$.
To deduce the value of $\phi(\vn)$, it remains to calculate the $\rr$-independent ratio $\phi_\qq(\rr)/\Phi_\qq(\rr)$. But for $r\gg b$ we can replace $\phi_\qq(\rr)$ and $\Phi_\qq(\rr)$ by their values within the zero-range model (since we also have $b\ll q^{-1}$) which we denote by $\phi_\qq^{\rm ZR}(\rr)$ and $\Phi_\qq^{\rm ZR}(\rr)$.
The two-body Schr\"odinger equation
\be
-\frac{\hbar^2}{m}\Delta\Phi_\qq^{\rm ZR} = E\,\Phi_\qq^{\rm ZR},\ \forall r>0
\ee
implies that
\be
\Phi_\qq^{\rm ZR}=e^{i\qq\cdot\rr}+\mathcal{N} H_0^{(1)}(q r)
\ee
where $\mathcal{N}$ is a constant and $H_0^{(1)}$ is an outgoing Hankel function.
The contact condition 
\be
\exists A/\ \Phi_\qq^{\rm ZR}(\rr)\underset{r\to0}{=} A\ln(r/a)+O(r)
\ee
together with the known short-$r$ expansion of the Hankel function~\cite{Lebedev} then gives
\be
A=\frac{-1}{\ln(q a e^\gamma/2)-i\pi/2}.
\ee
Of course we also have $\Phi_\qq^{\rm ZR}/\phi_\qq^{\rm ZR}=A$, which gives (\ref{eq:phi0_vs_g0_2D}).

Finally, Eqs.(\ref{eq:phi_tilde_3D},\ref{eq:phi_tilde_2D}) are obtained from (\ref{eq:phi0_vs_g0},\ref{eq:phi0_vs_g0_2D})
using the relations
$d(m/(4\pi\hbar^2a))/d(1/g_0) = 1$ in $3D$
and
$d(1/g_0)/d(\ln a)=-m/(2\pi\hbar^2)$ in $2D$,
which are direct consequences of
the relations (\ref{eq:g0_3D},\ref{eq:g0_2D}) between $g_0$ and $a$.

\section{Derivation of a lemma} \label{app:lemme}
In this Appendix, we derive the lemma~(\ref{eq:lemme_3D}) in three dimensions, as well as its two-dimensional version~(\ref{eq:lemme_2D}).

\noindent{\underline{\it Three dimensions:}}
\\
By definition we have
\be
\langle \psi_1, H \psi_2 \rangle - \langle H \psi_1, \psi_2 \rangle = -\frac{\hbar^2}{2 m} \int' d^3 r_1 \ldots d^3 r_N  \sum_{i=1}^N
\left[ \psi_1^* \Delta_{\rr_i} \psi_2 - \psi_2 \Delta_{\rr_i} \psi_1^* \right].
\ee
Here the notation $\int'$ means that the integral is restricted to the set where none of the particle positions coincide~\footnote{In other words, the Dirac distributions originating from the action of the Laplacian onto the $1/r_{ij}$ divergences can be ignored.}.
We rewrite this as:
\be
\langle \psi_1, H \psi_2 \rangle - \langle H \psi_1, \psi_2 \rangle = -\frac{\hbar^2}{2 m} \sum_{i=1}^N \int' \Big( \prod_{k\neq i} d^3 r_k \Big)
\lim_{\epsilon\to0}
 \int_{\{\rr_i / \forall j\neq i, r_{ij}>\epsilon \}} d^3 r_i
\left[ \psi_1^* \Delta_{\rr_i} \psi_2 - \psi_2 \Delta_{\rr_i} \psi_1^* \right].
\label{eq:Toto_echange}
\ee
We note that this step is not trivial to justify mathematically.
The order of integration has been changed and the limit $\epsilon\to0$ has been exchanged with the integral over $\rr_i$.
We expect that this is valid in the presently considered case of equal mass fermions, and more generally provided the wavefunctions are sufficiently regular in the limit where several particles tend to each other.

Since the integrand is the divergence of $\psi_1^* \nabla_{\rr_i} \psi_2 - \psi_2 \nabla_{\rr_i} \psi_1^*$,
Ostrogradsky's theorem gives
\be
\langle \psi_1, H \psi_2 \rangle - \langle H \psi_1, \psi_2 \rangle = \frac{\hbar^2}{2 m} \sum_{i=1}^N \int' \Big( \prod_{k\neq i} d^3 r_k \Big)
\lim_{\epsilon\to0}
 \sum_{j, j\neq i} \ \oiint_{S_\epsilon(\rr_j)} \left[
 \psi_1^* \nabla_{\rr_i} \psi_2 - \psi_2 \nabla_{\rr_i} \psi_1^*
 \right] \cdot \mathbf{dS}
 \label{eq:ostro}
\ee
where the surface integral is for $\rr_i$ belonging to  the sphere $S_\epsilon(\rr_j)$ of center $\rr_j$ and radius $\epsilon$, and the vector area $\mathbf{dS}$ points out of the sphere.
We then expand the integrand by using the contact condition, in the limit $r_{ij}=\epsilon\to0$ taken for fixed $\rr_j$ and fixed $(\rr_k)_{k\neq i,j}$. Using $\RR_{ij}=\rr_j + \epsilon \uu /2$ with 
$\uu\equiv(\rr_i-\rr_j)/r_{ij}$ we get
\bea
\psi_n & \underset{\epsilon\to0}{=}& \left(\frac{1}{\epsilon}
 -\frac{1}{a_n} \right)
  A_{ij}^{(n)}+ \frac{1}{2} \uu \cdot \nabla_{\RR_{ij}} A_{ij}^{(n)}
 +O(\epsilon)
 \label{eq:psi_n_3D}
 \\
 \nabla_{\rr_i} \psi_n & \underset{\epsilon\to0}{=}&
 -\frac{\uu}{\epsilon^2} A_{ij}^{(n)}
 +\frac{1}{2\epsilon} \left[
 \nabla_{\RR_{ij}} A_{ij}^{(n)}
 - \uu \left(
  \uu \cdot \nabla_{\RR_{ij}} A_{ij}^{(n)} \right) \right] +O(1)
  \label{eq:grad_psi_n_3D}
 \eea
where $n$ equals $1$ or $2$, and the functions $A_{ij}^{(n)}$ and $\nabla_{\RR_{ij}} A_{ij}^{(n)}$ are taken at $\left( \rr_j , (\rr_k)_{k\neq i,j} \right)$.
This simply gives
\be
\oiint_{S_\epsilon(\rr_j)} \left[
 \psi_1^* \nabla_{\rr_i} \psi_2 - \psi_2 \nabla_{\rr_i} \psi_1^*
 \right] \cdot \mathbf{dS} \underset{\epsilon\to0}{=} 4\pi \left( \frac{1}{a_1}-\frac{1}{a_2}\right)
 A_{ij}^{(1)\,*} A_{ij}^{(2)} +O(\epsilon)
 \label{eq:int_surface_3D}
\ee
because the leading order term cancels and most angular integrals vanish.
Inserting this into (\ref{eq:ostro}) gives the desired lemma (\ref{eq:lemme_3D}).

\noindent{\underline{\it Two dimensions:}}
\\
The derivation is analogous to the $3D$ case.
In (\ref{eq:ostro}), the double integral on the sphere of course has to be replaced by a simple integral on the circle.
Instead of (\ref{eq:psi_n_3D},\ref{eq:grad_psi_n_3D}),
we now obtain,
from
the $2D$ contact condition~(\ref{eq:CL_2D}),
\bea
\psi_n & \underset{\epsilon\to0}{=}&  
\ln( \epsilon/a_n)\ 
A_{ij}^{(n)}
 +O(\epsilon \ln \epsilon)
 \label{eq:psi_n_2D}
 \\
 \nabla_{\rr_i} \psi_n & \underset{\epsilon\to0}{=}&
 \frac{\uu}{\epsilon} A_{ij}^{(n)}
  +O(\ln \epsilon),
  \label{eq:grad_psi_n_2D}
 \eea
which gives
\be
\oint_{S_\epsilon(\rr_j)} \left[
 \psi_1^* \nabla_{\rr_i} \psi_2 - \psi_2 \nabla_{\rr_i} \psi_1^*
 \right] \cdot \mathbf{dS} \underset{\epsilon\to0}{=} 2\pi 
 \ln(a_2/a_1)
  A_{ij}^{(1)\,*} A_{ij}^{(2)} +O(\epsilon \ln^2\epsilon)
 \label{eq:int_surface_2D}
\ee
and yields the lemma (\ref{eq:lemme_2D}).

\section{First and second order isentropic derivatives of the mean energy 
in the canonical ensemble}
\label{app:adiab}

One considers a system with a Hamiltonian $H(\lambda)$ depending 
on some parameter
$\lambda$, and at thermal equilibrium in the canonical ensemble
at temperature $T$, with a density operator $\rho=\exp(-\beta H)/Z$.
In terms of the partition function 
$Z(T,\lambda) = \mbox{Tr}\, e^{-\beta H(\lambda)}$, with $\beta=
1/(k_B T)$, one
has the usual relations for the free energy $F$, the mean energy $U
=\mbox{Tr}(\rho H)$
and the entropy $S=-k_B \mbox{Tr}(\rho\ln\rho)$:
\bea
\label{eq:def}
F(T,\lambda) &=& -k_B T \ln Z(T,\lambda) \\
\label{eq:utile}
F(T,\lambda) &=& U(T,\lambda)-T S(T,\lambda) \\
\partial_T F(T,\lambda) &=& -S(T,\lambda).
\eea
One now varies $\lambda$ for a fixed entropy $S$. The temperature
is thus a function $T(\lambda)$ of $\lambda$ such that
\be
S(T(\lambda),\lambda) = \mbox{ct}.
\ee
The derivatives of the mean energy
for fixed entropy are then:
\bea
\left(\frac{dU}{d\lambda}\right)_S &\equiv & \frac{d}{d\lambda}
[U(T(\lambda),\lambda)] \\
\left(\frac{d^2U}{d\lambda^2}\right)_S &\equiv & \frac{d^2}{d\lambda^2}
[U(T(\lambda),\lambda)].
\eea
Writing (\ref{eq:utile}) for $T=T(\lambda)$ and
taking the first order and the second order derivatives
of the resulting equation with respect to $\lambda$,
one finds
\bea
\label{eq:deriv1}
\left(\frac{dU}{d\lambda}\right)_S  &=& 
\partial_\lambda F(T(\lambda),\lambda) \\
\label{eq:deriv2}
\left(\frac{d^2U}{d\lambda^2}\right)_S &=&
\partial_\lambda^2F(T(\lambda),\lambda)
-\frac{\left[\partial_T\partial_\lambda F (T(\lambda),\lambda)\right]^2}
{\partial_T^2 F (T(\lambda),\lambda)}.
\eea
It remains to use (\ref{eq:def}) to obtain a microscopic
expression of the above partial derivatives of $F$, from
the partition function expressed as a sum
$Z=\sum_n e^{-\beta E_n}$  over the eigenenergies $n$ of the
Hamiltonian: 
\bea
\label{eq:dl}
\partial_\lambda F(T,\lambda) &=& \overline
{\frac{dE}{d\lambda}} \\
\label{eq:dldl}
\partial_\lambda^2 F(T,\lambda) &= &
\overline{\frac{d^2E}{d\lambda^2}}
-\beta\, \mbox{Var}\!\left(\frac{dE}{d\lambda}\right) \\
\label{eq:dtdt}
\partial_T^2 F(T,\lambda) &=& -\frac{\mbox{Var}E}{k_B T^3} \\
\label{eq:dtdl}
\partial_T \partial_\lambda F(T,\lambda) &= &
\frac{\mbox{Cov}(E,dE/d\lambda)}{k_B T^2}.
\eea
The expectation value $\overline{(\ldots)}$ stands for a sum
over the eigenenergies with the canonical probability weights, 
and $\mbox{Var}$ and $\mbox{Cov}$ are the corresponding variance and 
covariance. E.g.\ $\overline{E}=U$ and
\bea
\overline{\frac{d^2E}{d\lambda^2}}
&\equiv& \sum_n \frac{d^2E_n}{d\lambda^2}\frac{e^{-\beta E_n}}{Z} \\
\mbox{Cov}(E,dE/d\lambda) &\equiv&
\sum_n E_n \frac{dE_n}{d\lambda}\frac{e^{-\beta E_n}}{Z} 
-\overline{E}\  \overline{\frac{dE}{d\lambda}}.
\eea
Insertion of (\ref{eq:dl}) into (\ref{eq:deriv1})
gives (\ref{eq:relation_T}).
Insertion of (\ref{eq:dldl},\ref{eq:dtdt},\ref{eq:dtdl})
into (\ref{eq:deriv2}) gives (\ref{eq:d2us}).

\section{Normalized wavefunction of an Efimov trimer}\label{app:Efi_psi}

In this Appendix we recall the  wavefunction of an Efimov trimer and give the expression of its normalization constant.
We consider an Efimov trimer state for three spinless bosons of mass $m$ interacting {\sl via} a zero
range infinite scattering length potential.
In order to avoid formal normalisability problems, we imagine that the Efimov trimer
is trapped in an arbitrarily weak harmonic potential, that is with a ground state
harmonic oscillator length  $a_0$ arbitrarily larger than the trimer size.
In this case, the energy of the trimer is essentially the free space energy
$E_{\rm trim} = - \frac{\hbar^2 \kappa_0^2}{m}$, $\kappa_0>0$. According to Efimov's theory
\cite{Efimov.bkp}
\be
\label{eq:kappa0}
\kappa_0 = \frac{\sqrt{2}}{R_t} e^{\pi q/|s_0|} e^{\mbox{\scriptsize Arg}\,\Gamma(1+s_0)/|s_0|}
\ee
where $R_t>0$ is a length known as the three-body parameter, the quantum number
$q$ may take all values in $\mathbb{Z}$ and the purely imaginary number $s_0 = i |s_0|$
is such that
\be
\label{eq:def_ms0}
|s_0| \cosh(|s_0|\pi/2) = \frac{8}{\sqrt{3}} \sinh(|s_0|\pi/6),
\ee
so that $|s_0|=1.00623782510\ldots$
The corresponding three-body wavefunction $\Psi$ may be written as
\be
\label{eq:etat}
\Psi(\mathbf{r}_1,\mathbf{r}_2,\mathbf{r}_3) 
\simeq \psi_{\rm CM}(\mathbf{C}) \left[
\psi(r_{12},|2\mathbf{r}_3-(\mathbf{r}_1+\mathbf{r}_2)|/\sqrt{3})
+\psi(r_{23},|2\mathbf{r}_1-(\mathbf{r}_2+\mathbf{r}_3)|/\sqrt{3})
+
\psi(r_{31},|2\mathbf{r}_2-(\mathbf{r}_3+\mathbf{r}_1)|/\sqrt{3})
\right],
\ee
where $\mathbf{C}=(\mathbf{r}_1+\mathbf{r}_2+\mathbf{r}_3)/3$ is the center of mass
position of the three particles and the parameterization of $\psi$
is related to the Jacobi coordinates 
$\mathbf{r}=\mathbf{r}_2-\mathbf{r}_1$
and $\rhob=[2\mathbf{r}_3-(\mathbf{r}_1+\mathbf{r}_2)]/\sqrt{3}$.
In our expression of $\Psi$, $\psi_{\rm CM}$ is the Gaussian wavefunction of the single particle ground state in the harmonic
trap, normalized to unity, and $\psi$ is a Faddeev component of the free space trimer wavefunction.
The explicit expression of $\psi$ is known \cite{Efimov.bkp}:
\be
\label{eq:psi_F}
\psi(r,\rho)  = \frac{\mathcal{N}_\psi}{\sqrt{4\pi}}  
\frac{K_{s_0}(\kappa_0\sqrt{r^2+\rho^2})}{(r^2+\rho^2)/2}
\frac{\sin[s_0(\frac{\pi}{2}-\alpha)]}{\sin(2\alpha)}
\ee
where $K_{s_0}$ is a Bessel function and $\alpha=\mbox{atan}(r/\rho)$.
The normalization factor ensuring that $||\Psi||^2=1$ may be calculated explicitly: One first
performs the change of variables
$(\mathbf{r}_1,\mathbf{r}_2,\mathbf{r}_3)\to (\mathbf{C}, \mathbf{r},\rhob)$, whose Jacobian
is  $D(r_1,r_2,r_3)/D(C,\rho,r)=(-\sqrt{3}/2)^3$. To integrate over $\mathbf{r}$ and $\rhob$ one
introduces hyperspherical coordinates in which the wavefunction separates; one then faces known
integrals on the hyperradius \cite{Gradstein} and on the hyperangles \cite{Efimov93}.
This leads to \cite{WernerThese}:
\be
\label{eq:norma}
|\mathcal{N}_\psi|^{-2} =
\left(\frac{\sqrt{3}}{2}\right)^3
\frac{3\pi^2}{2\kappa_0^2 \cosh(|s_0|\pi/2)}
\left[\cosh(|s_0|\pi/2) +\frac{|s_0|\pi}{2} \sinh(|s_0|\pi/2) 
-\frac{4\pi}{3\sqrt{3}}\cosh(|s_0|\pi/6)\right].
\ee
We also recalled, as promised in the main text, the value of the hyperangular scalar product 
derived in \cite{WernerThese}:
\be
(\phi_{s_0}|\phi_{s_0}) = \frac{12\pi}{s_0} \sin(s_0\pi/2) \left[\cos(s_0\pi/2) - s_0 \frac{\pi}{2} \sin (s_0\pi/2) 
-\frac{4\pi}{3\sqrt{3}}\cos (s_0\pi/6)\right].
\ee

\section{A lemma for three bosons in the zero-range model}
\label{app:3b}

Here we prove the relation (\ref{eq:lemme_dEdRt}).
The first step is to express the Hamiltonian in hyperspherical coordinates [\thefnnumberthesechaptroissectrois]:
\bea
\la \psi_1, H \psi_2\ra-\la H\psi_1,\psi_2\ra&=&
-\frac{\hbar^2}{2m}\left(\frac{\sqrt{3}}{2}\right)^3 \int_0^\infty dR\,R^5 \int d\Oom \int d\CC
\nonumber
\\ & &
 \left\{\psi^*_1\left(\frac{\partial^2}{\partial R^2}+\frac{5}{R}\frac{\partial}{\partial R}
+\frac{T_\Oom}{R^2} +\frac{1}{3}\Delta_\CC\right)\psi_2
-\left[\psi^*_1 \leftrightarrow \psi_2
\right]
\right\}
\\
&=&
-\frac{\hbar^2}{2m}\left(\frac{\sqrt{3}}{2}\right)^3
\left\{\int dR\,R^5 \int d\Oom\,\cA_\CC(R,\Oom)
+\int d\Oom d\CC\,\cA_R(\Oom,\CC)
\right.
\nonumber
\\
& &
\left.
+\int dR\,R^5\int d\CC\,\cA_\Oom(R,\CC)\right\}
\eea
where
\bea
\cA_\CC(R,\Oom)&\equiv&\int d\CC\,\left\{\psi^*_1\,\frac{1}{3}\Delta_\CC\,\psi_2
-\left[\psi^*_1 \leftrightarrow \psi_2
\right]
\right\}
\label{eq:A_C}
\\
\cA_R(\Oom,\CC)&\equiv&\int dR\,R^5\left\{\psi^*_1\left(\frac{\partial^2}{\partial R^2}+\frac{5}{R}\frac{\partial}{\partial R}\right)\psi_2
-\left[\psi^*_1 \leftrightarrow \psi_2
\right]\right\}
\\
\cA_\Oom(R,\CC)&\equiv&\int d\Oom\left\{\psi^*_1 \frac{T_\Oom}{R^2}\psi_2
-
\psi_2 \frac{T_\Oom}{R^2}\psi^*_1\right\},
\eea
$T_\Oom$ being a differential operator acting on the hyperangles and called Laplacian on the hypersphere.

Clearly $\cA_C(R,\Oom)=\frac{1}{3}\int d\CC\, \nabla_\CC \cdot \left\{ \psi^*_1 \nabla_\CC \psi_2
- \psi_2 \nabla_\CC \psi^*_1 \right\}=0$, since the $\psi_i$'s are regular functions of $\CC$ for every $(R,\Oom)$ except on a set of measure zero.

In what follows we will use the following simple lemma: if $\Phi_1(R)$ and $\Phi_2(R)$ are functions which decay quickly at infinity and have no singularity except maybe at $R=0$, then
\be
\int dR\,R^5\left\{\Phi_1^*\left(\frac{\partial^2}{\partial R^2}
+\frac{5}{R}\frac{\partial}{\partial R}\right)\Phi_2
-\left[\Phi_1^* \leftrightarrow \Phi_2\right]\right\}
=
-\lim_{R\to0} R\left\{ \mathcal{F}_1^* \frac{\partial \mathcal{F}_2}{\partial R} - \mathcal{F}_2 \frac{\partial \mathcal{F}_1^*}{\partial R} \right\}
\label{eq:lemmeR}
\ee
where $\mathcal{F}_i(R)\equiv R^2\,\Phi_i(R)$. 

We now show that
\be
\cA_\Oom(R,\CC)=0\  {\rm for\ any}\ \CC\ {\rm and\ } R>0.
\label{eq:A_Om=0}
\ee
We will use the fact that $\psi_1$ and $\psi_2$ satisfy the two-body boundary condition with the same $a$, and apply lemma (\ref{eq:lemme_3D}). More precisely, we will show that for any smooth function $f(R,\CC)$ which vanishes in a neighborhood of $R=0$,
\be
\int dR\,R^5 \int d\CC \, f(R,\CC)^2\,\cA_\Oom(R,\CC)=0;
\label{eq:lemme_f}
\ee
this clearly implies (\ref{eq:A_Om=0}).
To show (\ref{eq:lemme_f}) we note that
\be
-\frac{\hbar^2}{2m}\left(\frac{\sqrt{3}}{2}\right)^3
\int dR\,R^5 \int d\CC \, f(R,\CC)^2\,\cA_\Oom(R,\CC)
=
-\frac{\hbar^2}{2m}\left(\frac{\sqrt{3}}{2}\right)^3
\int dR\,R^5\int d\Oom\int d\CC
\left\{ (f \psi_1^*)\frac{T_\Oom}{R^2} (f \psi_2)
-\left[\psi^*_1 \leftrightarrow \psi_2\right] \right\},
\ee
which can be rewritten as
\bea
\int d^3r_1 d^3r_2 d^3r_3
\left\{ (f \psi_1^*)H (f \psi_2)
-\left[\psi^*_1 \leftrightarrow \psi_2 \right] \right\}
& &
+\frac{\hbar^2}{2m}\left(\frac{\sqrt{3}}{2}\right)^3
\int dR\,R^5 \int d\Oom \int d\CC
\nonumber
\\ & &
\left\{ (f \psi_1^*)
\left(\frac{\partial^2}{\partial R^2}+\frac{5}{R}\frac{\partial}{\partial R}
 +\frac{1}{3}\Delta_\CC\right) (f \psi_2)
-\left[\psi^*_1 \leftrightarrow \psi_2 \right]\right\}.
\label{eq:expression}
\eea
The first integral in this expression vanishes, as a consequence of lemma (\ref{eq:lemme_3D}).
This lemma is indeed applicable to the wavefunctions $f\psi_i$:
They vanish in a neighborhood of $R=0$ (see the discussion below (\ref{eq:Toto_echange})), moreover
 they satisfy the two-body boundary condition for the same value of the scattering length $a$ (as follows from the fact that $R$ varies quadratically with $r$ for small $r$).
 The second integral in (\ref{eq:expression}) vanishes as well: The contribution of the partial derivatives with respect to $R$ vanishes as a consequence of lemma (\ref{eq:lemmeR}), and the contribution of $\Delta_\CC$ vanishes because the $f\psi_i$'s are regular functions of $\CC$.

Finally, $\cA_R$ can be computed using lemma (\ref{eq:lemmeR}) and the boundary condition (\ref{eq:danilov}), yielding (\ref{eq:lemme_dEdRt}).

\section{First two terms of the large-$k$ expansion of the momentum distribution of an Efimov trimer}
\label{app:efimov}

Here we show that the momentum distribution of an Efimov bosonic trimer state of energy
$-\hbar^2\kappa_0^2/m$ (at rest and for an infinite
scattering length) has the asymptotic expansion (\ref{eq:nk_efi}) with
\bea
\label{eq:Cexact}
C/\kappa_0&=&
\frac{8\pi^2 \sinh(|s_0|\pi/2) \tanh(|s_0|\pi)}{\cosh(|s_0|\pi/2)+\frac{\pi |s_0|}{2}
\sinh(|s_0|\pi/2) - \frac{4\pi}{3\sqrt{3}}\cosh(|s_0|\pi/6)}  =
53.09722846003081\ldots
\\
D/\kappa_0^2 &\simeq& -89.26260
\\
\varphi &\simeq& -0.8727976
\eea

\subsection{Three-body state in momentum space}

We start from the three-body wavefunction  in position space $\Psi$ given in Appendix~\ref{app:Efi_psi}.
To obtain the momentum distribution of the Efimov trimer, we need to evaluate the Fourier transformation
of $\Psi$. Rather than directly using (\ref{eq:psi_F}), we take advantage of the fact that
the Faddeev component $\psi$ obeys Schr\"odinger's equation with a source term. With the change to Jacobi coordinates,
the Laplacian operator in the coordinate space of dimension nine
reads
$\sum_{i=1}^{3} \Delta_{\mathbf{r}_i} = \frac{1}{3}\Delta_{\mathbf{C}}
+2\left[\Delta_{\mathbf{r}}+\Delta_{\rhob}\right]$ so that 
\be
\label{eq:Schr}
-\left[\kappa_0^2-\Delta_{\mathbf{r}}-\Delta_{\rhobs}\right]
\psi(r,\rho) = \delta(\mathbf{r}) B(\rho).
\ee
The source term in the right hand side originates from the fact that 
\be
\psi(r,\rho) \underset{r\to 0}{\sim} -\frac{B(\rho)}{4\pi r}
\ee
for a fixed $\rho$, this $1/r$ divergence coming from the replacement
of the interaction potential by the Bethe-Peierls contact condition.
Taking the Fourier transform of (\ref{eq:Schr}) over $\mathbf{r}$ and
$\rhob$ leads to
\be
\label{eq:psit}
\tilde{\psi}(\kk,\KK) = -\frac{\tilde{B}(K)}{k^2+K^2+\kappa_0^2},
\ee
where the Fourier transform is defined as 
$\tilde{B}(K) \equiv \int d^3\rho  e^{-i\mathbf{K}\cdot\rhobs} B(\rho)$.
$B(\rho)$ is readily obtained from (\ref{eq:psi_F}) by taking the limit $r\to 0$:
\be
B(\rho) = -\mathcal{N}_\psi (4\pi)^{1/2} i \sinh(|s_0|\pi/2) \frac{K_{s_0}(\rho)}{\rho}.
\ee
The Fourier transform of this expression is known, see relation 6.671(5)
in \cite{Gradstein}, so that
\be
\label{eq:Bt}
\tilde{B}(K) = -\mathcal{N}_\psi \frac{2\pi^{5/2}}{K (K^2+\kappa_0^2)^{1/2}}
\left\{
\left[\frac{(K^2+\kappa_0^2)^{1/2}+K}{\kappa_0}\right]^{s_0}
-
\left[\frac{(K^2+\kappa_0^2)^{1/2}+K}{\kappa_0}\right]^{-s_0}
\right\}.
\ee
What we shall need is the large $K$ behavior of $\tilde{B}(K)$. Expanding (\ref{eq:Bt})
in powers of $\kappa_0/K$ gives
\be
\label{eq:agk}
\tilde{B}(K) = \mathcal{N}_\psi \frac{2\pi^{5/2}}{K^2}
\left[(2K/\kappa_0)^{-s_0}-\mbox{c.c.}\right]
+O(1/K^4).
\ee
When necessary one may further use the relation
\be
\label{eq:ifnec}
(\kappa_0/2)^{s_0}= (-1)^q \left(R_t\sqrt{2}\right)^{-s_0} \frac{\Gamma(1+s_0)}{|\Gamma(1+s_0)|}
\ee
that can be deduced from (\ref{eq:kappa0}). 

The last step is to take the Fourier transform of (\ref{eq:etat}), using the appropriate Jacobi coordinates
for each Faddeev component (or simply by Fourier transforming the first Faddeev component
using the coordinates $(\mathbf{C},\mathbf{r},\rhob)$ given above
and by performing circular permutations on the particle labels). This gives
\bea
\label{eq:etatf}
\tilde{\Psi}(\kk_1,\kk_2,\kk_3) &= &\left(\frac{\sqrt{3}}{2}\right)^3
\tilde{\psi}_{\rm CM}(\kk_1+\kk_2+\kk_3) \left[
\tilde{\psi}(|\kk_2-\kk_1|/2,\sqrt{3}|\kk_3-(\kk_1+\kk_2)/2|/3)
\right. \nonumber \\
&+&
\left. \tilde{\psi}(|\kk_3-\kk_2|/2,\sqrt{3}|\kk_1-(\kk_2+\kk_3)/2|/3)
+
\tilde{\psi}(|\kk_1-\kk_3|/2,\sqrt{3}|\kk_2-(\kk_3+\kk_1)/2|/3)
\right].
\eea

\subsection{Formal expression of the momentum distribution}

To obtain the momentum distribution, it remains to integrate over
$\kk_3$ and $\kk_2$ the modulus square of (\ref{eq:etatf}).
In the limit $\kappa_0 a_{\rm ho}\to +\infty$, one can set
\be
|\tilde{\psi}_{\rm CM}(\kk_1+\kk_2+\kk_3)|^2 = (2\pi)^3 \delta(\kk_1+\kk_2+\kk_3).
\ee
Integration over $\kk_3$ is then straightforward:
\be
n(\kk_1) = 3(\sqrt{3}/2)^6 \int\frac{d^3 k_2}{(2\pi)^3} 
\left|
\tilde{\psi}(|\kk_2-\kk_1|/2,\sqrt{3}|\kk_1+\kk_2|/2)
+
\tilde{\psi}(|\kk_2+\kk_1/2|,\sqrt{3}k_1/2)
+
\tilde{\psi}(|\kk_1+\kk_2/2|,\sqrt{3}k_2/2)
\right|^2.
\ee
The factor $3$ in the right hand side results from the fact that,
in this article, we normalize the momentum distribution $n(\kk)$
to the total number of particles (rather than to unity).
One further realizes that the sum of the squared moduli of the 
arguments of $\tilde{\psi}$ is constant and equal to $k_1^2+k_2^2+\kk_1\cdot\kk_2$
for each term in the right hand side.
One uses (\ref{eq:psit}), thus putting the denominator in (\ref{eq:psit}) as a common denominator,
to obtain
\be
\label{eq:nkb}
n(\kk_1) = 3(\sqrt{3}/2)^6 \int\frac{d^3 k_2}{(2\pi)^3}
\frac{\left[\tilde{B}(\sqrt{3}|\kk_1+\kk_2|/2)
+\tilde{B}(\sqrt{3}k_1/2) + \tilde{B}(\sqrt{3}k_2/2)
\right]^2}{(k_1^2+k_2^2+\kk_1\cdot\kk_2+\kappa_0^2)^2}.
\ee
For simplicity, we have assumed that the normalization factor
$\mathcal{N}_\psi$ is purely imaginary, so that $\tilde{B}(K)$ is a
real quantity.

In the above writing of $n(\kk_1)$, the only ``nasty" contribution is $\tilde{B}(\sqrt{3}|\kk_1+\kk_2|/2)$;
the other contributions are ``nice" since they only depend on the moduli $k_1$ and $k_2$.
Expanding the square in the numerator of (\ref{eq:nkb}), one gets six terms, three squared terms
and three crossed terms.
The change of variable $\kk_2=-(\kk'_2+\kk_1)$ allows, in one of the squared term and in one
of the crossed term, to transform a nasty term into a nice term.
What remains is a nasty crossed term that cannot be turned into a nice one; in that term,
as  a compromise, one performs the change of variable $\kk_2=-(\kk_2'+\kk_1/2)$.
We finally obtain the momentum distribution as the sum of four contributions,
\be
\label{eq:decomp}
n(\kk_1) = n_I(\kk_1) + n_{II}(\kk_1) + n_{III}(\kk_1) + n_{IV}(\kk_1),
\ee
with
\bea
\label{eq:nI}
n_I(\kk_1) &=&  
3(\sqrt{3}/2)^6 \int\frac{d^3 k_2}{(2\pi)^3}
\frac{\tilde{B}^2(\sqrt{3}k_1/2)}{(k_1^2+k_2^2+\kk_1\cdot\kk_2+\kappa_0^2)^2} 
\\
\label{eq:nII}
n_{II}(\kk_1) &=& 
3(\sqrt{3}/2)^6 \int\frac{d^3 k_2}{(2\pi)^3}
\frac{2\tilde{B}^2(\sqrt{3}k_2/2)}{(k_1^2+k_2^2+\kk_1\cdot\kk_2+\kappa_0^2)^2} \\
\label{eq:nIII}
n_{III}(\kk_1) &=& 
3(\sqrt{3}/2)^6 \int\frac{d^3 k_2}{(2\pi)^3}
\frac{4 \tilde{B}(\sqrt{3}k_1/2) \tilde{B}(\sqrt{3}k_2/2)}{(k_1^2+k_2^2+\kk_1\cdot\kk_2+\kappa_0^2)^2} \\
\label{eq:nIV}
n_{IV}(\kk_1) &=& 
3(\sqrt{3}/2)^6 \int\frac{d^3 k_2}{(2\pi)^3}
\frac{2\tilde{B}(\sqrt{3}|\kk_2+\kk_1/2|/2)\tilde{B}(\sqrt{3}|\kk_2-\kk_1/2|/2)}
{(\kappa_0^2+k_2^2+3k_1^2/4)^2}.
\eea
We shall now take the large $k_1$ limit, or equivalently formally the $\kappa_0\to 0$ limit
for a fixed $k_1$. From the asymptotic behavior (\ref{eq:agk})  we see that $\tilde{B}(k_1)^2$
involves a sum of ``oscillating" terms involving $k_1^{2s_0}$ or $k_1^{-2 s_0}$, and of
``non-oscillating" terms. We shall calculate first the resulting non-oscillating contribution, then
the resulting oscillating one, up to order $1/k_1^5$ included.

\subsection{Non-oscillating contribution up to $O(1/k_1^5)$}

We consider the small $\kappa_0$ limit successively for each of the four components
of $n(k_1)$ in (\ref{eq:decomp}).

\noindent{\bf Contribution $I$:} Taking directly  $\kappa_0\to 0$ in the integral defining $n_I$, replacing
$\tilde{B}(k_1)$ by its asymptotic behavior (\ref{eq:agk}) and averaging out the oscillating
terms $k_1^{\pm 2 s_0}$ gives the leading behavior
\be
\langle n_I(\kk_1)\rangle \simeq \frac{3\sqrt{3}}{8\pi} |\mathcal{N}_\psi|^2 \frac{4\pi^5}{k_1^5}.
\ee

\noindent{\bf Contribution $II$:}
In the integrand of (\ref{eq:nII}), we use the splitting
\be
(k_1^2+k_2^2+\kk_1\cdot\kk_2+\kappa_0^2)^{-2}=k_1^{-4} + 
\left[(k_1^2+k_2^2+\kk_1\cdot\kk_2+\kappa_0^2)^{-2}-k_1^{-4}\right].
\ee
The first term in the right hand side gives a contribution exactly scaling as $1/k_1^4$.
In the contribution of the second term in the right hand side, one may take the limit
$\kappa_0\to 0$ and replace $\tilde{B}^2(\sqrt{3} k_2/2)$ by its asymptotic expression to get
the subleading $1/k_1^5$ contribution. Performing the change of variable $\kk_2 = k_1 \qq$
in the integral and averaging out the oscillating terms
$k_1^{\pm 2 s_0}$ gives
\be
\langle n_{II}(k_1) \rangle = \frac{C}{k_1^4} 
-\frac{3\sqrt{3}}{2\pi} |\mathcal{N}_\psi|^2 \frac{4\pi^5}{k_1^5} + o(1/k_1^5),
\ee
with
\be
C = 3 (\sqrt{3}/2)^6 \int\frac{d^3 k_2}{(2\pi)^3} 2\tilde{B}^2(\sqrt{3}k_2/2).
\label{eq:Cint}
\ee
We calculate $C$ from the exact expression (\ref{eq:Bt}) of $\tilde{B}$: We integrate over solid
angles and we use the change of variables $\frac{\sqrt{3}}{2} k_2 = \kappa_0 \sinh\alpha$,
where $\alpha$ varies from zero to $+\infty$, to take advantage of the fact that
\be
\tilde{B}(\kappa_0\sinh\alpha) = -\mathcal{N}_\psi \frac{2\pi^{5/2}}{\kappa_0^2 \sinh\alpha\cosh\alpha}
\left(e^{s_0\alpha}-e^{-s_0\alpha}\right).
\label{eq:sacdv}
\ee
This leads to
\be
C =  12\pi^3 (\sqrt{3}/2)^3 \frac{|\mathcal{N}_\psi|^2}{\kappa_0} 
\int_0^{+\infty} d\alpha\, \frac{2-(e^{2s_0\alpha}+\mbox{c.c.})}{\cosh\alpha},
\ee
where we used the fact that $\mathcal{N}_\psi^2=-|\mathcal{N}_\psi|^2$.
The resulting integral over $\alpha$ may be extended over the whole real axis because
the integrand is an even function of $\alpha$; it may then be evaluated by using the general result
(that we obtained with contour integration)
\be
\label{eq:Utile}
K(\theta,s) \equiv \int_{-\infty}^{+\infty} d\alpha\, \frac{e^{is\alpha}}{\cosh\alpha +\cos\theta}
= \frac{2\pi}{\sin\theta} \frac{\sinh(s\theta)}{\sinh(s\pi)}
\ee
where $s$ is a real number and $\theta\in]0,\pi[$. One simply has to take $\theta=\pi/2$,
$s=0$ and $s=|s_0|$ respectively. We get
\be
\label{eq:Cprov}
C = \frac{24 \pi^4}{\kappa_0}  \left(\frac{\sqrt{3}}{2}\right)^3 
\frac{2\sinh^2(|s_0|\pi/2)}{\cosh(|s_0|\pi)}\,
|\mathcal{N}_\psi|^2.
\ee
This, together with (\ref{eq:norma}), leads to the explicit expression (\ref{eq:Cexact}) for $C$.

\noindent{\bf Contribution $III$:}
We directly take the limit $\kappa_0\to 0$ and we replace the factors $\tilde{B}$ by
their asymptotic expressions in (\ref{eq:nIII}). After the change of variable
$\kk_2=k_1 \qq$, angular integration and averaging  out of the oscillating
terms $k_1^{\pm 2 s_0}$, this gives
\be
\langle n_{III}(\kk_1)\rangle =
\frac{9}{2\pi^2} \frac{4\pi^5|\mathcal{N}_\psi|^2}{k_1^5} \int_0^{+\infty} dq \, \frac{q^{s_0}+q^{-s_0}}{q^4+q^2+1}
+o(1/k_1^5).
\ee
In this result, we change the integration variable setting $q=e^\alpha$, where $\alpha$ varies from $-\infty$ to
$+\infty$. The odd component of the integrand (involving $\sinh\alpha$) gives a vanishing contribution.
The even component of the integrand involves a rational fraction of $\cosh\alpha$ to which we apply
a partial fraction decomposition. Then we use (\ref{eq:Utile}) to obtain
\be
n_{III}(k_1) = 
\frac{4\pi^5 |\mathcal{N}_\psi|^2}{k_1^5} \frac{3\sqrt{3}}{2\pi}
\frac{\sinh(\pi|s_0|/3)+\sinh(2\pi|s_0|/3)}{\sinh(\pi|s_0|)} + o(1/k_1^5).
\ee

\noindent{\bf Contribution $IV$:}
We directly take the limit $\kappa_0\to 0$ and we replace the factors $\tilde{B}$ by
their asymptotic expressions in (\ref{eq:nIV}). We perform the change of variable
$\kk_2=(k_1/2)\qq$, we average out the oscillating terms $k_1^{\pm 2 s_0}$.
The angular integration in spherical coordinates of axis the direction of $\kk_1$
may be performed using
\be
\int dv\, \left(\frac{1+v}{1-v}\right)^{s_0/2}(1-v^2)^{-1} = \left(\frac{1+v}{1-v}\right)^{s_0/2}/s_0,
\ee
where the variable $v$ is restricted to the interval $(-1,1)$. This leads to
\be
\langle n_{IV}(\kk_1)\rangle =
\frac{4\pi^5 |\mathcal{N}_\psi|^2}{k_1^5}
\frac{36}{\pi^2}
\int_0^{+\infty} dq\, \frac{q}{q^2+1} (q^2+3)^{-2} \left[s_0^{-1} \left(\frac{q+1}{|q-1|}\right)^{s_0}+
\mbox{c.c.}\right]
+o(1/k_1^5).
\ee
Calculating this integral directly is not straightforward because of the occurrence of the absolute
value $|q-1|$. We thus split the integration domain in two intervals.
For $q\in [0,1]$ we set $q=(X-1)/(X+1)$ (an increasing function of $X$,
where $X$ spans $[1,+\infty]$).
For $q\in [1,+\infty]$ we set $q=(X+1)/(X-1)$ (a decreasing function of $X$,
where $X$ here also spans $[1,+\infty]$).
Then
\be
\langle n_{IV}(\kk_1)\rangle=
\frac{4\pi^5 |\mathcal{N}_\psi|^2}{k_1^5}
\frac{9}{2\pi^2}
\int_1^{+\infty} \frac{dX}{X}\, \frac{(X^2-1+X^{-2})(X-X^{-1})}{(X^2+1+X^{-2})^2}\left[s_0^{-1} X^{s_0} - s_0^{-1} X^{-s_0}\right] + o(1/k_1^5).
\ee
We then set $X=e^\alpha$, where $\alpha$ ranges from zero to $+\infty$,
and we use the fact that the resulting integrand is an even function of $\alpha$
to extend the integral over the whole real axis.
We integrate by parts, integrating the factor $\sin(\alpha|s_0|)$,
and we perform a partial fraction decomposition of the resulting rational fraction of
$\cosh\alpha$.
Using (\ref{eq:Utile}) and its derivatives with respect to $\theta$, we get
\be
\langle n_{IV}(\kk_1)\rangle =
-\frac{12\pi^5|\mathcal{N}_\psi|^2}{k_1^5}\times
\frac{-6[\cosh(2\pi|s_0|/3)-\cosh(\pi|s_0|/3)]
+\sqrt{3} |s_0| [\sinh(2\pi|s_0|/3)+\sinh(\pi|s_0|/3)]}
{2\pi|s_0|\sinh(\pi|s_0|)} + o(1/k_1^5).
\ee

\noindent{\bf Sum of the four contributions:}
Summing up the terms in $1/k_1^5$ of the contributions $n_I$, $n_{II}$, $n_{III}$
and $n_{IV}$, we obtain as a global prefactor the quantity
\be
\mathcal{S} = -\frac{\sqrt{3}}{8} +
\frac{\cosh(2\pi|s_0|/3)-\cosh(\pi|s_0|/3)}{|s_0|\sinh(\pi|s_0|)}.
\ee
Multiplying (\ref{eq:def_ms0}) on both sides by $\sinh(|s_0|\pi/2)$ 
and using 
\be
2\sinh a\sinh b = \cosh(a+b)-\cosh(a-b), \ \ \ \forall a,b
\ee
we find that $\mathcal{S}$ is exactly zero. As a consequence,
the non-oscillating part of the momentum distribution of an infinite
scattering length Efimov trimer behaves at large $k$ as
\be
\langle n(\kk_1)\rangle = \frac{C}{k_1^4} + o(1/k_1^5).
\ee

\subsection{Oscillating contribution at large $k_1$}

In the large $k_1$ tail of the momentum distribution, we now include
{\sl oscillating} terms, having oscillating factors such as $k_1^{\pm 2 s_0}$.
The calculation techniques are the same of in the previous subsection, so that
we give here directly the result.
We find that the leading oscillating terms scale as $1/k_1^5$:
\be
n(\kk_1)-\langle n(\kk_1)\rangle 
=
-\frac{12\pi^5}{k_1^5} |\mathcal{N}_\psi|^2 
\left[\mathcal{A}\left(\frac{k_1 \sqrt{3}}{\kappa_0}\right)^{2s_0}+\mbox{c.c.}
\right]
+o(1/k_1^5)
\ee
where the complex amplitude $\mathcal{A}$ is the sum of the contributions
coming from each of the four components (\ref{eq:nI},\ref{eq:nII},\ref{eq:nIII},\ref{eq:nIV})
of the moment distribution,
\be
\mathcal{A} = \mathcal{A}_I + \mathcal{A}_{II} + \mathcal{A}_{III}
+ \mathcal{A}_{IV}.
\ee
We successively find
\bea
\mathcal{A}_I &=&  
\frac{3}{8\pi^2} \int_0^{+\infty} \!\!\!\! dq\,
\frac{q^2}{q^4+q^2+1} = \frac{\sqrt{3}}{16\pi}, \\
\mathcal{A}_{II} &=& 
\frac{3}{4\pi^2} \int_0^{+\infty}\!\!\!\!  dq\,
\frac{q^{2s_0}}{q^2} \left[(q^4+q^2+1)^{-1}-1\right] 
= -\frac{\sqrt{3}}{4\pi} \frac{\sinh(4\pi|s_0|/3)+\sinh(2\pi|s_0|/3)}{\sinh(2\pi|s_0|)}, \\
\mathcal{A}_{III} &=& 
\frac{3}{2\pi^2} \int_0^{+\infty}\!\!\!\!  dq\,
\frac{q^{s_0}}{q^4+q^2+1} 
= \frac{\sqrt{3}}{4\pi}
\frac{\sinh(2\pi|s_0|/3)+\sinh(\pi|s_0|/3)-i\sqrt{3}
[\cosh(2\pi|s_0|/3)-\cosh(\pi|s_0|/3)]}{\sinh(\pi |s_0|)}, \\
\label{eq:dur}
\mathcal{A}_{IV} &=&
\frac{12}{\pi^2}\ 2^{-2 s_0}
\int_0^{+\infty} \!\!\!\! dq\, \frac{q (1+q^2)^{s_0}}{(q^2+3)^2(q^2+1)}
\int_0^{2q/(1+q^2)}\!\!\!\!   dv\, \frac{(1-v^2)^{s_0/2}}{1-v^2}
\simeq  0.0243657158 - 0.0698680251 i.
\eea
We have calculated analytically all these integrals, except for (\ref{eq:dur})
where the angular integration gives rise to the integral over $v$ and thus
to a difficult hypergeometric function. We used numerical integration for (\ref{eq:dur}).
Finally
\be
\mathcal{A} \simeq 0.1022397786 - 0.1218775240 i.   
\ee

\subsection{Momentum distribution at the origin}

The contribution $n_I(\kk_1)$ is straightforward to calculate at
all $\kk_1$:
\be
n_I(\\k_1) = \frac{\sqrt{3}}{4\pi\kappa_0}
\left(\frac{\sqrt{3}}{2}\right)^6
\frac{\tilde{B}^2(\sqrt{3}k_1/2)}{(k_1^2+4\kappa_0^2/3)^{1/2}}.
\ee
The contribution $n_{II}(\kk_1)$ is also exactly calculable
by performing the change of variable
$k_2=(2/\sqrt{3})\sinh\alpha$ and using the generalization of
(\ref{eq:Utile}):
\be
\int_{-\infty}^{+\infty} d\alpha\, \frac{e^{is\alpha}}{\cosh\alpha 
-\cosh\alpha_0}
= \frac{2\pi\sin[s(i\pi-\alpha_0)]}{\sinh\alpha_0 \sinh(s\pi)},
\ee
where $\alpha_0$ is a complex number with non-zero imaginary part.
This allows to obtain an exact expression of $n_{III}(\kk_1)$
if one further applies integration by part, integrating the
factor $\sin(|s_0|\alpha)$. 

We do not give here the resulting
expressions.
Contrarily to these first three contributions to $n(\kk_1)$, the contribution
$n_{IV}(\kk_1)$ in (\ref{eq:nIV}) indeed seems difficult to calculate
analytically for an arbitrary $\kk_1$, and blocked our attempt to
calculate $n(\kk_1)$ exactly. For $\kk_1=\mathbf{0}$
however it becomes equal to the contribution $n_{II}$ and
may be evaluated exactly.
We have thus calculated the value of $n(\kk_1=0)$:
\bea
n_{I}(\mathbf{0}) &=& 
\frac{3\tilde{B}^2(0)}{8\pi\kappa_0} \left(\frac{\sqrt{3}}{2}\right)^6
\\
n_{II}(\mathbf{0}) &=& 
\frac{6\sqrt{3}\tilde{B}^2(0)}{\pi|s_0|^2\kappa_0} \left(\frac{\sqrt{3}}{2}\right)^6
\left\{1-\frac{1}{\cosh(|s_0|\pi)}\right. \nonumber\\
&& \left. +\frac{|s_0|}{3}
\,\frac{\cosh(|s_0|2\pi/3)-\cosh(|s_0|4\pi/3)}{\sinh(|s_0|\pi)\cosh(|s_0|\pi)}
+\frac{5\sqrt{3}}{18}\,\left[\frac{\sinh(|s_0|4\pi/3)+\sinh(|s_0|2\pi/3)}
{\sinh(|s_0|\pi)\cosh(|s_0|\pi)}-2\right]\right\}
\\
n_{III}(\mathbf{0}) &=& 
\frac{6\sqrt{3}\tilde{B}^2(0)}{\pi|s_0|\kappa_0} \left(\frac{\sqrt{3}}{2}\right)^6
\,
\frac{\cosh(|s_0|\pi/3)-\cosh(|s_0|2\pi/3)+(|s_0|/\sqrt{3})[\sinh(|s_0|2\pi/3)+\sinh(|s_0|\pi/3)]}{\sinh(|s_0|\pi)}
\\
n_{IV}(\mathbf{0}) &=&  n_{II}(\mathbf{0}),
\eea
with $\tilde{B}(0)= -i\mathcal{N}_\psi 4\pi^{5/2}|s_0|/\kappa_0^2$ according to (\ref{eq:Bt}).
This leads to (\ref{eq:n_ori}).

\section{A direct calculation of $E(\Lambda)$}
\label{app:brutale}

To calculate the cut-off dependent energy $E(\Lambda)$ defined in (\ref{eq:elam})
for an infinite scattering length Efimov trimer, the method consisting
in calculating the momentum distribution $n(k)$ and then 
integrating (\ref{eq:elam}) is numerically demanding: a double integral
has to be performed to obtain $n(k)$, see (\ref{eq:nIV}), so that
the evaluation of $E(\Lambda)$ results in a triple integral.
A more direct formulation, involving only a double integration, is proposed here.
One simply rewrites (\ref{eq:elam}) as
\be
E(\Lambda) = \int_{\mathbb{R}^3} \frac{d^3\!k}{(2\pi)^3}  f(k)\frac{\hbar^2 k^2}{2m}
\left[n(k)-\frac{C}{k^4}\right]
\label{eq:wcof}
\ee
where the function $f(k)$ is equal to unity for $0\leq k\leq \Lambda$
and is equal to zero otherwise.
Then one plugs in (\ref{eq:wcof}) the expression (\ref{eq:decomp}) of $n(k)$, also
replacing $C$ with its integral expression (\ref{eq:Cint}). An integration over two
vectors in $\mathbb{R}^3$ appears:
\bea
E(\Lambda) &=& E_{\rm easy}(\Lambda) + E_{\rm hard}(\Lambda) \\
E_{\rm easy}(\Lambda) &=& 
3\left(\frac{\sqrt{3}}{2}\right)^6
\int \frac{d^3k}{(2\pi)^3} f(k)\frac{\hbar^2 k^2}{2m}
\int \frac{d^3q}{(2\pi)^3}
\left[
\frac{\tilde{B}^2(\frac{\sqrt{3}}{2}k) + 2 \tilde{B}^2(\frac{\sqrt{3}}{2}q)
+ 4 \tilde{B}(\frac{\sqrt{3}}{2}k) \tilde{B}(\frac{\sqrt{3}}{2}q)}{(k^2+q^2+
\kk\cdot\qq + \kappa_0^2)^2}
-\frac{2\tilde{B}^2(\frac{\sqrt{3}}{2}q)}{k^4}
\right] \\
E_{\rm hard}(\Lambda) &=& 
3\left(\frac{\sqrt{3}}{2}\right)^6 
\int \frac{d^3k}{(2\pi)^3} f(k)\frac{\hbar^2 k^2}{2m} 
\int \frac{d^3q}{(2\pi)^3}
\frac{2\tilde{B}(\frac{\sqrt{3}}{2}|\qq+\kk/2|) \tilde{B}(\frac{\sqrt{3}}{2}|\qq-\kk/2|)}
{(q^2+\frac{3}{4} k^2 +\kappa_0^2)^2}.
\label{eq:grosse_exp}
\eea
The first part $E_{\rm easy}$ of this expression originates from the bits $n_{I}$, $n_{II}$,
$n_{III}$ of the momentum distribution and from $C$; angular integrations
may be performed, one is left with a double integral over the moduli
$k$ and $q$. Taking $\kappa_0$ as a unit of momentum and $\hbar^2 \kappa_0^2/m$
as a unit of energy in what follows:
\be
E_{\rm easy}(\Lambda) = 3\left(\frac{\sqrt{3}}{2}\right)^6 
\left(\frac{4\pi}{(2\pi)^3}\right)^2
\int_0^{\Lambda} dk\, \frac{k^4}{2} \int_0^{+\infty} dq\, q^2
\left[
\frac{\tilde{B}^2(\frac{\sqrt{3}}{2}k) + 2 \tilde{B}^2(\frac{\sqrt{3}}{2}q)
+ 4 \tilde{B}(\frac{\sqrt{3}}{2}k) \tilde{B}(\frac{\sqrt{3}}{2}q)}{(k^2+q^2+1)^2-k^2q^2}
-\frac{2\tilde{B}^2(\frac{\sqrt{3}}{2}q)}{k^4}
\right]
\ee
that we integrate numerically.
The second part $E_{\rm hard}(\Lambda)$ in (\ref{eq:grosse_exp})  
originates from the bit $n_{IV}$ 
of the momentum distribution. Performing the change of variables
$\qq=(\kk_1-\kk_2)/2$ and $\kk=\kk_1+\kk_2$ ensures that the factors $\tilde{B}$
are now functions of the moduli $k_1$ and $k_2$ only, 
\be
E_{\rm hard}(\Lambda) = 3\left(\frac{\sqrt{3}}{2}\right)^6 \int \frac{d^3k_1}{(2\pi)^3}
\int \frac{d^3k_2}{(2\pi)^3}
\frac{1}{2} (\kk_1+\kk_2)^2 f(|\kk_1+\kk_2|) 
\frac{2\tilde{B}(\frac{\sqrt{3}}{2}k_1)\tilde{B}(\frac{\sqrt{3}}{2}k_2)}
{(k_1^2+k_2^2+\kk_1\cdot\kk_2 + 1)^2}
\ee
so that angular integrations
may again be performed, involving the integral
\bea
I(k_1,k_2) &=& \frac{1}{2} \int_{-1}^{1} du\, \frac{k_1^2+k_2^2+2k_1k_2 u}
{(k_1^2+k_2^2+k_1 k_2 u +1)^2} f\left(\sqrt{k_1^2+k_2^2+2k_1k_2 u}\right) \\
&=& 
\frac{1}{k_1 k_2}\,\left[ \ln(1+k_1^2+k_2^2+k_1 k_2 u)
+ \frac{1+(k_1^2+k_2^2)/2}{1+k_1^2+k_2^2+k_1 k_2 u}
\right]_{-1}^{\mathrm{max}[-1,\mathrm{min}(1,U)]}
\eea
where $u$ is the cosine of the angle between the vectors $\kk_1$ and $\kk_2$,
$U=[\Lambda^2-(k_1^2+k_2^2)]/(2 k_1 k_2)$, $\mathrm{max}(a,b)$ (resp. $\mathrm{min}(a,b)$)
is the largest (resp. smallest)
of the two numbers $a$ and $b$, 
and the notation $[F(u)]_a^b$
stands for $F(b)-F(a)$ for any function $F(u)$.
We also used the fact that $|\kk_1+\kk_2|\leq \Lambda$ if and only if
$u\leq U$.  This leads to
\be
E_{\rm hard}(\Lambda) = 3\left(\frac{\sqrt{3}}{2}\right)^6 
\left(\frac{4\pi}{(2\pi)^3}\right)^2
\int_0^{+\infty} dk_1 \, k_1^2 
\int_0^{+\infty} dk_2 \, k_2^2 \,
I(k_1,k_2) \tilde{B}(\frac{\sqrt{3}}{2}k_1) \tilde{B}(\frac{\sqrt{3}}{2}k_2).
\ee
Further simplifications may be performed. One can map the integration to the
domain $k_1\geq k_2$ since the integrand is a symmetric function of $k_1$ and
$k_2$. Then performing the change of variable $k_1=q+k/2$ and $k_2=q-k/2$,
and using the fact that $I(k_1,k_2)=0$ if $k_1-k_2>\Lambda$, we obtain the useful form
\be
E_{\rm hard}(\Lambda) = 3\left(\frac{\sqrt{3}}{2}\right)^6 \left(\frac{4\pi}{(2\pi)^3}\right)^2
2\int_0^{\Lambda} dk\, \int_{k/2}^{+\infty} dq\,
(q^2-k^2/4)^2 I(q+k/2,q-k/2) \tilde{B}\left[\frac{\sqrt{3}}{2}(q+k/2)\right]
\tilde{B}\left[\frac{\sqrt{3}}{2}(q-k/2)\right],
\ee
that we integrate numerically.
A useful result to control the numerical error due to the truncation of the integral
over $q$ to a value $\gg \Lambda$ and $\gg 1$ is
\be
I(q+k/2,q-k/2) \underset{q\to +\infty}{\sim} \frac{k^4-\Lambda^4}{8 q^6}.
\ee

\section{Validity of the energy-momentum relation for $\eta\to 0^+$ with a smooth regularisation}
\label{app:pour_les_sceptiques}

Here we prove (\ref{eq:emrfsr}) for the skeptics. We take $\kappa_0$ as a unit of wavevector
and $\hbar^2 \kappa_0^2/m$ as a unit of energy, so that the energy of the infinite
scattering length bosonic Efimov trimer is $E_{\rm trim}=-1$.

First we obtain an integral expression for $E_\eta$ for a non-zero $\eta$, using
the same technique as in Appendix \ref{app:brutale}, after having
replaced $\tilde{B}$ by $\tilde{B}_\eta$ in (\ref{eq:decomp}). 
After angular integration we obtain
\bea
E_\eta &=& 3 \left(\frac{\sqrt{3}}{2}\right)^6\left(\frac{4\pi}{(2\pi)^3}\right)^2
\int_0^{+\infty} dk\, k^2 \int_0^{+\infty} dq\, q^2
\left\{\frac{k^2}{2} \, \frac{\tilde{B}_\eta^2(\frac{\sqrt{3}}{2}k)+ 
2 \tilde{B}_\eta^2(\frac{\sqrt{3}}{2}q) + 
4 \tilde{B}_\eta(\frac{\sqrt{3}}{2}k) \tilde{B}_\eta(\frac{\sqrt{3}}{2}q)}{(k^2+q^2+1)^2-k^2 q^2}
-\frac{\tilde{B}_\eta^2(\frac{\sqrt{3}}{2}q)}{k^2} \right.\nonumber \\
&+& \left. \left[
\frac{1}{kq}\ln\frac{1+k^2+q^2+kq}{1+k^2+q^2-kq} - \frac{2+k^2+q^2}{(k^2+q^2+1)^2-k^2 q^2}
\right] \tilde{B}_\eta\left(\frac{\sqrt{3}}{2}k\right) \tilde{B}_\eta\left(\frac{\sqrt{3}}{2}q\right)
\right\}.
\eea
We collect all the squared terms in $\tilde{B}_\eta^2$, transforming 
$\tilde{B}^2_\eta\left(\frac{\sqrt{3}}{2}q\right)$ into
$\tilde{B}^2_\eta\left(\frac{\sqrt{3}}{2}k\right)$ by an exchange of the integration variables
$q$ and $k$. As a consequence the integral over $q$ can be performed explicitly for these terms:
\be
\int_0^{+\infty} dq\, q^2 \left[\frac{q^2+k^2/2}{(k^2+q^2+1)^2-k^2 q^2}-\frac{1}{q^2}\right]
= -\frac{3\pi}{4} \frac{k^2+2}{(3k^2+4)^{1/2}}.
\ee
Using the same exchange trick we that some simplification occurs
among the crossed terms 
in $\tilde{B}^2_\eta\left(\frac{\sqrt{3}}{2}q\right) \tilde{B}^2_\eta\left(\frac{\sqrt{3}}{2}k\right)$.
For convenience we split the final result in three pieces:
\be
E_\eta = \frac{}{}E^{(1)}_\eta + E^{(2)}_\eta + E^{(3)}_\eta 
\ee
with
\bea
E^{(1)}_\eta &=&  3 \left(\frac{\sqrt{3}}{2}\right)^6\left(\frac{4\pi}{(2\pi)^3}\right)^2 
\left(-\frac{3\pi}{4}\right) \int_0^{+\infty} dk\, k^2 \tilde{B}_\eta^2\left(\frac{\sqrt{3}}{2}k\right)
\frac{k^2+2}{(3k^2+4)^{1/2}} \\
\\
E^{(2)}_\eta &=& 3 \left(\frac{\sqrt{3}}{2}\right)^6\left(\frac{4\pi}{(2\pi)^3}\right)^2
\int_0^{+\infty} dk\, k^2
\int_0^{+\infty} dq\, q^2
 \tilde{B}_\eta\left(\frac{\sqrt{3}}{2}k\right) \tilde{B}_\eta\left(\frac{\sqrt{3}}{2}q\right)
\frac{1}{kq} \ln\frac{1+k^2+q^2+kq}{1+k^2+q^2-kq} \\
E^{(3)}_\eta &=& 3 \left(\frac{\sqrt{3}}{2}\right)^6\left(\frac{4\pi}{(2\pi)^3}\right)^2(-2)
\int_0^{+\infty} dk\, k^2\int_0^{+\infty} dq\, q^2
\frac{\tilde{B}_\eta\left(\frac{\sqrt{3}}{2}k\right) \tilde{B}_\eta\left(\frac{\sqrt{3}}{2}q\right)}
{(1+k^2+q^2+kq)(1+k^2+q^2-kq)}.
\eea

In a second step we use the property
\be
\tilde{B}_\eta (\sinh\alpha) = \frac{\tilde{B}(0)}{|s_0|} 
\frac{\sin(|s_0|\alpha)}{\sinh\alpha\cosh\alpha}\, e^{-\eta\alpha}.
\ee
Hence we perform the change of variable $k=(2/\sqrt{3})\sinh\alpha$ and
$q=(2/\sqrt{3})\sinh\beta$.
The first piece is transformed into
\be
E^{(1)}_\eta = 3 \left(\frac{\sqrt{3}}{2}\right)^3\left(\frac{4\pi}{(2\pi)^3}\right)^2\left(-\frac{3\pi}{4}\right)
\left(\frac{\tilde{B}(0)}{|s_0|}\right)^2
\int_0^{+\infty} d\alpha\, \sin^2(|s_0|\alpha) e^{-2\eta\alpha} \left(\frac{2}{3}+
\frac{1}{3\cosh^2\alpha}\right).
\ee
Taking the limit $\eta\to 0^+$ in $E^{(1)}$, we see that
\be
E^{(1)}_\eta = 3 \left(\frac{\sqrt{3}}{2}\right)^3\left(\frac{4\pi}{(2\pi)^3}\right)^2\left(-\frac{3\pi}{4}\right)\left(\frac{\tilde{B}(0)}{|s_0|}\right)^2
\left[\frac{1}{6\eta}+ \int_0^{+\infty} d\alpha\, \frac{\sin^2(|s_0|\alpha)}{3\cosh^2\alpha}
+O(\eta)\right].
\ee
The second piece is transformed into
\be
E_\eta^{(2)} = 3  \left(\frac{4\pi}{(2\pi)^3}\right)^2\left(\frac{\tilde{B}(0)}{|s_0|}\right)^2 \frac{3}{4}
\int_0^{+\infty}\!\!\!\! d\alpha \int_0^{+\infty}\!\!\!\! d\beta\, 
\sin(|s_0|\alpha) \sin(|s_0|\beta) e^{-\eta(\alpha+\beta)}
\ln\frac{\frac{3}{4}+\sinh^2\alpha+\sinh^2\beta+\sinh\alpha \sinh\beta}
{\frac{3}{4}+\sinh^2\alpha+\sinh^2\beta-\sinh\alpha \sinh\beta}.
\ee
Calculation of this double integral looks hopeless. However with the natural change
of variables
\bea
\label{eq:cvxy1}
\alpha &=& \frac{y+x}{2} \\
\beta &=& \frac{y-x}{2},
\label{eq:cvxy2}
\eea
of Jacobian equal to $1/2$, one can use the magic identity
\be
\frac{3}{4}+\sinh^2\alpha+\sinh^2\beta+\sinh\alpha \sinh\beta =
\left(\frac{1}{2}+\cosh x\right)
\left(-\frac{1}{2}+\cosh y\right).
\label{eq:magic}
\ee
Using the fact that the integrand is a symmetric function of $\alpha$ and $\beta$,
we can restrict the integration domain to $\alpha\geq\beta$ that is $x\geq 0$ so that,
with the well known relation $\sin a\sin b = [\cos(a-b)-\cos(a+b)]/2$, we obtain
\be
E_\eta^{(2)} = 3  \left(\frac{4\pi}{(2\pi)^3}\right)^2\left(\frac{\tilde{B}(0)}{|s_0|}\right)^2 \frac{3}{4}
\int_0^{+\infty}\!\!\!\! dx \int_{x}^{+\infty}\!\!\!\!  dy\, 
\frac{1}{2}\left[\cos(|s_0|x)-\cos(|s_0|y)\right] e^{-\eta y}
\left[
\ln\left(\frac{\cosh x+1/2}{\cosh x -1/2}\right)+\ln\left(\frac{\cosh y-1/2}{\cosh y +1/2}\right)
\right].
\ee
Since this integrand is now a sum of factorized terms, one of the integrals may be
calculated (in some cases, one needs to exchange the order of integration
over $x$ and $y$). We are left with a single integration, in which we may take
the limit $\eta\to 0^+$:
\be
E_\eta^{(2)} = 3  \left(\frac{4\pi}{(2\pi)^3}\right)^2\left(\frac{\tilde{B}(0)}{|s_0|}\right)^2 \frac{3}{8\eta}
\int_0^{+\infty} dx\, \cos(|s_0|x)\ln\left(\frac{\cosh x+1/2}{\cosh x -1/2}\right)
+O(\eta).
\ee
The third piece is transformed into
\begin{multline}
E_\eta^{(3)} = 3  \left(\frac{4\pi}{(2\pi)^3}\right)^2\left(\frac{\tilde{B}(0)}{|s_0|}\right)^2
\left(-2\right) \times \\
\int_0^{+\infty}\!\!\!\! d\alpha \int_0^{+\infty}\!\!\!\! d\beta \,
\frac{\frac{9}{16}\sin(|s_0|\alpha)\sin(|s_0|\beta)\sinh\alpha\sinh\beta\, e^{-\eta(\alpha+\beta)}}
{(\frac{3}{4}+\sinh^2\alpha+\sinh^2\beta+\sinh\alpha \sinh\beta)
(\frac{3}{4}+\sinh^2\alpha+\sinh^2\beta-\sinh\alpha \sinh\beta)}.
\end{multline}
Using the change of variables (\ref{eq:cvxy1}), (\ref{eq:cvxy2})
and the magic relation (\ref{eq:magic}), we obtain the simpler form
\be
E_\eta^{(3)} = 3  \left(\frac{4\pi}{(2\pi)^3}\right)^2\left(\frac{\tilde{B}(0)}{|s_0|}\right)^2
\left(-2\right)
\int_0^{+\infty} \!\!\!\! dx \int_x^{+\infty} \!\!\!\! dy \,
\frac{9}{64}\,\frac{[\cos(|s_0|x)-\cos(|s_0|y)](\cosh y-\cosh x) e^{-\eta y}}
{(\cosh^2x-1/4)(\cosh^2y-1/4)}.
\ee
We can take directly the limit $\eta\to 0^+$ without producing diverging terms in $E_\eta^{(3)}$.
The integrand is a sum of factorized terms; when a term involves the factor $\cos(|s_0|x)$,
we calculate the integral over $y$; when a term involves the factor $\cos(|s_0|y)$,
we calculate the integral over $x$. We are thus left with single integration:
\be
E_\eta^{(3)}= 3 \left(\frac{4\pi}{(2\pi)^3}\right)^2\left(\frac{\tilde{B}(0)}{|s_0|}\right)^2
\left(-2\right)
\int_0^{+\infty} dx\, \frac{\pi\sqrt{3}}{16} \cos(|s_0|x) \frac{3-2\cosh x}{4\cosh^2x-1}
+O(\eta).
\ee

Finally, it remains to collect all the three pieces in $E_\eta$. The  terms proportional to $1/\eta$ 
in $E_\eta^{(1)}$ and $E_\eta^{(2)}$ 
can be checked to cancel exactly: one uses integration by part to eliminate
the logarithmic function in the integrand of the coefficient of $1/\eta$
in $E_\eta^{(2)}$, and the resulting integrals in $E_\eta^{(1)}$ and $E_\eta^{(2)}$ may
be calculated using (\ref{eq:Utile}). What remains is
\be
\lim_{\eta\to 0^+} E_\eta = -\frac{9\sqrt{3}}{128\pi^3} \left(\frac{\tilde{B}(0)}{|s_0|}\right)^2
\int_0^{+\infty} dx\, \left[
\frac{\sin^2(|s_0|x)}{\cosh^2 x}+
\cos(|s_0|x) \frac{1-\frac{2}{3}\cosh x}{\cosh^2x-\frac{1}{4}}\right].
\ee
This may be calculated using again (\ref{eq:Utile}). From the value of $\tilde{B}(0)$ given
in the Appendix \ref{app:efimov}, we finally obtain the expected result
\be
\lim_{\eta\to 0^+} E_\eta = -1.
\ee

\section{Momentum distribution asymptotics of an Efimov trimer in presence of a smooth regularisation}
\label{app:douce}

To understand the deviation between the true Efimov trimer energy $E_{\rm trim}=-\hbar^2\kappa_0^2/m$
and the value $\bar{E}$ predicted by a at first sight convincing application
of an energy-momentum relation, see (\ref{eq:elam}), we suggested in the main text
to apply the regularisation procedure (\ref{eq:replace}) depending on a parameter $\eta$
that one eventually sets to $0^+$. Here we give the expressions
of the coefficients $C_\eta$ and $\bar{D}_\eta$ of the asymptotics (\ref{eq:nketaasymp}) of the corresponding
momentum distribution $n_\eta(k)$.

The calculations are similar to the one of Appendix \ref{app:efimov}. We shall need simply
the asymptotic behavior of $n_\eta(k)$ after having averaged out the $O(1/k^5)$ contributions
involving oscillating terms in $k^{\pm 2i|s_0|}$:
\be
\langle n_\eta(\kk)\rangle = 
\frac{C_\eta}{k^4} + \frac{\bar{D}_\eta}{k^5} e^{-2\eta\ln(\sqrt{3}k)}
+ O(1/k^6).
\ee
The coefficient $\bar{D}_\eta$ vanishes for $\eta\to 0^+$ but it is non zero
for $\eta>0$:
\be
\bar{D}_\eta = \frac{9}{2\pi^2} |\mathcal{N}_\psi|^2 4\pi^5
\left[\frac{\pi}{4\sqrt{3}} + I_\eta + J_\eta + K_\eta\right]
\ee
with
\bea
I_\eta &=& \int_0^{+\infty} dq\, \frac{-(1+q^2)}{1+q^2+q^4} \, e^{-2\eta\ln q}
\\
J_\eta &=& \int_0^{+\infty} dq\, \frac{q^{s_0}+q^{-s_0}}{1+q^2+q^4} e^{-\eta\ln
q} \\
K_\eta &=& \int_0^{+\infty} dq\,
\frac{8q}{1+q^2} \frac{e^{-\eta \ln\left(\frac{1+q^2}{4}\right)}}{(q^2+3)^2}
\int_0^{2q/(1+q^2)} dv\, \frac{e^{-\eta\ln\sqrt{1-v^2}}}{1-v^2}
\left[\left(\frac{1+v}{1-v}\right)^{s_0/2} + \mbox{c.c.}\right] 
\eea
The contributions $I_\eta$, $J_\eta$ and $K_\eta$ originate respectively
from the bits $n_{II}$, $n_{III}$ and $n_{IV}$ in (\ref{eq:decomp}).
Their expressions allow a numerical calculation of $\bar{D}_\eta$ if desired.
At first sight, the calculation of $K_\eta$ is more difficult because it involves
a double integration; since the inner integral is from $0$ to a function of $q$,
it may however be advanced step by step in parallel with the evaluation of the outer
integral, so that the complexity remains the same as for a single integral.
Anyway, such a numerical calculation for a finite $\eta$ is not necessary,
what matters is the knowledge of the derivative $d\bar{D}_\eta/d\eta$ in $\eta=0$,
see (\ref{eq:tec}). We obtain for the derivatives:
\bea
\frac{dI_\eta}{d\eta}|_{\eta=0}  & =& 0 \\
\frac{dJ_\eta}{d\eta}|_{\eta=0}  &= & \int_0^{+\infty} dq\, 
(-\ln q)\, \frac{q^{s_0}+q^{-s_0}}{1+q^2+q^4} \\
&=& -0.2456950243427\ldots \\
\frac{dK_\eta}{d\eta}|_{\eta=0}  &= &
-\int_0^{+\infty} dq\, \frac{16q}{1+q^2}
(q^2+3)^{-2}
\ln\left(\frac{1+q^2}{4}\right)
|s_0|^{-1}
\sin\left[|s_0| \ln \left(\frac{1+q}{|1-q|}\right)\right]\nonumber \\
&-& \int_0^{+\infty} dq\, 2 \ln\left(\frac{1+q^2}{|1-q^2|}\right)
\left[\left(\frac{1+q}{|1-q|}\right)^{s_0}+\mbox{c.c.}\right]
\left[\frac{1}{2(q^2+3)}
+\frac{\ln\left[\frac{2(1+q^2)}{q^2+3}\right]}{1-q^2}\right] 
\label{eq:sis}\\
&=& 0.04934911139697\ldots
\eea
Remarkably, in (\ref{eq:sis}) a single integral is obtained, after integration
by part, taking the derivative of the bit $\int_0^{2q/(1+q^2)}dv\ldots$.
All the integrals may be calculated numerically to a high precision with Maple,
resulting in
\be
\frac{d\bar{D}_\eta}{d\eta}|_{\eta=0} = -8.3720476291291\ldots \times\kappa_0^2
\ee
and (\ref{eq:ebar_analy}).

\bibliography{felix}
\end{document}